\documentclass[twocolumn,astrosymb]{aastex631}

\newcommand{\aicmod}{\mathrm{AIC_{p}}}
\newcommand{\ngh}{n_\mathrm{GH}}
\newcommand{\fsmooth}{\alpha_S}
\newcommand{\rhostar}{\mathrm{\rho_{*}}}
\newcommand{\rhodm}{\mathrm{\rho_{DM}}}
\newcommand{\mbh}{\mathrm{M_{\bullet}}}
\newcommand{\rhonorm}{\mathrm{\rho_{0}}}

\newcommand{\slpin}{\mathrm{\gamma_{in}}}
\newcommand{\slpout}{\mathrm{\gamma_{out}}}
\newcommand{\sclrad}{r_{\mathrm{s}}}
\newcommand{\qdm}{\mathrm{q_{DM}}}
\newcommand{\reff}{r_\mathrm{e}}

\newcommand{\centredis}{\Delta r_\mathrm{sky,M87}}
\newcommand{\dmfrac}{\mathrm{f_{DM}}}
\newcommand{\paperrefeDARKMATTER}{VW--\MakeUppercase{\romannumeral 2}}
\newcommand{\paperrefeDARKMATTERsp}{VW--\MakeUppercase{\romannumeral 2} }
\newcommand{\paperrefemethods}{LT}

\shortauthors{Lipka et al.}

\begin{document}

\title{The VIRUS-dE Survey I: Stars in dwarf elliptical galaxies - 3D  dynamics and radially resolved stellar initial mass functions}

\correspondingauthor{Mathias Lipka}
\email{mlipka@mpe.mpg.de}

\author[0000-0002-0730-0351]{Mathias Lipka}
\affiliation{Max-Planck-Institut für extraterrestrische Physik, Giessenbachstrasse, D-85748 Garching}
\affiliation{Universitäts-Sternwarte München, Scheinerstrasse 1, D-81679 München, Germany}

\author[0000-0003-2868-9244]{Jens Thomas}
\affiliation{Max-Planck-Institut für extraterrestrische Physik, Giessenbachstrasse, D-85748 Garching}
\affiliation{Universitäts-Sternwarte München, Scheinerstrasse 1, D-81679 München, Germany}

\author[0000-0003-0378-7032]{Roberto Saglia}
\affiliation{Max-Planck-Institut für extraterrestrische Physik, Giessenbachstrasse, D-85748 Garching}
\affiliation{Universitäts-Sternwarte München, Scheinerstrasse 1, D-81679 München, Germany}

\author[0000-0001-7179-0626]{Ralf Bender}
\affiliation{Max-Planck-Institut für extraterrestrische Physik, Giessenbachstrasse, D-85748 Garching}
\affiliation{Universitäts-Sternwarte München, Scheinerstrasse 1, D-81679 München, Germany}

\author[0000-0002-7025-6058]{Maximilian Fabricius}
\affiliation{Max-Planck-Institut für extraterrestrische Physik, Giessenbachstrasse, D-85748 Garching}
\affiliation{Universitäts-Sternwarte München, Scheinerstrasse 1, D-81679 München, Germany}

\author[0000-0001-6717-7685]{Gary J. Hill}
\affiliation{McDonald Observatory, The University of Texas at Austin, 2515 Speedway Boulevard, Stop C1402, Austin, TX 78712, USA}
\affiliation{Department of Astronomy, The University of Texas at Austin, 2515 Speedway Boulevard, Stop C1400, Austin, TX 78712, USA}

\author[0000-0002-9618-2552]{Matthias Kluge} 
\affiliation{Max-Planck-Institut für extraterrestrische Physik, Giessenbachstrasse, D-85748 Garching} 
\affiliation{Universitäts-Sternwarte München, Scheinerstrasse 1, D-81679 München, Germany}

\author[0000-0003-1838-8528]{Martin Landriau}
\affiliation{Lawrence Berkeley National Laboratory, 1 Cyclotron Rd, Berkeley 94720, CA, USA}

\author{Ximena Mazzalay}
\affiliation{Max-Planck-Institut für extraterrestrische Physik, Giessenbachstrasse, D-85748 Garching}

\author{Eva Noyola}
\affiliation{Department of Astronomy, The University of Texas at Austin, 2515 Speedway Boulevard, Stop C1400, Austin, TX 78712, USA}

\author[0000-0002-0621-6238]{Taniya Parikh}
\affiliation{Max-Planck-Institut für extraterrestrische Physik, Giessenbachstrasse, D-85748 Garching}

\author[0000-0003-4044-5357]{Jan Snigula}
\affiliation{Max-Planck-Institut für extraterrestrische Physik, Giessenbachstrasse, D-85748 Garching}
\affiliation{Universitäts-Sternwarte München, Scheinerstrasse 1, D-81679 München, Germany}

\begin{abstract}

We analyse the stellar structure of a sample of dwarf ellipticals (dE) inhabiting various environments within the Virgo cluster. Integral-field observations with a high spectral resolution allow us to robustly determine their low velocity dispersions ($\sim25$ km~s$^{-1}$) and higher-order kinematic moments out to the half-light radius. We find the dEs exhibit a diversity in ages with the younger dEs being less enhanced than the older, suggesting a complex star formation history for those dEs that recently entered Virgo while others have been quenched shortly after reionization. Orbit-superposition modeling allowed us to recover viewing angles, stellar mass-to-light ratios (with gradients), as well as the intrinsic orbit structure. We find that the angular momentum of the dEs is strongly suppressed compared to ordinary early-type galaxies and correlates with the environment. Flattened dEs are so because of a suppressed kinetic energy perpendicular to their equatorial plane. Combining population and dynamical modeling results, we find an age-dependent stellar initial mass function (IMF) or, alternatively, evidence for a more extended star formation history for those galaxies that have had higher initial mass and/or inhabited lower density environments. dEs appear to have a spatially homogeneous stellar structure but the state they were `frozen' in as they stopped forming stars varies dramatically according to their initial conditions. 

\end{abstract}


\section{Introduction}
\label{sec:intro}
Dwarf galaxies with $\log_{10}(L_{B}/L_{\sun})\in [8-10]$ dominate the galaxy census in the local Universe \citep[][]{Ohlson_2024}. Within dense galaxy clusters, like Virgo or Fornax, the \textit{quiescent} dwarf elliptical galaxies (dE) in this luminosity range are the most abundant type of dwarfs \citep[][]{Sandage_1985,Paudel_2023}. Still their origin, evolution, cosmological role and relation to other galaxy classes still poses many questions. Like their namesakes, the more massive elliptical galaxies, dEs appear to be well approximated by featureless ellipsoidals without any substantial substructures which have ceased star formation lacking sufficient gas reservoirs to form new stars. Furthermore, similar to these giant early-type galaxies (ETGs) \citep[][]{Dressler_1980}, the dwarf early-types seem to follow a strong morphology--environment dichotomy: dEs are found predominately in dense cluster and group environments with fewer dEs interspersed in the field. In contrast, late-type dwarfs of comparable mass avoid these denser environments and inhabit almost exclusively lower density field environments \citep[][]{Binggeli_1987,Geha_2012}.

Substantial evidence suggests that dEs ($M_{B}\gtrsim-18$) are physically distinct from the brighter giant ETGs and should be thought of as a separate galaxy class with its own unique formation channel. For example, dEs differ in their location on the fundamental plane \citep[][]{Bender_1992}, have almost exponential light profiles \citep[][]{Kormendy_1985}, and when investigated in detail often exhibit substructure such as faint spiral arms, blue centers, or signs of tidal harassment \citep[][]{Jerjen_2000,Lisker_2006_a,Lisker_2007,Paudel_2014}. In this context dEs appear to be much more closely related to the smallest galaxies in the ETG sequence: the dwarf spheroidal galaxies (dSphs) with $M_{B}\gtrsim-13$ which are found as satellites of the Milky Way and M31 in the Local Group.

These discrepancies to the giant galaxies in the ETG sequence suggests a distinct formation scenario for dwarf galaxies: dEs similar to their even less massive associates dSphs, are believed to be the remnants of transformed late-type galaxies (LTG), which have lost most gas through quenching \citep[][]{Kormendy_2009}. In this context some of the dEs may be better classified as Sphs which form the extension of the S0-branch in parallel to their irregular dwarf progenitors \citep[][]{Kormendy_2012}. 

This transformation from late-type to early-type may have happened through a combination of processes induced by interactions with the local environment such as ram-pressure stripping \citep[][]{Gunn_1972,Lin_1983}, starvation \citep[][]{Larson_1980}, and galaxy harassment \citep[][]{Moore_1998}, while some of the smallest dwarf galaxies may even directly originate from tidal interactions of more massive galaxies \citep[][]{Barnes_1992,Yang_2014}. Which and by how much these processes are important is still a point of contention as it is difficult to disentangle their effects. Furthermore, the driving quenching mechanism may vary with the total mass, age and environment. 

To improve our understanding of where, when, and how dEs have formed their \textit{intrinsic} properties need to be studied in more detail. For example, little is known about the orbital composition of dwarf elliptical galaxies. Do they follow a similar correlation between flattening and anisotropy as intermediate-mass or very massive early-type galaxies do \citep{Cappellari_2007,Thomas_2009_b,Santucci_2022}? Likewise, we do not know much about the dark-matter (DM) content of these galaxies. Do they follow the scaling relations of disk galaxies \citep[][]{Kormendy_2016} or do they have denser DM halos, similar to ETGs \citep{Gerhard_2001,Thomas_2009}? Finally, new insight might come from studying the shape of the initial stellar mass function (IMF). In particular, comparison of the IMF of dEs to that of LTGs could shed light on the question whether the star-formation conditions in dEs and LTGs were different from the beginning or whether their evolutionary paths diverged only later after most of the stars had already formed.

This study is part of a pair of papers with the overarching goal to recover the intrinsic mass and kinematic structure of the dEs, investigate their formation, evolution, and relation to other galaxy classes and what it implies for cosmological structure formation in a broader context. We have targeted a sample of dEs in the intermediate magnitude range $M_{B}\sim -17$, i.e. fainter than giant ETGs but still brighter than dSphs. All of them are part of the Virgo cluster, but inhabit locally different environments: ranging from the very center, over sub-clumps, to the outskirts of the cluster. Primarily, we analysed spatially resolved spectra obtained for these dEs. Similar spectroscopic studies that are focused on cluster dEs, studying their stellar population properties and projected velocities and dispersions, have been conducted in the past \citep[e.g.][]{Geha_2002,van_Zee_2004,Paudel_2010,Rys_2013,Toloba_2014,Scott_2020,Bidaran_2020}. However, due to the low velocity dispersions and low surface brightnesses of dEs previous spectra often had too low resolution, covered the kinematics only along slits, or had insufficient signal-to-noise to infer higher-order kinematic moments beyond the rotation velocity and velocity dispersion. Here, instead, we analyse spectra obtained with the VIRUS-W spectrograph which achieves sufficient spectral resolution ($R=7900-9000$) and is able to exploit the full 2D kinematic information available on the sky. The high signal-to-noise of our data allows us to study the higher-order kinematic moments beyond rotation and velocity dispersion in an unbiased manner. This novel information is a requirement to go beyond an analysis of the on-sky structure of the dEs and infer their 3D structure with dynamical models. We use sophisticated, orbit-based dynamical models to recover the intrinsic structure of the galaxies (e.g. dark matter halos, 3D orbit structure, black holes). 

The current study presents the observational data we obtained and outlines the information extraction techniques we applied. In the second half of the paper we present the first results which are focused on the \textit{stellar} component of the dEs. Among other things we examine their stellar mass, stellar mass-to-light ratio gradients, projected and intrinsic kinematic structure, ages and metallicities as well as the form of the IMF. In a companion paper (in the following \paperrefeDARKMATTER) we will present the corresponding \textit{dark} components of the dEs, i.e. the dark matter halos and black holes, and discuss the implications of our results in the broader context of $\Lambda \rm CDM$ cosmology and galaxy evolution.  

The current paper is organized as follows: In Section~\ref{sec:data} we present the photometric and spectroscopic data sets and describe how we processed them to obtain the input for the dynamical models: the 3D luminosity distribution $\nu$ and the spatially resolved line of sight velocity distributions (LOSVDs). We explain our SSP modeling approach and we outline the dynamical modeling technique we employed, the \citet{Schwarzschild_1979} method, and motivate our choice of its implementation and sampling strategy. Then in Section~\ref{sec:empirics} we present the on-sky kinematic structure and SSP properties and investigate their relation to the cluster environment. The 3D stellar mass, mass-to-light ratio gradients, and intrinsic anisotropy structure that was inferred using the dynamical models are presented in Section~\ref{sec:dwarf_modelling}. Finally, in Sec.~\ref{subsec:SSP_comparision} we use both the dynamical and SSP constraints in combination to discuss their physical implications regarding their IMF and star-formation history. The paper concludes with a summary in Section~\ref{sec:conclusions}. In App.~\ref{append:kindata} we show an example of a typical LOSVD recovery from a VIRUS-W spectrum. We compile and compare existing kinematic and SSP results from the literature with our results in App.~\ref{append:Literature_comparison}. In App.~\ref{append:ML_check} and App.~\ref{append:IMF_metal} we discuss the robustness of the mass estimates from the SSP models and show a supplementary IMF-metallicity relation. In App.~\ref{append:galdiscussion} we discuss each dE individually based on the photometric and kinematic data we obtained. Two alternative Figures to those in Section~\ref{sec:empirics} are located in App.~\ref{append:Lambda}.

\section{Galaxy sample and data extraction}
\label{sec:data}
\begin{figure*}
	\centering
	\includegraphics[width=1.0\textwidth]{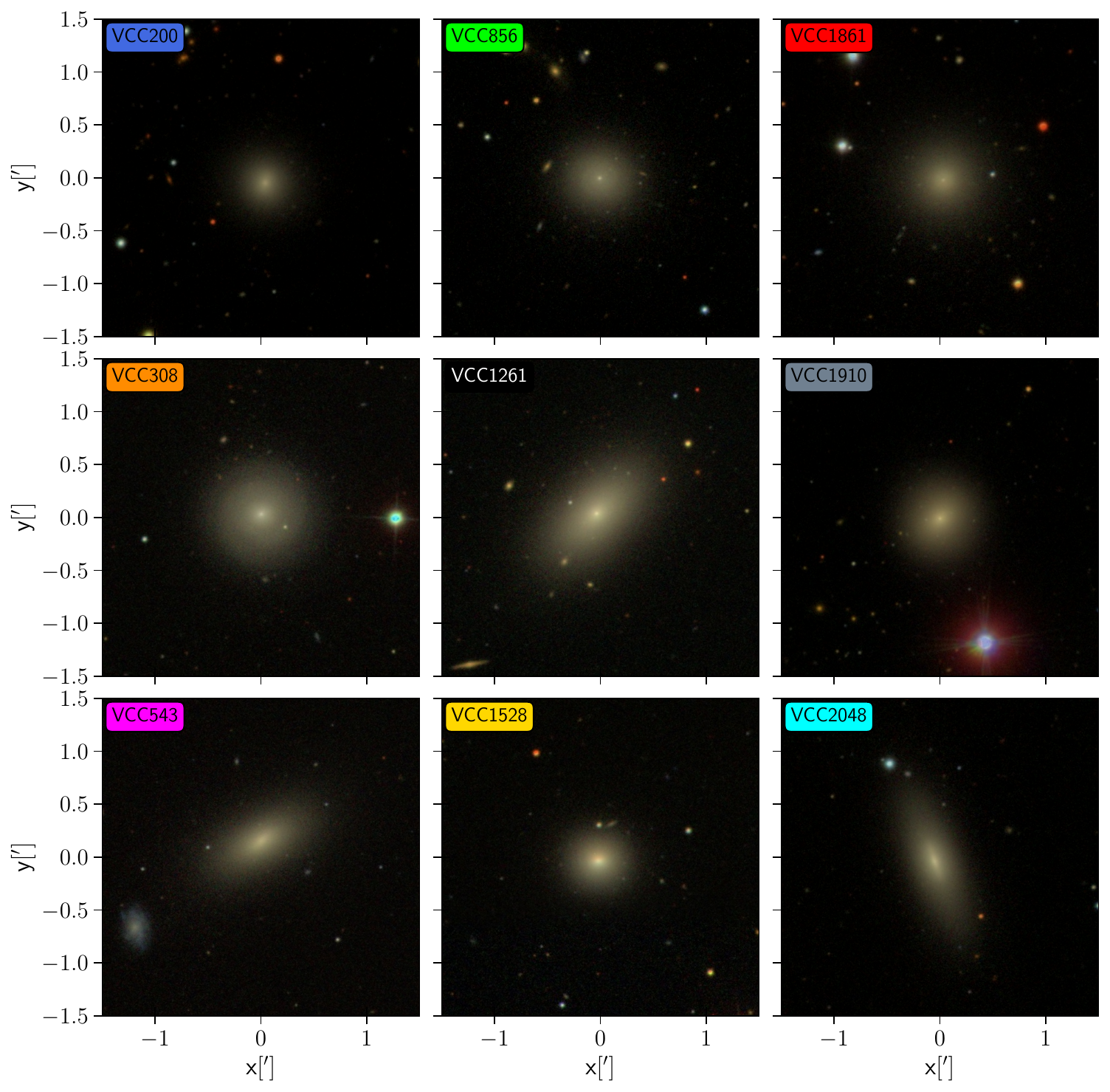}
    \caption{A imaging preview of the dE sample investigated in this study. Shown is the $g,r,i$ image for each galaxy based on publicly available SDSS data. We obtained the images with the help of \textit{Aladin} \citep[][]{Boch_2014}.}
    \label{fig:gallery}
\end{figure*}

We investigate a total of 9 dwarf elliptical (dE) galaxies located in the Virgo Cluster which occupy the small apparent magnitude range of $m_{B} \in [15.0,13.5]$ mag or in absolute magnitudes $M_{B} \in [-16.3,-17.8]$ mag. Fig.~\ref{fig:gallery} is a thumbnail gallery of the dE galaxies obtained from the $g,r,i$ images captured by the Sloan Digital Sky Survey \citep[][]{Ahn_2012}. We follow the general dE classification of \citet{Lisker_2007}, i.e. galaxies that are sometimes classified as dS0 or have faint spiral arms are also included in the dE class. The galaxies were chosen such that (i) the sample includes as many sub-classes of dEs as possible and (ii) different environments within Virgo are probed. Specifically, the sample includes the more common nucleated dE(N) and non-nucleated dE(nN) and the less common blue-centered dE(bc) and disky dE(di) \citep[][]{Lisker_2007}. This choice was done purposefully with the goal to identify possibly different formation and evolution mechanisms that may correlate with environment or morphological substructure. Notably missing are visibly merging or tidally disrupted dEs which are much rarer \citep[][]{Paudel_2023}. The sample dEs explore a range of $g$-band surface brightnesses from $\mu_{e}\sim22.2~\mathrm{mag}/\mathrm{arcsec}^2$ down to $\mu_{e}~\sim23.3\mathrm{mag}/\mathrm{arcsec}^2$ measured at the stellar effective radius $\reff$. The galaxy sizes range from $\reff= 0.8~\mathrm{kpc}$ to $\reff = 1.6~\mathrm{kpc}$ \citep[e.g.][]{Ferrarese_2006} which is fairly representative of dEs in our magnitude range. Our sample covers mostly the brighter end of the dE distribution \citep[e.g.][]{Kormendy_2012,Paudel_2023}, missing fainter and more diffuse dEs, and, notably, does not contain the even fainter ultra-diffuse galaxies (UDGs) which sometimes are regarded as their own subclass. 

Table~\ref{tab:galaxy_table} lists some of the basic galaxy properties we adopted for the dynamical modeling. The input data required for the construction of the dynamical models (Sec.~\ref{subsec:technique}) are the 3D stellar luminosity density $\nu$ and the spatially resolved line-of-sight velocity distributions (LOSVDs) of the stars. These need to be obtained from photometric and spectroscopic data sets respectively. In the following section we present the data sets we used and outline the procedure to retrieve the luminosity density $\nu$ (Sec.~\ref{subsec:photometry}) and the LOSVDs (Sec.~\ref{subsec:spectroscopy}). In Sec.~\ref{subsec:population_analysis} and Sec.~\ref{subsec:technique} we briefly describe the (stellar population and dynamical modeling) techniques we employed to obtain the information about the structure of the dwarf ellipticals. 

In App.~\ref{append:galdiscussion} we provide a summary of the morphological peculiarities of each galaxy. To ease the identification of individual galaxies we keep the same color-coding scheme throughout the paper (see Tab.~\ref{tab:galaxy_table}, Fig.~\ref{fig:kinmap}, Fig.~\ref{fig:ml_dyn_vs_ml_pop}, or Fig.~\ref{fig:kinematic_literature} for the connection between color and Virgo Cluster Catalog (VCC) labels). The basic properties we derive throughout this work are tabulated in Tab.~\ref{tab:results_table}. 

\definecolor{royalblue}{HTML}{4169E1}
\definecolor{darkorange}{HTML}{FF8C00}
\definecolor{fuchsia}{HTML}{FF00FF}
\definecolor{lime}{HTML}{00FF00}
\definecolor{black}{HTML}{000000}
\definecolor{gold}{HTML}{FFD700}
\definecolor{red}{HTML}{FF0000}
\definecolor{slategrey}{HTML}{708090}
\definecolor{aqua}{HTML}{00FFFF}

\begin{table*}
	\centering
	\caption{Basic properties of the galaxy sample. From left to right: The Virgo Cluster Catalog (VCC) identification number \citep[][]{Binggeli_1985}. The color identification used throughout this article. The morphological sub-classification following \citet{Lisker_2007} (bc: blue center, di: disk feature, N: classical dE with nucleus, nN: classical dE without nucleus). The distance adopted from \citet{Blakeslee_2009}, if there are no distances available (other than from redshifts) we use the average Virgo Cluster distance of 16.5 Mpc. The heliocentric velocity $cz$ we measured from the spectra. The projected distance to M87 (assuming $d=16.5 \rm Mpc$). The stellar mass we obtained. The photometric band we use for the modeling of the light distribution. Effective radius (in the $z$-band), Sersic-index, and position angle all adopted from \citep[cf.][]{Ferrarese_2006} who investigated all galaxies in our sample except for VC~308 (for which we consulted Hyperleda \citep[][]{Makarov_2014} instead). VCC~308 and VCC~1861 appear almost perfectly round, therefore a robust PA is not given.}
	\label{tab:galaxy_table}
	\begin{tabular}{ccccccccccc} 
    VCC ID & Color & Classification & $d$ [Mpc] & $cz$[km/s] & $\centredis$[Mpc] & $\log_{10}(M_{*}/M_{\sun})$ & Band & $\reff["]$ & $n_{s}$ & PA $[\degr]$ \\ \hline
     VCC 200  & \cellcolor{royalblue} & dE(N)  & 18.3 & 16   & 1.018 & 8.85 & F850LP & 13.12 & 1.933 & $-7$ \\
     VCC 308  & \cellcolor{darkorange}& dE(bc) & 16.5 & 1527 & 1.556 & 8.88 & F814W  & 11.40 & 1.340 & -  \\
     VCC 543  & \cellcolor{fuchsia}   & dE(nN) & 15.8 & 962  & 0.905 & 9.37 & F850LP & 18.29 & 1.716 & $-53$ \\ 
     VCC 856  & \cellcolor{lime}      & dE(di) & 16.9 & 1016 & 0.755 & 8.88 & F850LP & 16.70 & 1.317 & $80$ \\
     VCC 1261 & \cellcolor{black}     & dE(N)  & 18.2 & 1861 & 0.466 & 9.44 &\textit{V} & 20.13 & 2.135 & $-47$ \\
     VCC 1528 & \cellcolor{gold}      & dE(nN) & 16.3 & 1614 & 0.343 & 9.01 & F850LP & 9.88 & 2.101 & $84$ \\
     VCC 1861 & \cellcolor{red}       & dE(N)  & 16.1 & 636  & 0.795 & 8.88 & F850LP & 18.24 & 1.593 &  -  \\
     VCC 1910 & \cellcolor{slategrey} & dE(di) & 16.0 & 241  & 0.818 & 8.85 & F850LP & 12.01 & 1.564 & $-49$ \\ 
     VCC 2048 & \cellcolor{aqua}      & dE(di) & 16.5 & 1096 & 1.320 & 9.38 & \textit{V} & 12.64 & 1.973 & $19$ \\ 
     \hline
	\end{tabular}
\end{table*}
 
\begin{table*}
	\centering
	\caption{Some of the basic properties derived in this study (see text for details). Dark matter and super massive black hole properties will be tabulated in \paperrefeDARKMATTER. From left to right: The VCC identification of the galaxy. The stellar population age (brackets indicate that value is the \textit{average} of the two population ages we derived at around $r=2.5\arcsec$ and $r=7.5\arcsec$, respectively). Similarly the \textit{average} metallicity $[Z/\mathrm{H}]$, abundance ratio $\mathrm{[Mg/Fe]}$ and IMF parameter $\alpha_{\mathrm{IMF}}=\Upsilon_{\mathrm{dyn}}/\Upsilon_{\mathrm{Kroupa}}$. The population values for VCC~1910 are highlighted with a star because they may be compromised (cf. Sec.~\ref{subsec:population_analysis}). The angular momentum parameters $\lambda_{e/2}$ and $\lambda_{e}$ within half and one stellar effective radius respectively. Values with a star had to be extrapolated beyond the last kinematic data point. The total stellar angular momentum $j_{*}$. The inclination $i$ derived via dynamical modeling. The ellipticity $\epsilon_{e/2}$ of the isophotes within $\reff/2$. The average cylindrical anisotropy parameters $\overline{\beta_{z}}$ and $\overline{\gamma}$.}
	\label{tab:results_table}
	\begin{tabular}{cccccccccccc} 
     VCC ID   & $\left<\mathrm{Age[Gyr]}\right>$ & $\left<Z/\mathrm{H}\right>$ & $\left<\mathrm{[Mg/Fe]}\right>$ & $\left<\alpha_{\mathrm{IMF}} \right>$ & $\lambda_{e/2}$ & $\lambda_{e}$  & $\log \left( j_{*}[\mathrm{kpc~ km~s^{-1}} ] \right)$ & $i[\degr]$ & $\epsilon_{e/2}$ & $\overline{\beta_{z}}$ & $\overline{\gamma}$ \\ \hline
     VCC 200  & 11.6 & $-0.67$ & 0.23 & 0.485 & 0.208 & 0.211 & 0.957 & 90 & 0.189 & $-0.038$ & $-0.193$ \\
     VCC 308  & 4.3  & $-0.39$ & 0.11 & 0.775 & 0.222 & 0.233 & 1.014 & 46 & 0.138 & 0.019 & 0.060 \\
     VCC 543  & 7.3  & $-0.39$ & 0.24 & 1.319 & 0.375 & $0.352^*$ & 1.488 & 67 & 0.456 & 0.342 & $-0.005$ \\
     VCC 856  & 8.8  & $-0.55$ & 0.25 & 0.574 & 0.330 & $0.301^*$ & 1.719 & 32 & 0.093 & 0.333 & $-0.156$ \\
     VCC 1261 & 7.5  & $-0.37$ & 0.15 & 0.767 & 0.046 & $0.041^*$ & 1.085 & 65 & 0.275 & 0.410 & $-0.075$ \\
     VCC 1528 & 6.0  & $-0.24$ & 0.16 & 0.897 & 0.028 & 0.027 & 0.307 & 90 & 0.229 & 0.030 & 0.013 \\
     VCC 1861 & 9.6  & $-0.25$ & 0.16 & 0.492 & 0.109 & 0.113 & 1.176 & 44 & 0.053 & 0.020 & 0.095 \\
     VCC 1910 & $3.6^*$  & $0.25^*$ & $0.12^*$ & $1.84^*$ & 0.126 & 0.101 & 1.020 & 90 & 0.199 & 0.135 & 0.012 \\
     VCC 2048 & 3.6  & $-0.20$ & 0.22 & 2.80  & 0.256 & $0.225^*$ & 1.366 & 70 & 0.554 & 0.372 & $-0.401$ \\ 
     \hline
	\end{tabular}
\end{table*}

\subsection{Photometry \& Deprojection}
\label{subsec:photometry}
To obtain a model for the deprojected luminosity distribution $\nu$ we draw on publicly available \textit{HST} data. For the majority of the galaxies we use photometry observed with ACS/WCS in the F850LP and F475W filters calibrated in the AB system \citep[cf.][]{Sirianni_2005}. One galaxy, VCC 308, has no recorded ACS data, and instead we used WFPC2 observations available in F814W and F555W filters. In the following we will abbreviate these two filters with their very similar counterparts in the Sloan Digital Sky Survey (SDSS) filter system: The \textit{g}- and \textit{z}-band respectively. The \textit{HST} data are not always as deep as one may wish, resulting in deprojections that must be extrapolated to larger radii to cover the full radial extent of a typical orbit model. Therefore, in the case of VCC 1261 and VCC 2048, we chose to re-utilize the more extended photometry of \citet{Kormendy_2009} and \citet{Kormendy_2012} instead of the ACS data from the archives. These isophotes are calibrated in the \textit{V}-band and were extracted using a combination of ACS and SDSS images.

Since the field of view (FoV) of ACS is significantly larger than the size of the dEs, we were able to estimate the background sky value within the science image by calculating the median count in several sky boxes around the galaxies. A good sky estimation is essential for the recovery of the dark matter halos of the dEs. Since the outskirts of the dwarfs are barely brighter than the typical sky background, a bad subtraction may significantly distort the shape and slope of the resulting luminosity distribution $\nu$. In that case dark mass components of the model will have to compensate for any missing or excessive luminous mass in order to reproduce the correct combined gravitational potential. Furthermore an accurate recovery of $\nu$ is important because it serves as a boundary constraint for the stellar mass density in the dynamical models (Sec.~\ref{subsec:technique}).

After subtracting the sky and masking problematic regions we performed elliptical isophote fits for each galaxy\footnote{We implemented the isophote fitting using python, the code made use of astropy routines \citep[][]{astropy_2022}. While the isophote models allowed deviations from an ellipse, the $a_4$, $a_6$ profiles were often quite noisy but consistent with zero for the majority of the dEs.}. Where available we compared our isophotes with previous measurements \citep{Ferrarese_2006} and confirmed that we were reliably reproducing them. 

The dynamical models we employ are axisymmetric and therefore they require axisymmetric luminosity distributions that can reproduce the observed photometry. We deprojected each galaxy using the Metropolis-Algorithm of \citet{Magorrian_1999}, which allows us to explore the full range of axisymmetric solutions for any given inclination $i$. For our sample, external inclination estimates (e.g. using gas disks) are not possible and we need to probe different inclinations in the dynamical modeling itself. In practice one is limited to testing a small number of deprojections otherwise the modeling quickly becomes computationally unfeasible. To make the modeling as efficient as possible we decided to choose the inclinations $i$ such that the corresponding intrinsic flattening $q(i)=b/a$ of the resulting luminosity densities are (roughly) spaced linearly in the stellar axis ratio $q$. This maximizes the \textit{intrinsic} physical differences of the densities rather than the absolute difference in the viewing angle itself. The strategy to sample the intrinsic shapes instead of viewing angles is not novel and has found frequent use in triaxial modeling \citep[][]{Bosch_2010,Walsh_2012,Zhu_2018,Jin_2019,Poci_2019}. For a thorough discussion of this sampling choice see \citet[][]{Bosch_2009,Quenneville_2022}. We also took advantage of the fact that the range of possible viewing angles is a priori limited by the observed photometry \citep[e.g.][]{deNicola_2020}. In the axisymmetric case this means we only need to sample the inclination starting from some minimum allowed inclination $i_{\mathrm{min}}$ (consistent with a flat disk). 

It is well established that axisymmetric deprojections generally do not have a unique solution for a given viewing angle unless the inclination is exactly edge-on \citep[cf.][]{Rybicki_1987} meaning different boxy and disky deprojections can result in the same projected photometry \citep[cf.][]{Gerhard_1996}. We decided to probe \textit{one} deprojection per inclination as the dynamical differences between boxy and disky deprojections for our dwarf galaxies are typically much smaller than they are between deprojections at different inclinations. The deprojections were chosen to be sufficiently smooth and close to elliptical while still fitting the data adequately.

Table~\ref{tab:galaxy_table} shows the filter band we worked with to obtain the deprojections for the dynamical modeling. Throughout this paper we state all mass-to-light ratios, luminosities, etc. in this corresponding band.

\subsection{Spectroscopy \& LOSVD extraction}
\label{subsec:spectroscopy}
We took the spectra of the dwarf galaxies with the integral-field-unit (IFU) spectrograph VIRUS-W mounted to the Harlan J. Smith telescope at the McDonald Observatory. Table~\ref{tab:obs_table} summarises the various observation runs. The 267 fibres of VIRUS-W, which each cover a 3.2\arcsec diameter on the sky, combine to a field of view (FoV) of 105\arcsec x 55\arcsec. Since the velocity dispersion of dEs is generally very low, the instrument was operated in its high resolution mode achieving a spectral resolution of $R=7900$ to $9000$ (or 14 km~s$^{-1}$ to 16 km~s$^{-1}$) within the optical wavelength range from 4850\r{A} to 5475\r{A} \citep{Fabricius_2008,Fabricius_2012}. In App.~\ref{append:Literature_comparison} we demonstrate that such a high resolution is essential to study the low dispersions of the dEs. Previous studies with lower resolutions $R<5000$ are often biased towards too high dispersions (in some cases by up to 50\%). 

For each galaxy we obtained multiple dithered exposures with a median seeing of 2.0\arcsec. The offsets between the different exposures in the dither pattern range from 1.7\arcsec to 3.7\arcsec. The small size of the dEs compared to the FoV allowed us to perform the sky correction using the science frames themselves. To reduce the data we used a pipeline based on the `Cure' and `Fitstools' package which was designed for the HETDEX project \citep[][]{Goessl_2002,Hill_2004,Goessl_2006,Hill_2021}.

\begin{table*}
	\centering
	\caption{Basic Information about IFU observation runs which we attained with the Harlan J. Smith telescope at the McDonald Observatory (Sec.~\ref{subsec:spectroscopy}). From left to right: The Virgo Cluster Catalog (VCC) identification. The month and year of the observation runs. The total exposure time. The total number of Voronoi-binned spectra that survived our restrictions (cf. Sec.~\ref{subsec:spectroscopy}) and were dynamically modelled. The mean Signal-to-noise ratio $S/N$ of said Voronoi bins.}
	\label{tab:obs_table}
	\begin{tabular}{ccccc} 
     VCC ID & Runs (Month and year) & Total exposure time[min] & $N_{\mathrm{bin}}$ & $S/N_{\mathrm{bin}}$ \\ \hline
     VCC 200  & Apr 2016, May 2017           & 1620 & 18 & 21.2 $\pm$ 7.1 \\
     VCC 308  & Feb 2018, Mar 2018, Apr 2018 & 990  & 15 & 17.8 $\pm$ 3.3 \\
     VCC 543  & Feb 2018                    & 1410 & 16 & 27.0 $\pm$ 6.8 \\ 
     VCC 856  & Feb 2014, Mar 2014          & 1050 & 18 & 21.9 $\pm$ 4.8 \\
     VCC 1261 & Feb 2013, Apr 2013          & 630  & 45 & 33.2 $\pm$ 7.6 \\
     VCC 1528 & Mar 2018                    & 450  & 16 & 19.6 $\pm$ 5.1 \\
     VCC 1861 & Mar 2014, Apr 2016          & 1050 & 29 & 22.0 $\pm$ 6.0 \\
     VCC 1910 & Jun 2012                    & 510  & 26 & 25.0 $\pm$ 7.0 \\
     VCC 2048 & Apr 2012, Feb 2013          & 1050  & 36 & 42.9 $\pm$ 9.0 \\
     \hline
	\end{tabular}
\end{table*}

\begin{figure*}
	\centering
	\includegraphics[width=1.0\textwidth]{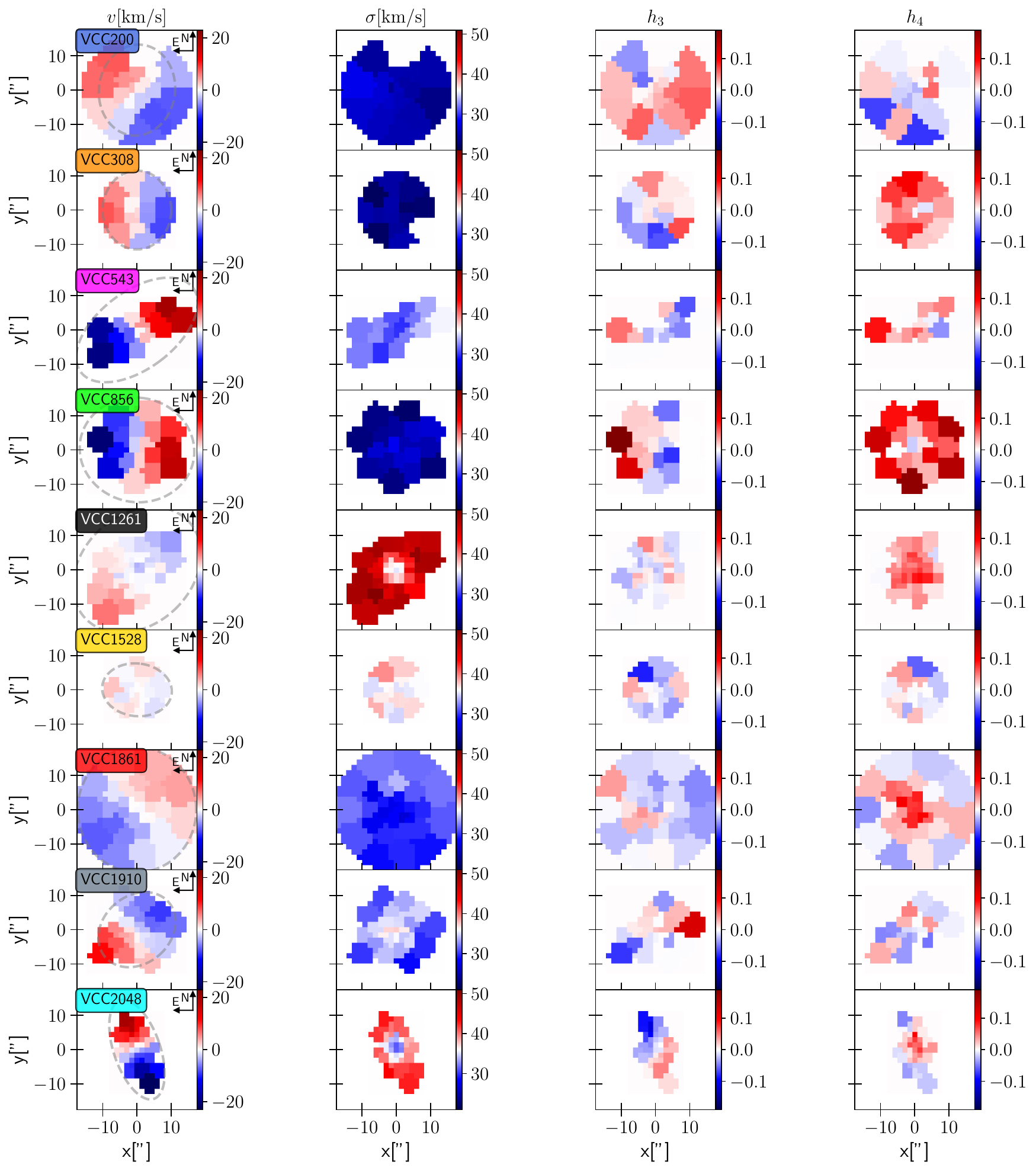}
    \caption{The kinematic maps of the LOSVDs approximated by Gauss--Hermite expansion with fixed $\ngh=4$, i.e. this does not show potentially higher order deviations. In a few bins the recovered LOSVDs are essentially Gaussian because the optimum $\ngh$ was 2. This makes them appear white in the $h_{3,4}$-maps (the northern most bin of VCC 200 was excluded due to its poor spectrum). We use a \textit{common} spatial and color scale (see colorbars on the right sides of the panels) to enable an easy comparison between \textit{different} sample galaxies. Spatial variations within a given galaxy are more discernible in the radial profiles (Fig.~\ref{fig:kinprofiles}). 
    In the velocity panels we indicate the North-East directions and the photometric $1.0~\reff$ aperture (\textit{dashed ellipse}).}
    \label{fig:kinmap}
\end{figure*}

From the dithered exposures a 3D data cube was generated, forming a regular grid of 1.6\arcsec pseudo spaxels. We binned the pixels further using Voronoi tesselation \citep[][]{Cappellari_2003} with the goal of obtaining spatial bins with an approximately uniform and sufficiently large signal-to-noise ratio $S/N$. In the case of our dwarf galaxies this proves to be difficult because they are relatively small compared to the size of the fibres which leads to a relatively large bin-to-bin $S/N$ gradient even after the Voronoi Binning. We found that bins with a $S/N \lesssim 15$ rarely provide stable and reliant LOSVD shapes and exclude them from the further analysis. The mean $S/N$ of the useful Voronoi bins is listed in Table~\ref{tab:obs_table}. The stellar kinematics cover one effective radius $\reff$, the sample median FoV size being $1.02~\reff$ (see also Fig.~\ref{fig:kinmap}). 

We derived LOSVDs from the binned spectra using the spectral-fitting code WINGFIT (Thomas et al. in prep.) which allows both a fully non-parametric and also a Gauss--Hermite description of the LOSVDs (see below). The LOSVDs are extracted from each of the Voronoi binned spectra by convolving a model of the LOSVD with a weighted sum of stellar templates. To match the high-resolution of VIRUS-W we used the ELODIE library of template stars in its low resolution version as they cover wavelengths from 3900\r{A} to 6800\r{A} with a resolution of $R=10000$ \citep{Prugniel_2001,Prugniel_2007}. The library consists of spectra including stellar atmosphere parameters with temperatures $T \in [3000K, 60000K]$, surface gravity $\log \left(g\right) \in [-0.3,5.9]$, and $[\mathrm{Fe/H}]\in [-3.2,1.4]$. Individual abundance ratios are not resolved in the library but expected to match the solar neighbourhood patterns. This may result in template mismatch if both metallicity and e.g. $[\alpha/\mathrm{Fe}]$ differ significantly from solar \citep[][]{Prugniel_2007_B}. Fortunately dEs are expected to have abundance patterns similar to those of LTGs \citep[][]{Sen_2018}. Reassuringly, even the most \textit{non}-solar stellar population in our sample, e.g. the central population of VCC~200 with $[\alpha/\mathrm{Fe}] = 0.29$ can still be fitted well using the ELODIE library (see for example Fig.~\ref{fig:spectrum_example}). To minimize potential effects of template mismatch on the LOSVD recovery we employed some of the strategies discussed by \citet{Mehrgan_2023}, which includes optimizing the number of polynomial orders for the continuum fit of the spectra and a preselection of template spectra. For all dwarf galaxies we only fit absorption features as we did not find any significant emission lines within the VIRUS-W spectral range. We inspected each spectrum individually and masked noise contaminated regions near the edges of the spectral wavelength ranges and near cosmic ray hits and sky lines that survived the data reduction pipeline. Fig.~\ref{fig:spectrum_example} in App.~\ref{append:kindata} shows a typical Voronoi-binned spectrum obtained with VIRUS-W and the corresponding fit we obtained with the convolved model. 

The aforementioned model of the LOSVD can be characterized in two different forms. The first option is a description using a suitable parametric function, typically a Gauss--Hermite series truncated at some highest non-zero order $\ngh$ \citep[cf.][]{Marel_1993}. The second option is using a more general non-parametric model with some smoothing penalty (e.g. a second derivatives penalty). However, both cases suffer from a similar problem: For Gauss--Hermite models it is unclear for which order $\ngh$ the series should be truncated. While for the non-parametric models it unclear how strong the smoothing penalty strength $\fsmooth$ should be. Both issues are essentially the same issue of finding the right balance between overfitting and underfitting the spectra: If $\ngh$ is too large (the penalty strength $\fsmooth$ too small) the model LOSVD can become arbitrarily complicated, and the model will be overfitting the noise in the spectra. In contrast, if $\ngh$ is too small (the smoothing penalty $\fsmooth$ too strong) the LOSVDs will be overly smooth and will not emulate the, perhaps more complex, structure of the underlying stellar motions. Often this problem is passed over by choosing a (hopefully) suitable $\ngh$ or $\fsmooth$ or calibrating these factors with mock simulations \citep[e.g.][]{Marel_1993,Ocvirk_2006,Liepold_2020,Falcon_Barroso_2021}. To avoid this we derived a generalized information criterion $\aicmod$ \citep[cf.][]{Thomas_2022} which provides a more systematic approach by minimizing the statistical information loss. Defined as $\aicmod=\chi^2+2m_{eff}$ it penalizes the goodness-of-fit $\chi^2$ a model achieves with its effective flexibility $m_{eff}$ \citep[see][]{Lipka_2021,Thomas_2022}. One finds the optimum amount of model complexity by comparing the $\aicmod$ of model with different degrees of smoothing $\fsmooth$ (or $\ngh$) with one another. The $\aicmod$ criterion is very general and allows us to compare the performance of a Gauss--Hermite model directly with non-parametric model descriptions, thus, allowing us decide what the best way to describe the LOSVD for each individual spectrum. 

With this tool in hand we decided to employ the following strategy: For every Voronoi binned spectrum we perform the kinematic extraction using both a Gauss--Hermite parametrization and a non-parametric model with various values of $\ngh$ and $\fsmooth$ respectively. We then calculate and compare the corresponding $\aicmod$ for each combination and choose the model that achieves the minimum $\aicmod$ to represent the LOSVD in the given Voronoi Bin \citep[cf.][]{Thomas_2022}. To provide a coherent input for the subsequent dynamical modeling we describe all models in the same velocity-binned form no matter if the optimum model is a Gauss--Hermite series or a non-parametric description. We estimated the errors in the velocity bins for both model types by re-fitting 100 Monte-Carlo realizations based on the flux noise (assumed to be gaussian) in each wavelength bin of the spectra. The errors in each velocity bin are quantified as the standard deviation of these realizations. However, since a Gauss--Hermite expansion (by construction) suppresses the error in the high-velocity wings (cf. Fig.~\ref{fig:LOSVD_comparison}) we decided on the conservative approach to always adopt the larger error estimate of the two (Gauss--Hermite error or non-parametric error) for each of the velocity bins.  

\begin{figure*}
	\centering
	\includegraphics[width=1.0\textwidth]{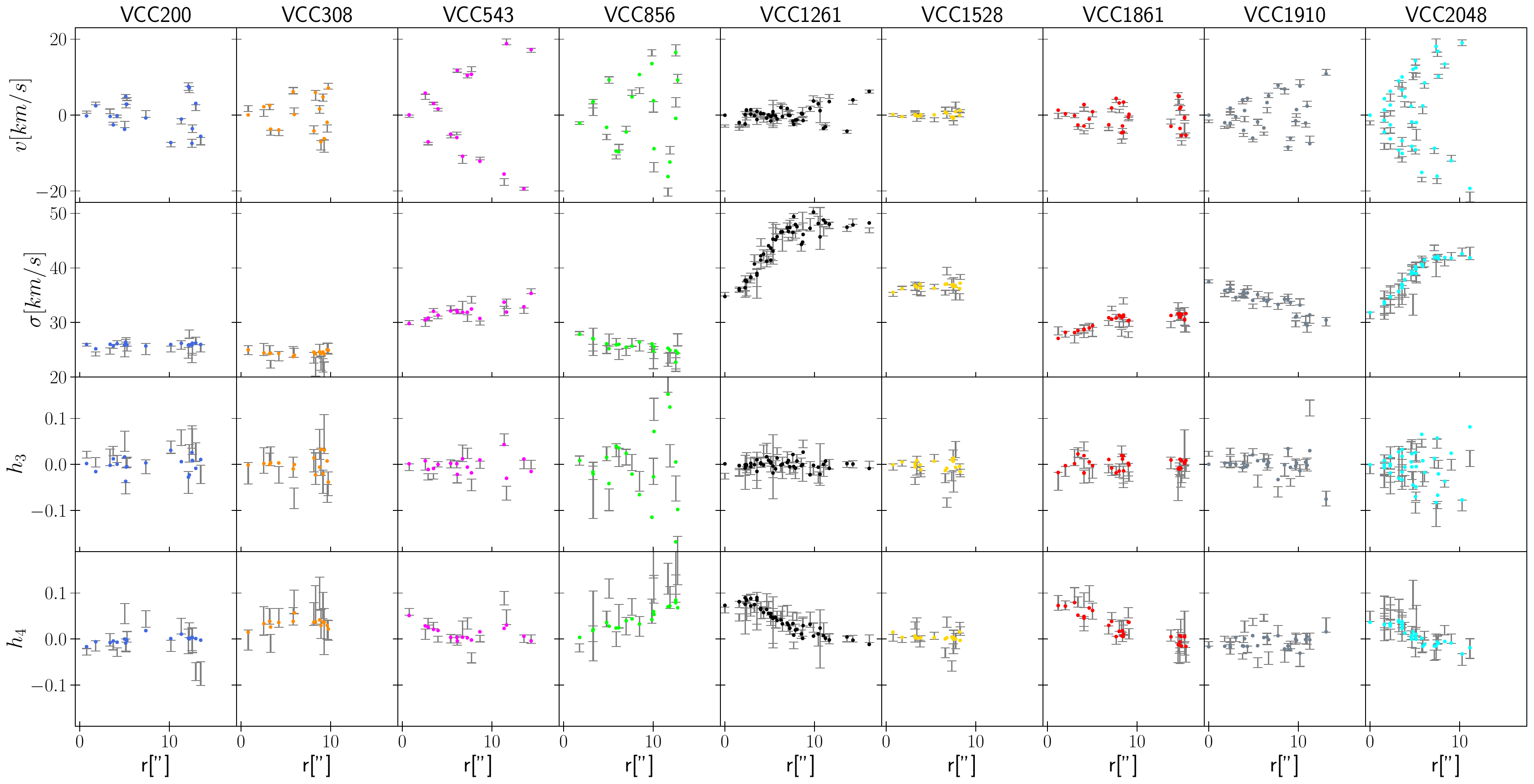}
    \caption{The radial profiles of the Gauss--Hermite representations (up to $h_4$) with $1\sigma$-errors as derived from the spectroscopic data (\textit{gray}) and the corresponding fit of the best dynamical orbit model we found for each galaxy (\textit{colored}). The $1\sigma$-errors are quantified by fitting Gauss--Hermites to 100 Monte-Carlo realizations on the velocity binned LOSVDs. The models fit the observed kinematics quite well considering deviations from axisymmetry in the data cannot be emulated by axisymmetric models.}
    \label{fig:kinprofiles}
\end{figure*}

When comparing the non-parametric LOSVDs with the corresponding ones derived from the optimized Gauss--Hermite expansion in the same Voronoi Bin we find that both are fairly consistent with each other, thus, strengthening our confidence in the derived kinematics. A comparison of two typical LOSVDs derived from the same spectrum once by fitting a Gauss--Hermite series and once by fitting a non-parametric description is shown in Fig.~\ref{fig:LOSVD_comparison}. Only in the high-velocity tails, where (depending on the maximum order $\ngh$ of the Gauss--Hermites) the LOSVD signal tends to be suppressed, we find minor differences between the models. This was already recognized by \citet{Mehrgan_2019}. Broadly speaking we find that in Voronoi bins with a relatively high $S/N$ the spectra are often preferred to be modelled non-parametrically, while in bins with lower $S/N$ the $\aicmod$ criterion tends to favor Gauss--Hermite expansions with a smaller maximum order $\ngh$ (sometimes even gaussian, i.e. $\ngh=2$). This is to be expected as $\aicmod$ is designed to prevent overfitting noise, i.e. if the noise is more dominant than any underlying LOSVD substructure it is preferable to obtain a smoother description of the main peak of the LOSVD than it is to fit the noise. In Figs.~\ref{fig:kinmap} and \ref{fig:kinprofiles} we show the Gauss--Hermite representations of the final LOSVD data if one approximates each LOSVDs a posteriori as a Gauss--Hermite expansion\footnote{We provide a supplementary table for each dE containing these Gauss--Hermite moments, their errors, and pixel location.} with a fixed $\ngh=4$. This allows for an easier comparison to literature values (App.~\ref{append:Literature_comparison}) which usually only consist of the first two moments: the rotation $v$ and velocity dispersion $\sigma$. We note that the full non-parametric LOSVDs that we used for the dynamical modeling might differ slightly since moments higher than $h_4$ are not displayed. Similarly, Figure~\ref{fig:kinprofiles} shows the Gauss--Hermite description of the LOSVD data when compared to the corresponding kinematics of the best dynamical model we found for each galaxy with the technique as described in Sec.~\ref{sec:dwarf_modelling}. Within the errors, the dynamical models we constructed emulate the observations well. 

\subsection{Single Stellar Population modeling}
\label{subsec:population_analysis}
The stellar population analysis generally requires a higher $S/N$ spectrum than the kinematic extraction we used to obtain the LOSVDs (Sec.~\ref{subsec:spectroscopy}). Since the VIRUS-W Voronoi bins we used for the kinematics have fairly low $S/N$ (Tab.~\ref{tab:obs_table}) we decided to re-bin the spectra into two radial annulli centered around $r=2.5\arcsec$ and $r=7.5\arcsec$. This allows reliable global estimates of the population quantities while still being able to notice a possible radial variation in the stellar population if present. We can also compare the mass-to-light ratio gradients obtained from the population analysis with the mass-to-light ratio gradients derived from the dynamical models and identify a possible radial variation of the IMF \citep[e.g.][]{Mehrgan_2024,Parikh_2024}. 

The majority of the Virgo dEs is known to host small blue nuclei with core sizes of $0.2\arcsec-0.4\arcsec$\citep[][]{Ferrarese_2006,Hamraz_2019}. \citet{Paudel_2011} separated the nuclei from the galaxy's main body and found that the nuclei are often composed of a much younger, more metal-rich stellar population compared to the main body of the galaxy. For their nucleated dEs (including VCC~308, VCC~856, VCC~1261, VCC~1861) they found the contribution of the nucleus to the total light within the centre $r\leq 0.375\arcsec$ usually does not exceed 50\%. Even though we mask the central $0.8\arcsec$ before adding the spectra (i.e. the nuclei are not part of the $r=2.5\arcsec$ annulus) the large VIRUS-W fibres with a diameter of $3.2\arcsec$ and the atmospheric seeing of $\sim 2.0\arcsec$ could lead to a partial contamination of the inner annulus with light from the nuclei. We estimate an upper bound for this contamination by comparing the \textit{total} luminosity of the blue nucleus to the luminosity of the galaxy main body within the circular area covered by the $2.5\arcsec$ aperture (incl. the masked centre). The luminosities are obtained by fitting a two-component model to the \textit{HST} photometry, where we assume a King model for the nucleus and a Sersic model for the galaxy main body. For our dE sample we find a median light contribution of 1.3\% in the \textit{z}-band and 1.6\% in the \textit{g}-band with a maximum contribution of 10\% for VCC~856. This implies that, even if we were not masking the central $0.8\arcsec$, the contamination from the nucleus to the galaxy light within $r=2.5\arcsec$ would be small to insignificant. Hence, our annulli are not affected much by the central nuclei but are well suited to focus on detecting gradients within the main body of galaxy.

\begin{figure}
	\centering
	\includegraphics[width=1.0\columnwidth]{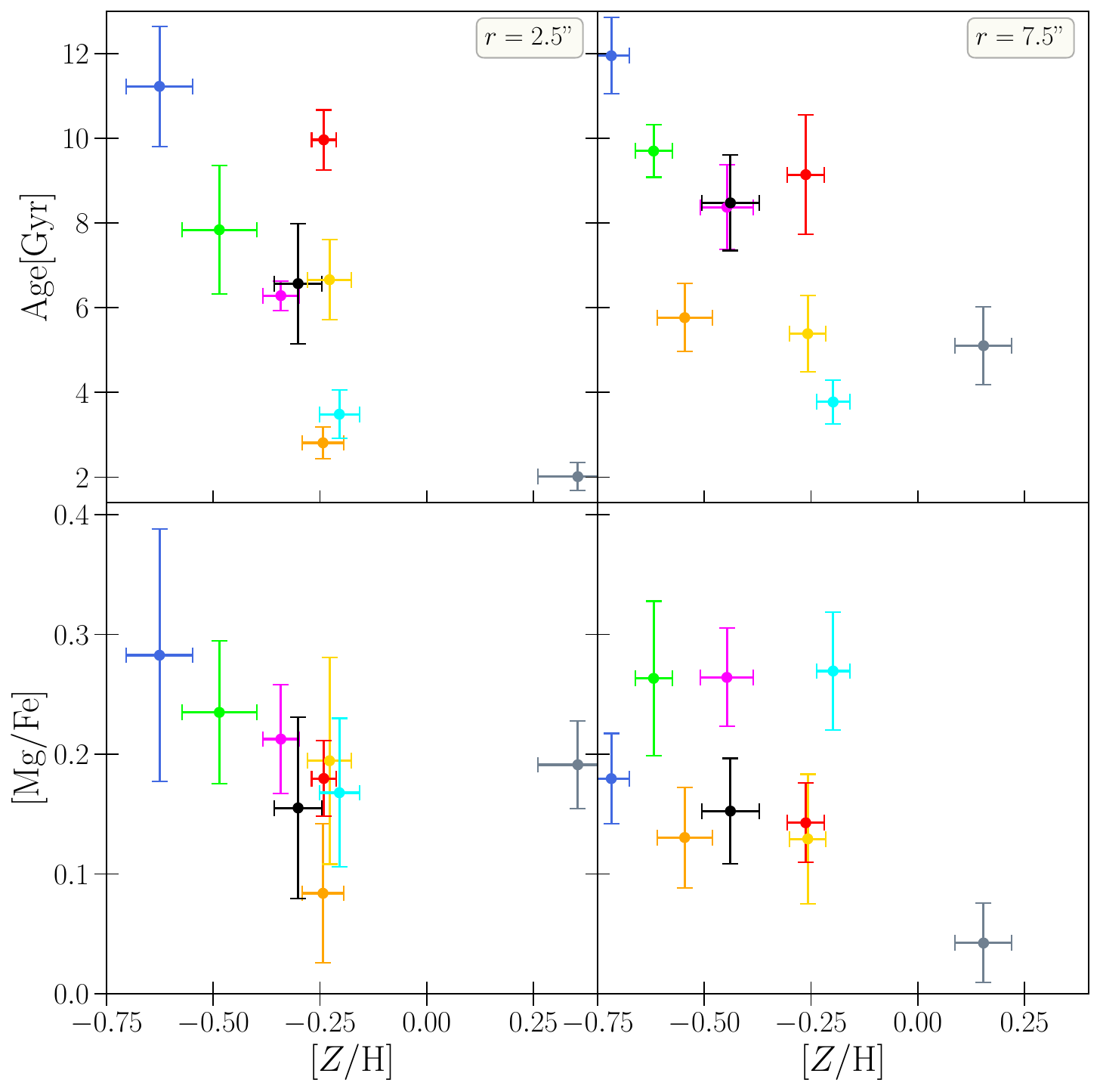}
    \caption{Correlations of the SSP parameters for the dE sample. Shown are the results for the spectra in the two apertures centered around $r=2.5\arcsec$ (\textit{left}) and $r=7.5\arcsec$ (\textit{right}). \textit{Top:} Age vs metallicity. \textit{Bottom:} [Mg/Fe] vs metallicity.}
    \label{fig:SSP_diagnostics}
\end{figure}

The SSP models were derived from the absorption line indices by modeling them with the index models of \citet{Thomas_Maraston_2011}. Within the VIRUS-W spectral range we model 4 notable Lick indices: H$\beta$, Mg$b$, Fe$5270$, Fe$5335$ to obtain the light-weighted age, metallicity $[Z/\rm H]$, and the abundance ratio [Mg/Fe] of the stellar population. The models were probed by sampling a grid spanned by these 3 parameters and their best fit values were derived by interpolating the grid in $\chi^2$. The population models are resolution corrected to match the data and the uncertainties are based on 100 Monte-Carlos simulations assuming gaussian noise for the flux in each wavelength bin \citep[more details in ][]{Parikh_2018}. The best fit parameters and their relation to each other are shown in Fig.~\ref{fig:SSP_diagnostics}. We find that old dEs are metal-poor while younger ones have nearly solar-like metallicity. Conversely the [Mg/Fe] ratios are slightly over-abundant and show no clear correlation with the metallicity. 

VCC~1910 has an exceptionally high, super-solar metallicity (see Fig.~\ref{fig:SSP_diagnostics}). We suspect that the SSP analysis for this galaxy is compromised (especially for the inner spectrum) because the most important age-sensitive Lick index within our spectral range (H$\beta$) lies just at the edge of the usable wavelength region which makes a robust derivation of the Lick continuum difficult. While we still show the results for VCC~1910 for the remainder of this work, one should take the SSP results for this galaxy with a grain of salt. We refer to the literature SSP values for VCC~1910 (App.~\ref{append:Literature_comparison}).

\subsection{The Dynamical modeling technique}
\label{subsec:technique}
The dynamical modeling code we employ is the current version of the axisymmetric orbit superposition code of \citet{Thomas_2004}, which is continuously updated and is based on the Schwarzschild orbit superposition approach \citep[][]{Schwarzschild_1979}. Schwarzschild modeling is a versatile numerical approach and can in principal be applied to any collisionless system. For a detailed description of the modeling we refer to \citet{Thomas_2004}.

To briefly summarize: The technique can be broken down into a few steps. A manifold of trial density mass models $\rho$ is constructed and for each a corresponding gravitational potential $\Phi$ is calculated. In each potential a set of representative orbits is integrated which densely sample the phase-space compatible with said potential $\Phi$. Each orbit is assigned an occupation weight and all orbits are superimposed and the model properties are determined by the weighted superposition. The best set of weights is found by fitting a model's observables to the corresponding observations, which in our case are the spatially resolved, non-parametric LOSVDs of the stars. The number of orbits, and thus the number of weights, is typically larger than the number of data constraints. Therefore a regularization term is required that prevents overfitting. In \citet{Lipka_2021,Thomas_2022} we presented and applied a novel data-driven approach to optimize the degree by which the models should be regularized to avoid both, over- and underfitting. This is the same $\aicmod$ approach we already utilized in the kinematic fitting procedure (cf. Sec.~\ref{subsec:spectroscopy}). It can generally be used to optimize penalty strengths in any penalized model fitting process. After fitting the weights for each of the trial mass models, the quality of each model can be compared by the $\aicmod$ it achieved: The single model which achieves the smallest $\aicmod$ is then deemed to be the best representation of the galaxy that's under investigation. In \citet{Lipka_2021} we demonstrated and discussed why the $\aicmod$ selection is more appropriate and unbiased in finding the best model out of all the probed trial models when compared to a simple $\chi^2$ selection. This is because $\aicmod$ accounts for the intrinsically varying model flexibility across different orbit models whereas $\chi^2$ only evaluates how well the data was fitted, making it prone to overfitting. 

A reasonable approach is to describe the mass model with a mass distribution consisting of 3 components: 
\begin{equation}
\rho \left( \mathbf{r}\right)=\Upsilon_{*} \cdot \nu+\rhodm+\mbh\cdot \delta \left( \mathbf{r}\right)
	\label{eq:Total_mass_model}
\end{equation}
Where $\mbh$ is a central supermassive central black hole (SMBH), $\rhodm$ is a description of the dark matter halo (usually given in a parametric form), and $\Upsilon_{*}$ is a stellar mass-to-light ratio which together with the 3D luminosity distribution $\nu$ describes the distribution of the baryonic matter. The luminosity distribution $\nu$ functions as a boundary condition for the models and is obtained from the photometry (Sec.~\ref{subsec:photometry}). 

For the construction of the trial mass models we follow this standard 3-component mass description. We find that to reliably decompose different mass components, a flexible and generic description of the \textit{dark matter} component is crucial. Otherwise the recovery of the other mass components (stars and SMBH) may be negatively affected as they may try and compensate for the inflexibility of the halo model. In Lipka \& Thomas (in prep.), in the following abbreviated as \paperrefemethods, we will lay out our arguments for why we believe that a sufficiently unbiased, yet still computationally efficient, decomposition is achieved if we parametrize the dark matter with a (flattened) Zhao-profile \citep[][]{Zhao_1996} where the transition width parameter is fixed. This means the halo component of our mass models is described by the following elliptical profile:
\begin{equation}
\rhodm(m,\theta)=\frac{\rhonorm}{\left(\frac{m}{\sclrad}\right)^{\slpin} \cdot \left(1+\frac{m}{\sclrad}\right)^{\slpout-\slpin}}
	\label{eq:dark_matter_model}
\end{equation}
where the ellipsoidal radius $m$ and angle $\theta$ are the (oblate) ellipsoidal coordinates within the meridional plane. This model has an inner logarithmic density slope $\slpin$ and a corresponding outer slope $\slpout$. The scale radius $\sclrad$ locates the transition between the two slopes while $\rhonorm$ sets the density scale. Together with the halo flattening $\qdm$, this means the halo is described by a 5-parameter halo model, much more than the standard 1- or 2-parameter models that are commonly used. Since the dark matter component is the subject of \paperrefeDARKMATTER, we discuss our halo parameter sampling strategy in more detail in that study. 

We do not know a priori by how much the galaxy is inclined, which not only affects the projection to the sky of the entire model but also the photometric boundary constraint $\nu$. Therefore we sample differently inclined orbit models and deprojections (see also Sec.~\ref{subsec:photometry}). In the past it was rather common to model only a single (edge-on) deprojection per galaxy. This practice was reinforced by the question whether it is even possible to dynamically constrain the inclination of axisymmetric galaxies because there appears to be a significant $\chi^2$ bias favoring edge-on models over less inclined models (see \citealt{Lipka_2021} for an in-depth explanation of the origin of this bias). In \citet{Lipka_2021} we demonstrated that the viewing angle of a galaxy can be well constrained using dynamical models directly as long as one follows the model selection approach we developed \citep[cf.][]{Lipka_2021,Thomas_2022} and described above.

Apart from the viewing angles, we also do not know whether the assumption of a constant stellar mass-to-light ratio is a reasonable one. For example the stellar mass-to-light ratio could vary with radius because the IMF \citep[e.g.][]{Mehrgan_2024} or the age and composition of the stellar population could exhibit spatial variations within the regions probed by our dynamical models. Similar to an inflexible dark matter component, an erroneous stellar component would also have effects on the recovery of all other mass components as those will try and compensate for this as much as possible to emulate the overall gravitational potential well. For this reason we allowed for another parameter that increases the flexibility of the \textit{stellar} model component. We allow for a radial variation in the stellar mass-to-light ratio $\Upsilon_{*}$. This is implemented by independently sampling an inner and outer mass-to-light ratio $\Upsilon_{i}$, $\Upsilon_{o}$ at two fixed ellipsoidal radii $m_{i}$ and $m_{o}$. Between the two radii the mass-to-light ratio is interpolated log-linearly, while outside of it ($m<m_{i}$ and $m>m_{o}$) the mass-to-light ratio is kept constant at $\Upsilon_{i}$ and $\Upsilon_{o}$ respectively. The ellipsoidal coordinates on which the mass-to-light ratios are stratified have the \textit{average} intrinsic axisymmetric flattening $q(i)$ inherited from the stellar luminosity deprojection $\nu(i)$ at inclination $i$. For a more detailed explanation of the gradient implementation see also \citet{Mehrgan_2024}.

The choice of radii at which one samples the two mass-to-light ratios is non-trivial and several issues have to be considered: Dynamical models have reduced constraining power at scales smaller than the spatial resolution and in the areas not covered by data (see \paperrefemethods). Therefore probing $\Upsilon_{i}$ at radii smaller than the resolution and $\Upsilon_{o}$ outside the FoV is dangerous and, in the worst case, biases the interpolation of $\Upsilon_{*}(r)$ between the two radii. This issue should be accounted for in the modeling of gradients for all types of galaxies. However, in the specific case of our dE sample we want to highlight two further issues that could lead to misleading gradients $\Upsilon_{*}(r)$ if not treated with care. Many of the dEs in our sample host distinct blue nuclei (cf. App.~\ref{append:galdiscussion} and \citealt{Ferrarese_2006}) whereas the extended host galaxies exhibit a $g-z$ color that is almost spatially constant. This suggests a distinct, younger stellar population in the nucleus embedded in a relatively homogeneous older population \citep[][]{Paudel_2011}. If one were to sample $\Upsilon_{i}$ within said nucleus, it could distort the results for the entire galaxy since a log-linear interpolation over the entire radial range would not be a good approximation of the locally much more concentrated young central population. This could become a significant problem if the nuclei are resolved (or just about resolved). Fortunately, in our case the typical size of the blue nuclei is much smaller than the fibre size of our observations and the nuclei only contribute a small portion of light to the central Voronoi bins (see discussion in Sec.~\ref{subsec:population_analysis}).

Therefore the mass-to-light ratio variation $\Upsilon_{*}(r)$ we derive should only reflect a (possible) stellar population variation within the larger galaxy envelope itself. The second issue we had to consider when modeling the stellar gradients of the dEs is their relatively faint surface brightness which reaches the level of the sky background for $r\gtrapprox 2~\reff$. A small systematic mis-estimate of the sky level will systematically increase/decrease the slope of the measured surface brightness as the relative contribution of the sky to the total light increases with radius. In turn, this affects the luminosity deprojection and the associated dynamical mass-to-light ratios. To be confident in the level of sky that we determined from \textit{HST} data (Sec.~\ref{subsec:photometry}), we carefully checked how well the reduced surface brightness profiles agree with the independently obtained profiles by \citet{Ferrarese_2006}, and avoided probing $\Upsilon_{o}$ at very large radii. Considering the above arguments we decided to sample the mass-to-light ratios of the dEs at $m_{i}=1.5\arcsec$ ($\sim 0.12 ~\rm kpc$) and $m_{o}$ between $9\arcsec$ to $14\arcsec$ ($\sim 0.7 ~\rm kpc$ -- $1.2 ~\rm kpc$) depending on the FoV size. That is at about the size of the spatial resolution limit and near the edge of the FoV, i.e. we allow the mass-to-light ratio to vary out to approximately 1 effective radius. This choice avoids the less constrained regions at very small and large radii, mitigates possible distortions by the nuclei or sky, and at the same time still allows for a stellar variation over the entire radial range that we believe is well constrained by the data. 

For each galaxy we calculated $10^{3}-10^{4}$ models on the 9-Dimensional grid of candidate mass models that is spanned by all probed parameters ($\Upsilon_{i}$, $\Upsilon_{o}$, $i$ , $\mbh$ , $\rhonorm$, $\sclrad$, $\slpin$, $\slpout$, $\qdm$) which determine the total mass distribution of the model. Since we sample about 5-20 values of each of these parameters, this huge grid can only be probed partially. Therefore we searched the grid efficiently by using the Nonlinear Optimisation by Mesh Adaptive Direct search NOMAD \citep{Audet_2006,Digabel_2022}. We conducted multiple \textit{independent} NOMAD iterations to avoid biases due to the search algorithm (see \paperrefemethods). We stopped the model calculation when the fest best $\aicmod$ models all had roughly congruent mass distributions within the regions that are well constrained by data, and when $\aicmod$ had essentially converged to a constant value. As we will discuss in \paperrefemethods, further search might change the nominal values of some of the (correlated) halo \textit{parameters} significantly, yet such a change in parameter values would only result in minuscule changes in the mass-distribution/composition due these correlations. Consequently such parameters should be thought of as \textit{nuisance} parameters only employed to describe the mass distribution.   

Error estimates in the data (here the LOSVDs) are often imperfect, non-gaussian, and correlated \citep[][]{Houghton_2006}, whereas the modeling implementations assume the errors to be independent and gaussian. As long as these issues are statistically unaccounted for, the \textit{absolute} values of $\chi^2$, and consequently $\aicmod$, are not meaningful statistically as they do not reflect the correct level of noise. For example, correlated raw data suppress the value of $\chi^2$ if independent gaussian noise is assumed. In such a situation, the often applied $\chi^2$ criteria to gauge the goodness of fit and to compute confidence intervals becomes meaningless.

Ideally one would model independent measurements of the same objects and evaluate the resulting scatter to obtain a realistic measure of the error. We can achieve this by modeling the quadrants of axisymmetric systems independently, which has the benefit of not relying on the perfect accounting of noise patterns and even includes possible \textit{systematic} uncertainties (e.g. deviation from axisymmetry, dust, etc.) in its calculation. While this is our preferred approach, the number of spatial bins $N_{\mathrm{bin}}$ which survived our $S/N$ limit of $15$ is fairly low for the majority of the dEs (Tab.~\ref{tab:obs_table}). Therefore we were forced to model the FoV as a whole instead of individual quadrants. This has two side-effects: 1) Deviations from axisymmetry across the different quadrants cannot be emulated by the axisymmetric orbit models. 2) We cannot use the scatter of the 4 independent quadrant modelings to estimate an error for the galaxy properties we infer from the modeling. 

Therefore, in the case where one can only model a single measurement with imperfect noise accounting, we need to come up with an alternative noise estimation (see details \paperrefemethods). Even if the absolute differences in $\aicmod$ (or $\chi^2$) between different models is not meaningful in itself, we can assume the model evaluation statistic (i.e. $\aicmod$ or $\chi^2$) is at least consistent and unbiased. In that case the relative ranking of the orbit models can still be used to gauge the significance of the resulting dynamical constraints. For example, if the differences in terms of $\aicmod$ between the best few $N$ models are low, yet these models differ significantly in their recovered $\mbh$, then the confidence in the $\mbh$ measurement is correspondingly bad. In order to not \textit{under}-estimate the errors this requires that enough models are included in the calculation of the scatter. And in order not to \textit{over}-estimate it, enough models have to have been probed globally for $\aicmod$ to have approximately converged. 

In practice, we tried to ensure an accurate error estimation by probing so many models that neither $\aicmod$ nor the scatter of the $N$ best models changes significantly anymore. We then implemented the error estimation by calculating all errors from the scatter between the best 25 models we found for each galaxy. Given the number of total models we calculated (see above) this translates to a significance criterion where the best $\sim0.5-1.0\%$ of all probed candidate models or analogously all models with roughly $\Delta \aicmod \lesssim 10$ are considered (see also Fig.~\ref{fig:dE_AIC_curves}). In conventional statistical modeling a rule of thumb is that models with $\Delta \mathrm{AIC}>10$ are so unlikely they can be excluded \citep[cf.][]{burnham_2002}.

We stress-tested the above dynamical modelling strategy and code on a mock VIRUS-W observation of an N-body dwarf simulation with stellar mass $M_{\rm star}=4 \cdot 10^8 \, \rm M_\odot$. The stress-test is set up such that stellar and dark matter are distributed similarly in large regions of space, which we expect to complicate the dynamical decomposition of the two mass components. This allows us to gauge how well we can recover dynamical stellar mass-to-light ratios, dark matter distributions, black holes, and the anisotropy structure even under particularly bad conditions and whether the model parameters could suffer from degeneracies. The results of the mock test are shown in \paperrefeDARKMATTERsp where we also present all the dynamically recovered properties of the dEs (here only the stellar component is discussed).

\section{Projected kinematic structure, stellar populations and their link to environment}
\label{sec:empirics}
Before we discuss the dynamical models and \textit{intrinsic} structure of the dEs we first discuss their \textit{on-sky} kinematic structure and stellar populations, and investigate if and how these stellar properties depend on the local environment the galaxies currently inhabit.  

\subsection{On-sky kinematic structure}
\label{sec:empirics_kinematic}
Overall, the kinematic moments of the dEs display a large diversity\footnote{To highlight this diversity we sum up and discuss the kinematic (and photometric) features of each galaxy individually in App.~\ref{append:galdiscussion}.} which suggests they took distinct evolutionary paths, or are at least at a different stage of their evolution which among other things could depend on their initial mass and/or past interactions with their environment. We find that many dEs have radially increasing velocity dispersions, while others have essentially flat profiles, only two galaxies exhibit a steadily decreasing $\sigma$ that peaks in the center. The degree to which the dEs are rotation- or pressure-supported also seems to vary widely from galaxies with $v/\sigma \sim 0$ (e.g. VCC~1261, VCC~1528) to galaxies with a substantial amount of rotation $v/\sigma \sim 0.5-0.8$ (e.g. VCC~543, VCC~856, VCC~2048). At least in the dwarfs where rotation is a significant factor and the signal-to-noise is high enough to reliably constrain higher moments, we find that the higher Gauss--Hermite moments of dwarf ellipticals follow the well established $v-h_{3}$ anti-correlation known from massive ellipticals \citep[][]{Bender_1994}. The $h_{4}$ profiles we observe show surprising individuality. About half the sample galaxies have a maximum $h_{4}$ in their center which steadily drops to 0 with increasing radius. Other galaxies show no clear $h_{4}$ trend with radius, i.e. they either scatter stochastically or are consistent with a gaussian LOSVD. Only VCC~856, and to some degree VCC~308, are outliers in having a \textit{rising} $h_4$ profile which is gaussian in the center. This peculiarity could be linked to the fact that both these galaxies are found to contain weak signatures of face-on spiral arms (cf. App.~\ref{append:galdiscussion}), and indeed we will find these two dEs to be the ones that are closest to face-on (Sec.~\ref{sec:dwarf_modelling}). 

`Ordinary' early-type galaxies\footnote{In the following the term `ordinary' ETG broadly refers to ETGs with stellar masses $\log_{10}(M_{*}/M_{\sun})\gtrsim 10$. Galaxies classified as dEs (or spheroidals) dominate the population of early types below this mass threshold, while `ordinary' (or classical) ETGs, which differ in their surface brightness distribution \citep[see][]{Kormendy_2012}, are usually found above this mass (there is only a small overlap of these two populations). The `ordinary' ETGs can be sub-classified further \citep[][]{Kormendy_1996} but this distinction is not considered here.} are known to come in two types \citep[][]{Kormendy_1996}, that are often separated into `Slow-Rotators' and `Fast-Rotators' according to their angular momentum parameter $\lambda$ as defined by \citet{Emsellem_2007}:
\begin{equation}
    \lambda=\frac{\langle r |v|\rangle}{\langle r\sqrt{v^2+\sigma^2} \rangle}
	\label{eq:Lambda_parameter}
\end{equation}
Here the brackets indicate the flux-weighted sum over all spatial bins within a given aperture (usually within half or one stellar effective radius).

Galaxies can then be classified by comparing their $\lambda$ with their apparent ellipticity\footnote{The ellipticity is measured within the same aperture as $\lambda$. The calculation (or definition) of $\epsilon$ is not always entirely consistent across different studies. We follow the definition of \citet{Emsellem_2007} which means we are stating the \textit{flux-weighted} ellipticity within the considered aperture.} $\epsilon$. This allows one to quantify whether the flattening of a galaxy is mostly supported by its angular momentum (i.e. ordered motion) or more by an anisotropy in its stellar velocity dispersion tensor (though anisotropy and rotation can easily go together, cf. \citealt{Thomas_2009}). 

\begin{figure}
	\centering
	\includegraphics[width=1.0\columnwidth]{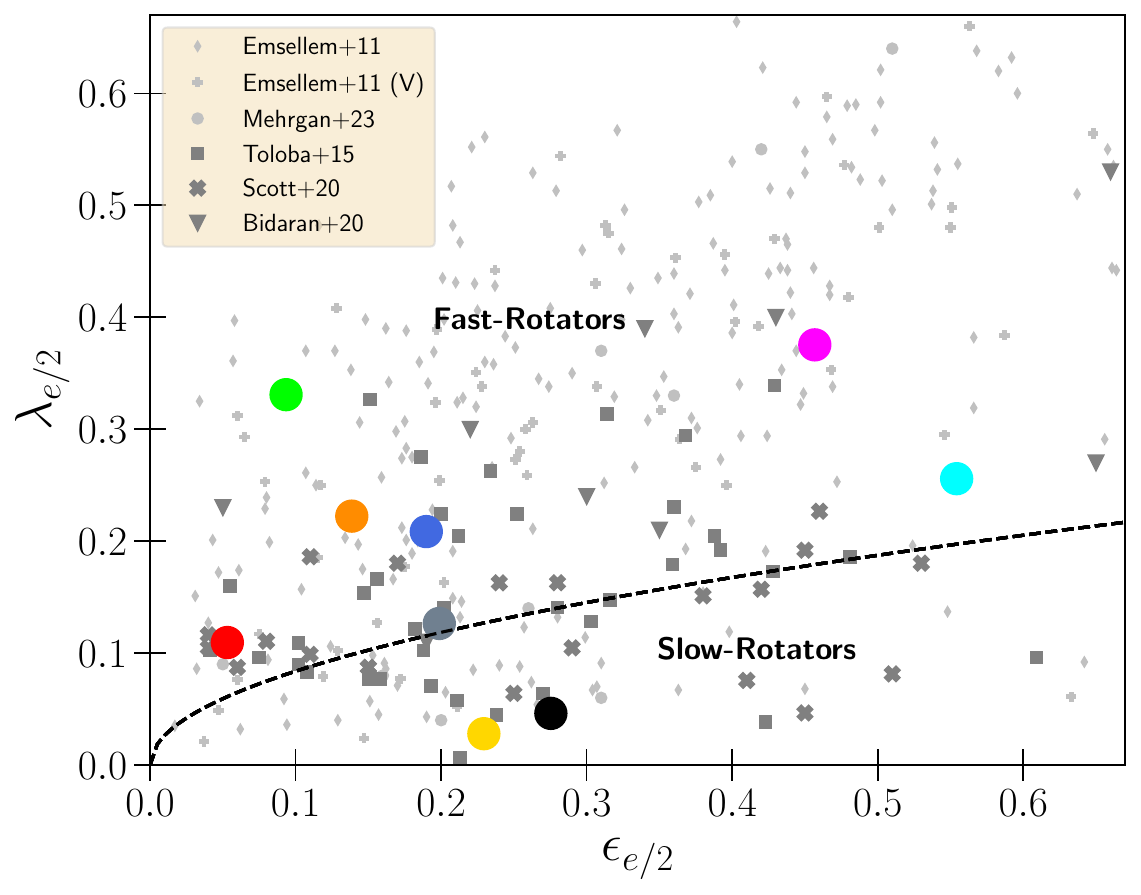}
    \caption{The projected angular momentum parameter $\lambda$ vs ellipticity $\epsilon_{e/2}$ within \textit{half} the effective radius. The \textit{dashed} curve indicates the dividing line between the Slow- and Fast-Rotator classification \citep[][]{Emsellem_2007}. \textit{Colored circles:} Our data. \textit{Gray squares:} The Virgo dEs of \citet{Toloba_2015}. \textit{Gray X-marks:} The Fornax dEs of \citet{Scott_2020}. \textit{Gray triangles}: The infalling Virgo dEs of \citet{Bidaran_2020}. \textit{Light gray crosses:} The massive Virgo cluster ETGs ($M_{*} \sim 10^{10}-10^{12}M_{\sun}$) of \citet{Emsellem_2011} which are part of the ATLAS$^{\rm 3D}$-survey \citep[][]{Cappellari_2011}. \textit{Light gray diamonds:} The massive ETGs of \citet{Emsellem_2011} that are not part of Virgo. \textit{Light gray dots:} The massive ETGs of \citet{Mehrgan_2023}.} 
    \label{fig:angular_momentum}
\end{figure}

Generally speaking the angular momentum parameter is highest for LTGs (particularly for those on the star-formation main-sequence, \citealt{Wang_2020}) as stars tend to form in rotating gaseous disks. However, even ETGs can have a high angular momentum support and several studies have found a systematic change of $\lambda$ in the ETG sequence depending on total stellar mass and environment (see below). Since the FoV of our dE sample does not always extend out to $1~\reff$, we decided to evaluate the angular momentum parameter within an aperture of $\reff/2$. Fig.~\ref{fig:angular_momentum} shows $\lambda_{e/2}$ vs $\epsilon_{e/2}$ for our dE sample together with samples of `ordinary' ETGs \citep[][]{Cappellari_2011,Mehrgan_2023} and three dE samples in a similar mass range as ours \citep{Toloba_2015,Scott_2020,Bidaran_2020}. For \citet{Scott_2020} and \citet{Bidaran_2020} the ellipticity was not given in the $\reff/2$ aperture. However, the differences between $\epsilon_{e/2}$ and $\epsilon_{e}$ can be expected to be small (e.g. from \citet{Toloba_2015} who show both, the mean difference between the two is $\sim 0.019$). An alternative to Fig.~\ref{fig:angular_momentum} for the larger $1~\reff$ aperture is shown in App.~\ref{append:Lambda}. While it relies on using extrapolated $\lambda_{e}$ values for some of the dEs when the FoV is too small, none of the following conclusions change.

We estimate the statistical errors of our $\lambda_{e/2}$ values using 100 Monte-Carlo realizations of the velocity and dispersions maps. For each realization the values in each Voronoi bin are perturbed according to their respective measurement error and a new $\lambda_{e/2}$ is calculated from the perturbed values. From the 100 $\lambda_{e/2}$ values we estimate the statistical errors. We find the statistical errors of $\lambda_{e/2}$ are quite small $\sim0.01$ (i.e. roughly the size of the markers in Fig.~\ref{fig:angular_momentum}). Differences between the $\lambda$ parameters from different studies are more likely to be driven by systematic effects due to different spatial binning, apertures and spectral resolution (App.~\ref{append:Literature_comparison}). We are able to estimate the impact of such systematics if we compare our values of those dEs that we have in common with the sample of \citet{Toloba_2015}. We find that our $\lambda_{e/2}$ are on average 0.077 higher than their 2D extrapolations\footnote{The values of \citet{Toloba_2015} are based on long-slit measurements that were extrapolated to a 2D estimate using a correction factor.}. Our $\lambda_{e/2}$ are systematically higher than those of \citet{Toloba_2015} because of the poorer spectral resolution of their spectra. Low resolution biases the velocity dispersions high (we show this in App.~\ref{append:Literature_comparison}), which itself results in an underestimation of the angular momentum parameter. While this spectral resolution effect results in a systematic bias we expect the impact of spatial binning to be of a more random nature. And indeed we find no correlation of $\lambda_{e/2}$ for our dEs with the spatial binning (the spatial binning is not homogeneous for all dEs, see Tab.~\ref{tab:obs_table}). By re-binning the spectra of VCC~2048 into coarser bins and calculating $\lambda_{e/2}$ again we estimate the spatial resolution systematics to be around $\pm 0.02$. Finally we gauge the systematics in the ellipticity $\epsilon_{e/2}$, again by comparing the dEs in common with \citet{Toloba_2015}, and find a mean difference of $\Delta \epsilon_{e/2}=-0.037$. 

Our dEs are distributed fairly similarly to the other 3 published dE studies, though our sample tends to have lower ellipticity compared to some of the other studies\footnote{Considering \citet{Lisker_2007} who analysed 413 Virgo dEs (complete in the regime of our dEs) we would expect a blindly selected Virgo dE to have $\epsilon\sim 0.25$ on average.}. However, we note that this comparison is not always straightforward. The $\lambda$ values of \citet{Toloba_2015} are based on long-slit data that were transformed to `integrated' $\lambda$ values using a 2D-correction factor (see \citet{Toloba_2015} for details). The values in \citet{Scott_2020} were given only within the larger aperture of $1~\reff$ and some of them are extrapolations using the aperture corrections of \citet[][]{van_de_Sande_2017}. We applied the inverse of this correction to estimate $\lambda_{e/2}$ from the $\lambda_{e}$ values stated in \citet{Scott_2020}. These $\lambda_{e/2}$ are shown in Fig.~\ref{fig:angular_momentum}. The dEs of \citep[][]{Bidaran_2020} may be selection biased as they are all part of the same, recently accreted, galaxy group and most of them are classified as dE(di). Such dEs can be expected to have a higher angular momentum parameter than the average Virgo dE (see Sec.~\ref{sec:empirics_environment}). Finally, as suggested above, the poorer spectral resolution of \citet{Toloba_2015} has likely biased the $\lambda_{e/2}$ low. In comparison the intermediate resolution of the samples of \citet{Scott_2020} with $R=4500$ and \citet{Bidaran_2020} with $R=3000$ may have been just enough to obtain unbiased estimates if the $S/N$ was high enough \citep[cf.][]{Eftekhari_2022}.

Compared to the global picture all the above differences between dE studies are minuscule. Taken together all 4 dE studies (\citet{Toloba_2015,Scott_2020,Bidaran_2020} and ours) suggest that the angular momentum parameter of dEs is systematically lower than those of the `ordinary' ETGs. This was first noted by \citet{Scott_2020} and even the high spectral resolution power of VIRUS-W does not change this result. Very few dEs have a high angular momentum parameter ($\lambda_{e/2} \gtrsim 0.3$) and the angular momentum parameter is almost independent of the ellipticity. In contrast, the vast majority of `ordinary' ETGs with $M_{*}>10^{10}M_{\sun}$ (excluding the most massive ones) are classified as fast-rotators \citep[][]{Emsellem_2011,Guo_2020,Santucci_2022} and $\lambda_{e/2}$ tends to be more strongly correlated with $\epsilon_{e/2}$. 

It is known that `ordinary' ETGs are not homogeneous in $\lambda$ either. Several studies of the massive ETGs have previously noted a \textit{dichotomy} in the angular momentum parameter as the total stellar mass increases further: The slow-rotator regime is mostly dominated by the most massive ETGs with $M_{*}\sim 10^{12}M_{\sun}$ whereas the intermediate mass ETGs ($10^{10}-10^{11}M_{\sun}$) tend to be more rotation supported \citep[e.g.][]{Emsellem_2011,Graham_2018,Santucci_2023}. This dichotomy in the ETG sequence has been well established for some time now as several properties change in this regime \citep[e.g.][]{Bender_1988_B,Bender_1989,Kormendy_1996,Kormendy_1999,Kormendy_2013,Dekel_2006,Nelson_2018}. 

This behaviour of a decreasing $\lambda$ with increasing stellar mass found for `ordinary' ETGs is opposite to what we observe going from the regime of the dEs to more massive ETGs: an increase of $\lambda$ with mass. Consequently, if we include the dEs in the ETG mass sequence (i.e. a sequence spanning 5 dex from $\log_{10}(M/M_{\sun})=8$ to $12$) we would not find a dichotomy but a `trichotomy' in $\lambda$ as the total stellar mass changes: The angular momentum parameter is low in the mass regime of the dEs but increases with stellar mass until it reaches a maximum for galaxies with $10^{10}-10^{11}M_{\sun}$ at which point the trend reverses and the angular momentum parameter decreases again with the most massive ETGs being slow rotators again. Unfortunately, we are not aware of a \textit{single} study/sample that covers the ETG sequence sufficiently well from $\log_{10}(M/M_{\sun})=8-10.5$ to observe the increase in $\lambda$ with mass directly and we rely on comparison across different galaxy samples with possibly different systematics. See for example \citealt{Scott_2020} or \citealt{Spavone_2022} who also observed this trichotomy, or `U-shape', by showing the $\lambda$ vs stellar mass using a number of different studies. How the relation between stellar mass and $\lambda$ continues for even lower masses is less clear. The results of \citet{de_los_Reyes_2023} suggest that rotation support decreases even further for galaxies below $\log_{10}(M/M_{\sun})\leq8$.

\begin{figure*}
	\centering
	\includegraphics[width=1.0\textwidth]{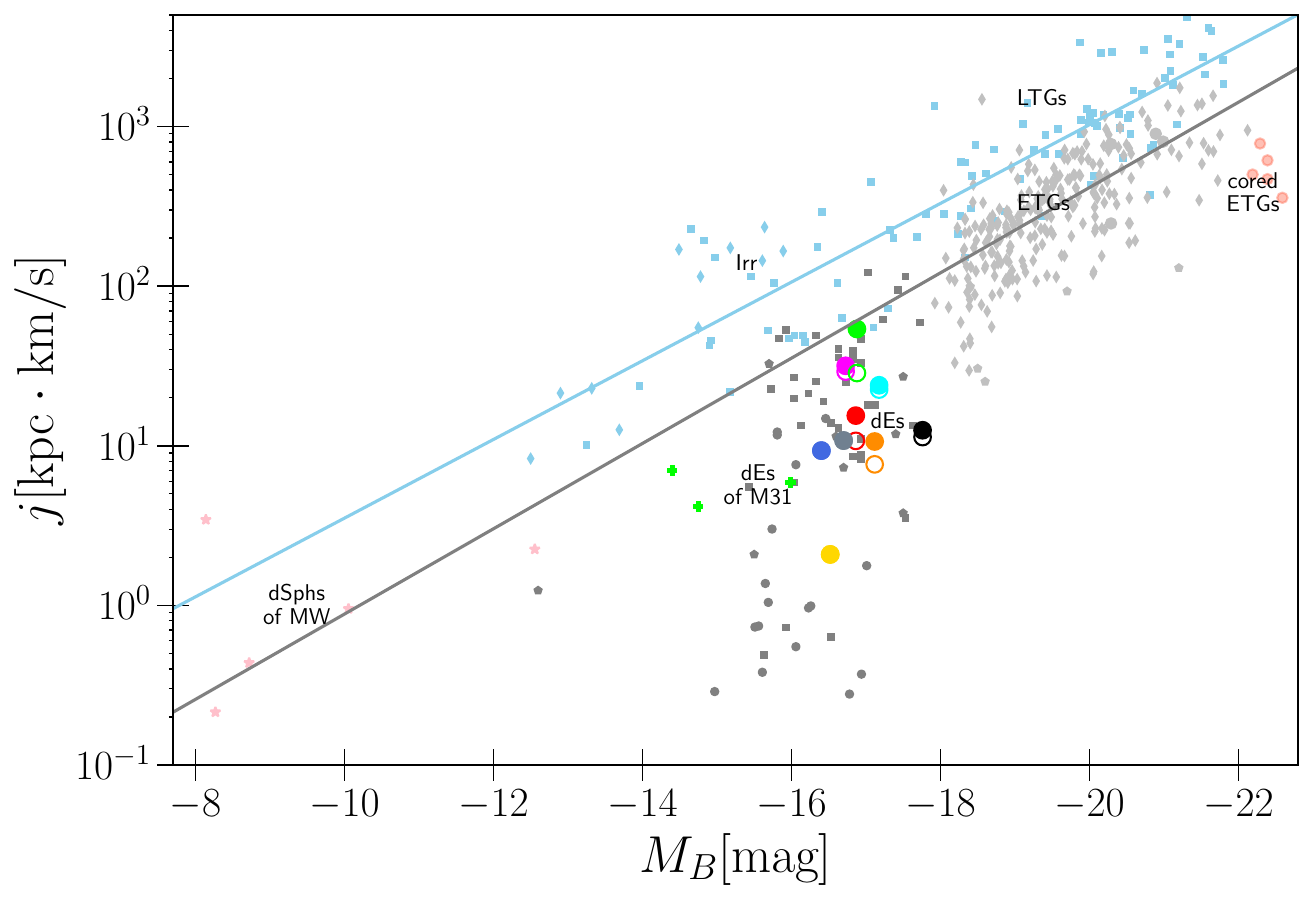}
    \caption{Estimate of the stellar specific angular momentum $j$ versus absolute magnitude (\textit{B}-band) for various galaxy types. For those studies where the magnitude was not stated we consulted Hyperleda \citep[][]{Makarov_2014}. For ETG samples we estimated $j$ from the the stated velocity curves (see text), and when velocity data were noisy we clipped outliers. For the LTG samples we used the $j$ directly from the source tables. \textit{Blue diagonal:} Fit to \textit{all} LTG points ($\log_{10} j=-0.2465 \cdot M_{B}+const.$). \textit{Gray diagonal:} Fit to `ordinary' ETGs with $M_{B}\in[18.5,21]\rm mag$ ($\log_{10} j=-0.2673 \cdot M_{B}+const.$). Massive ETG and LTGs ($M_{*}\gtrsim 10^{10}M_{\sun}$) are known to roughly follow the same relation ($j\propto M_{*}^{2/3}$), but LTGs are offset towards higher $j$ \citep[cf.][]{Fall_1983,Bender_1990,Romanowsky_2012}. \textit{Large, colored points:} Our Virgo dE sample (inclination-corrected: filled, not-corrected: open). \textit{Small green crosses:} The dE satellites of M31 (NGC 147, NGC 185, NGC 205) inferred from individual stars \citep[][]{Geha_2006,Geha_2010} with velocity curves out to $\sim 8~\reff$. Studies of integrated light (smaller FoV) suggest significantly smaller velocities \citet{Bender_1991, Simien_2002}. \textit{Gray squares:} Long-slit dE observations from \citet{Toloba_2015}. \textit{Gray dots:} dEs from \citet{Geha_2003} some of which have exceptionally low $j$ (presumably because of a very small FoV or a slit not aligned with the kinematic major-axis). \textit{Gray and light gray pentagons:} ETG sequence (dwarfs and `ordinary' ETGs) of \citet{Bender_1990}. \textit{Faint, red stars:} dSphs of the Milky Way from \citet{Martinez_2021}. For these, $j$ is from \textit{intrinsic} velocities as they derived their values from the Gaia proper motions and radial velocities of \textit{individual} stars.\textit{Light gray diamonds:} The ETGs of \citet{Emsellem_2011}. \textit{Light gray dots:} power-law ETGs of \citet{Mehrgan_2023}. \textit{Light, red dots:} Giant cored ETGs of \citet{Mehrgan_2023}. \textit{Light blue diamonds:} Dwarf irregulars of \citet{Kurapati_2018} ($j$ is of stars only). \textit{Light, blue squares:} Dwarf and massive spirals of \citet{Posti_2018}.} 
    \label{fig:specific_momentum}
\end{figure*}

The $\lambda$ correlation with mass could suggest that there is a process that suppressed the \textit{total} angular momentum $J$ of the dEs, or conversely, that dEs hide more angular momentum at $r>1~\reff$ than `ordinary' ETGs do. Fig.~\ref{fig:specific_momentum} shows the estimated stellar specific angular momentum $j=J/M$ (i.e. per mass) vs absolute magnitude for our Virgo dE sample together with other galaxy samples of various morphologies. An alternative version of the Figure vs stellar mass $M_{*}$ instead of magnitude is shown in App.~\ref{append:Lambda}. 

For LTGs such a diagram is a standard analysis tool, but for ETGs it is much more difficult to determine the total specific angular momentum $j=J/M$ due to the lack of (intrinsic) rotation measurements beyond $\sim 1-2~\reff$ and the frequent lack of information about the viewing angles and intrinsic shape. Still it has been done for `ordinary' ETGs \citep[e.g.][]{Romanowsky_2012,Pulsoni_2023} using careful approximations and information from non-stellar tracers at larger radii. However, where dEs (or dwarf ETGs in general) appear in these diagrams has not been addressed yet. 

One can obtain a rough estimate of a galaxy's angular momentum using $j\approx \kappa \cdot v_{m} \cdot \reff$ where $v_{m}$ is the velocity around the maximum of the rotation curve, $\reff$ the effective radius, and $\kappa$ a numerical factor that follows from assumptions about the density and velocity distributions of the stars. $\kappa$ can have a range of values $\in[1,4]$ depending on the galaxy morphology and assumptions going into it \citep[cf.][]{Fall_1983,Scorza_1995,Romanowsky_2012,Pulsoni_2023}.

Following \citet{Bender_1990,Bender_1991,Scorza_1995} we analyse \textit{all} ETGs using $\kappa=1.03$ which is an approximation of $j$ enclosed within $5~\reff$ derived for de Vaucouleurs galaxies (Sersic index $n=4$). More intricate approximations of $j$ exist that, for example, account for variation of $\kappa$ with Sersic index \citep[][]{Romanowsky_2012}. However, such 2nd order corrections make it difficult to compare a large number of galaxy samples of different morphologies and our intention here is to investigate the transitional behaviour of $j$ as one goes from `ordinary' ETGs to dEs as `fair' as possible. Relying on different a priori assumptions and corrections for the two galaxy types could distort and complicate this comparison. Still, we stress that an inclusion of such 2nd order corrections would strengthen the evidence for the following conclusion rather than weakening it. For example, dEs generally have lower Sersic indices $n_{s}\sim2$ than `ordinary' ETGs (see Tab.~\ref{tab:galaxy_table} or \citealt{Kormendy_2009}) and the correction factor $\kappa$ is rising monotonically with Sersic index. From the correction factor of \citet{Romanowsky_2012} we expect dEs to have $j$ overestimated by about $50\%$ \textit{relative} to `ordinary' ETGs which have a higher Sersic index ($n_{s}\sim4$). Other sources of uncertainty are the poorly known inclinations and the small FoV (i.e. the maximum of the circular velocity curve may not be sampled yet). These are errors expected to be of similar order as the $\kappa$ correction \citep[][]{Scorza_1995,Romanowsky_2012} but we can expect those to affect all ETGs alike (i.e. relative differences remain mostly unaffected). 

Nonetheless a comparison across different samples and morphologies is not straight-forward. For the ETG samples shown in Fig.~\ref{fig:specific_momentum} we obtained the maximum of the velocity curve $v_{m}$ and $\reff$ as stated in these studies\footnote{For the ATLAS$^{\rm 3D}$ ETGs we used their kinematic maps \citep[][]{Cappellari_2011} and estimated $v_{m}$ from the median of the bins around the maximum of the measured velocity.}, but this does not mean the maximum of velocity curve has actually been probed fully. For example, for our dEs Fig.~\ref{fig:kinprofiles} suggests we are close to the maximum for some of the dEs but others may still rise beyond the FoV. In Fig.~\ref{fig:specific_momentum_vs_mass} we show one of the, currently most definite $j-M_{*}$ scaling relations \citep[][]{Pulsoni_2023} established for `ordinary' ETGs. \citet{Pulsoni_2023} measured the specific angular momentum of ETGs out to large radii using planetary nebulae. 

For our dE sample we are in the position to correct the velocity curves for the inclination using the information derived from dynamical models (see Sec.~\ref{subsec:inclination}). Fig.~\ref{fig:specific_momentum} shows both, inclination-corrected (filled circles) and uncorrected (open circles) estimates for the angular momentum of our dE sample. For LTGs with an exponential disk profile, a different approximation is used $j\approx2\cdot v_{max}\cdot r_{disk}$ with the disk scale length \citep[e.g.][]{Fall_1983,Scorza_1995,Romanowsky_2012}. However, for LTGs $j$ is also often inferred more directly from outer gas disks which extend to several $\reff$ and enable inclination correction.  

Fig.~\ref{fig:specific_momentum} shows, as has long been known, that \textit{bright} ETGs and LTGs follow two near \textit{parallel}, but \textit{offset}, sequences \citep{Fall_1983,Romanowsky_2012,Pulsoni_2023}. This so-called `Fall-relation' is usually expressed in terms of stellar mass $M_{*}$ and not $M_{B}$ (see Fig.~\ref{fig:specific_momentum_vs_mass}). It can be physically motivated from $\Lambda \rm CDM$ structure formation as its slope is remarkably close to what is expected from the angular momentum acquisition of dark matter halos, $j\propto M^{2/3}$ \citep[][]{Peebles_1969,Efstathiou_1979}. Initially the gas (which later forms a galaxy's stars) is expected to follow this halo relation, but the subsequent evolution, star formation, environment, etc. may modify how much is inherited from the stars (see also \paperrefeDARKMATTER). 

If we were to assume a magnitude-independent mass-to-light ratio, the theoretical Fall-relation $j\propto M_{*}^{2/3}$ translates to $\log_{10}(j) \propto -0.2667\cdot M_{B}$ in magnitude space. In Fig.~\ref{fig:specific_momentum} we show linear fits to all LTG galaxies (blue) and to the `ordinary' ETGs with magnitudes $\forall M_{B} \in[-18.5,-21] ~\rm mag$ (black). Most of the latter were classified `Fast-rotators' (see above Fig.~\ref{fig:angular_momentum}) and they follow the halo-momentum relation very closely with $j \propto -0.2673\cdot M_{B}$. The LTGs are not far off either with a slope of $-0.2465$. The fact that LTGs are distributed on a slightly shallower curve may be because of the inclusion of dwarf LTGs in the fit. Dwarf spirals tend to fall above versions of the Fall relation fitted to massive spirals only \citep{Chowdhury_2017,Butler_2017,Kurapati_2018}. However, the full LTG sequence including dwarfs can still be described well by a single power-law with a slightly shallower slope \citep{Posti_2018,Mancera_Pina_2021}.

In contrast, ETGs do not follow a single universal scaling relation. The most luminous (cored) ETGs tend to fall below the standard Fall-relation that fits the bulk of `ordinary' ETGs much better. And, akin to what was noticed for the suppressed angular momentum parameter $\lambda$, nearly the entirety of dEs (and the dE satellites of Andromeda) fall \textit{below} what one would expect from the Fall relation of `ordinary' ETGs. At the low mass-end the two data points for the biggest Milky Way dSph (Fornax) still fall below the relation (similar to the dEs) but the situation becomes more ambiguous for the smallest Milky Way dSphs which on average have a angular momentum in line with the Fall relation. 

In parts this angular momentum suppression can be explained by the smaller mass-to-light ratios that dEs tend to have compared to `ordinary' ETGs, but the dEs remain in tension with the standard Fall-relation as shown in the $j$ vs $M_{*}$ diagram (Fig.~\ref{fig:specific_momentum_vs_mass}). The decrease in $j$ for the dEs could also be because the estimate for dwarfs may not be 1:1 comparable to that of the `ordinary' ETGs, particularly if the total angular momentum of these galaxy types is distributed very differently in space. A large portion of the angular momentum of the dwarfs could `hide' at larger radii. However, we argue the difference in $j$ is too big to be explained by this \textit{alone}. While the velocity profiles of our dEs only extend out to $\sim 1~\reff$ and likely still rise beyond that, none of our dEs show signs of a rapidly rising velocity profile. The velocity curves we found either start to plateau or rise mildly at the FoV edge (Fig.~\ref{fig:kinprofiles}). In comparison the `ordinary' ETGs, to which the Fall relation in Fig.~\ref{fig:specific_momentum} is fitted to, are actually much more suggestive of a rotation curve maximum that has not been reached yet and a detailed analysis of planetary nebulae tracers \citep[][]{Pulsoni_2023} shows that these galaxies already hide a lot of angular momentum at large radii. This means for the dEs to fall on the same relation they would have to hide even more $j$ at large radii than the `ordinary' ETGs, all the while having lower angular momentum parameters $\lambda_e$ in the centre. 

To answer with more certainty whether dEs live below the standard Fall relation, rotation curves sampled out to much larger radii will be needed, e.g. by using non-stellar tracers as was done for the brighter ETGs. For the dEs of M31 the velocity profiles from stars alone were probed out to $8~\reff$ \citep[][]{Geha_2006,Geha_2010}, which showed that rotation velocities reach their maximum significantly beyond $1~\reff$. However, even these higher velocities are still not sufficient to elevate the angular momentum $j$ of these dEs to the level that is expected from the Fall relation of bright ellipticals (Fig.~\ref{fig:specific_momentum}). For some dEs an analysis of their globular clusters suggests rapidly rising profiles beyond $1~\reff$. \citet{Beasley_2009} for example have analyzed the two slow rotators VCC~1261 and VCC~1528 in our sample using globular clusters as kinematic tracers. Their results suggest much higher velocities which would be able bring the dEs onto the standard Fall relation. However, the reported uncertainties in the GC analysis are high and there is some tension with the stellar rotation profiles, which suggest a milder increase in velocity (Fig.~\ref{fig:kinprofiles}). 

In conclusion, while at least some of the initial angular momentum of dEs may have just been redistributed to larger radii (i.e. it is hidden at larger radii at $2-8~\reff$) we may still expect that a significant portion has been lost entirely. The processes responsible for this can be manifold and in superposition. They can be \textit{internal} (e.g. supernova outflows) or \textit{external} (e.g. galaxy harassment) as discussed below, but the comparison to the `ordinary' ETGs suggests an onset (of importance) of these mechanisms for galaxies with stellar masses below $\sim 10^{9.5}M_{\sun}$.  

A comparison with studies of LTGs is also interesting. The finding that the average angular momentum parameter $\lambda$ \textit{increases} as one is moving along the sequence of ETGs with stellar masses from $10^{9}M_{\sun}$ up to $10^{10}M_{\sun}$ appears to be traced even by the star-forming analogs (dwarf spirals). While the highest mass LTGs ($\gtrsim 10^{10}M_{\sun}$) have a relatively high $\lambda$ that is more or less independent of the total galaxy mass (i.e. unlike ETGs with $\gtrsim 10^{10}M_{\sun}$), the amount of rotation seems to start to drop for LTGs with masses below $\log_{10}(M_{*}/M_{\sun})=9.3$ \citep[][]{Falcon_Barroso_2019,Wang_2020}, similar to the low-mass behaviour in the ETGs sequence. In other words: the process responsible for the change in $\lambda$ that occurs somewhere between $10^{9}M_{\sun}$ and $10^{10}M_{\sun}$ could be a \textit{common} one for both passive and star-forming galaxies. However, a comparison of the total angular momentum $j$ in the LTG sequence shows this behaviour in $\lambda$ is not accompanied by a similar behaviour in total specific angular momentum $j$. This may suggest that the suppression of $\lambda$ for LTGs is more so because of a redistribution of the angular momentum to larger radii without changing $J$ globally. This discrepancy between the behaviour of $j$ in late-type dwarfs and early-type dwarfs could suggest that there is an additional mechanism removing the angular momentum entirely. We take up this discussion again in \paperrefeDARKMATTER, where we analyse the relation between LTG and ETG dwarfs in conjunction with our results for their dark matter structure. 

\subsection{Correlations of angular momentum and stellar population with environment}
\label{sec:empirics_environment}
Beyond the above dependence of the angular momentum on stellar mass, the environment could affect the amount of ordered motion in galaxies. In fact, it may even be the driving factor behind the suppressed $j$ (and $\lambda$) of dEs (compared to `ordinary' ETGs) we found, since they have lower stellar (or total) mass, their potential well and their ability to withstand their environment is reduced \citep[][]{Romero_Gomez_2024}. Similarly, the environment is often thought to be the main reason why dEs have stopped forming stars as they lost their gas reservoir due to external influences. In the following we investigate whether the angular momentum parameter and stellar population properties (age and metallicity) are a function of the galaxy environment.

For the purpose of a quantitative analysis of past environmental influences one must find a measure that quantifiably traces the degree to which each galaxy has experienced interaction with its host cluster. Commonly used tracers are: The projected cluster-centric distance, the local galaxy number/luminosity density, distance to nearest large neighbour, or the local density of the intracluster medium. A priori it is not clear which tracer is best used as a proxy for past environmental influences. We argue one should treat all environment indicators with some caution. Firstly because of the large uncertainties in the distance along the line of sight and consequently its actual 3D position within the cluster. And secondly because a galaxy's \textit{current} location may not be very representative of the \textit{past} interactions it had with the cluster and other galaxies. Some dEs might have experienced multiple pericenter passages already, or pre-processing in groups \citep[][]{Fujita_2004} which may not be reflected in its current day position. For the Virgo cluster this consideration might be especially crucial because it may be fairly young dynamically and still unrelaxed \citep[][]{Binggeli_1987}. \citet{Sybilska_2017} have tested different environment proxies for a sample of Virgo dEs (including some of our sample galaxies). While the different proxies they tested may differ quantitatively, they agree at least qualitatively for the most part, i.e. if a strong correlation was present, it is detected in all proxies. Therefore we decided to use the most common and simple proxy for our study: The projected distance $\centredis$ to the central cluster galaxy of Virgo, M87.

\begin{figure}
	\centering
	\includegraphics[width=1.0\columnwidth]{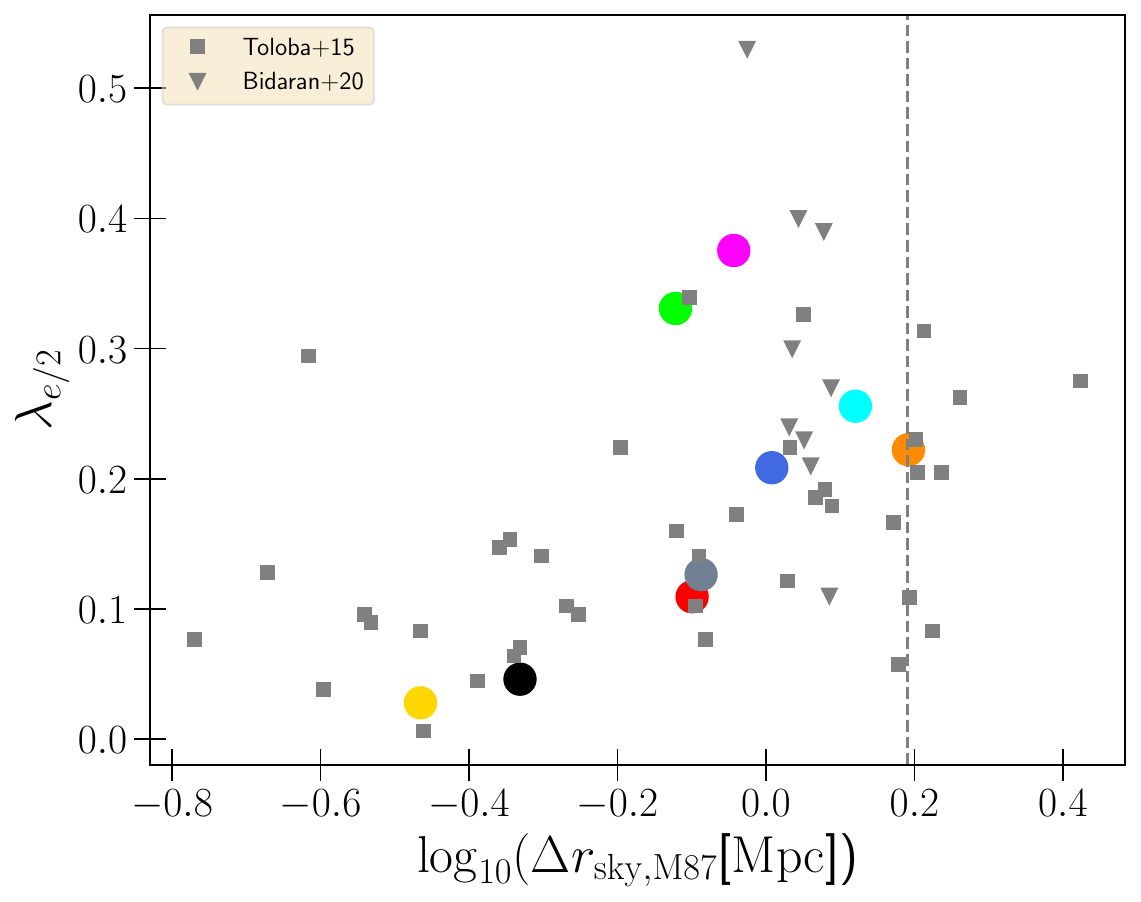}
    \caption{The angular momentum parameter $\lambda$ within half the effective radius vs the environment proxy (the projected cluster-centric distance to M87). To convert the projected distance to Mpc we assumed a constant distance to Virgo of 16.5 Mpc, i.e. the position of the galaxy along the line of sight is not considered. The angular momentum parameter of dEs appears to decrease for the galaxies that have experienced stronger and longer-lasting interactions with the cluster environment, which confirms previous trends of angular momentum-environment relations in the Virgo-cluster \citep{Toloba_2015} and Fornax-cluster \citep{Scott_2020}. Shown here are only the dEs of \citet{Toloba_2015} and \citet{Bidaran_2020} since these also inhabit the Virgo-cluster. \textit{Dashed line:} The virial radius of the Virgo-cluster \citep[][]{Ferrarese_2012}} 
    \label{fig:angular_momentum_vs_environment}
\end{figure}

Using this environment proxy we find a strong correlation of the angular momentum parameter $\lambda$ with the Virgo environment (Fig.~\ref{fig:angular_momentum_vs_environment}). The galaxies near the cluster center are slow rotators whereas galaxies in the cluster outskirts tend to posses more angular momentum. Similar results were found by \citet{Toloba_2015} and \citet{Scott_2020} who studied Virgo and Fornax dEs, respectively. Conforming with this environment correlation are also the dEs of the recently accreted galaxy group analyzed by \citet{Bidaran_2020}. As shown in Fig.~\ref{fig:angular_momentum_vs_environment} their angular momentum parameters tend to be higher than that of dEs deep in the cluster, although their $\lambda_{e/2}$ display a wide range of values which they attribute to the pre-processing of some of their dEs within the group.

This dependence of the rotational support vs environment may well extend from cluster scales down to the scales of individual galaxies. For example, within the Local Group, the dE satellites of M31 (which are a few magnitudes fainter than the Virgo dEs) that are closest to M31 show signs of tidal heating and reduced rotation due to their interaction with M31 \citep[][]{Bender_1991,Geha_2006,Geha_2010}. For the even fainter dSphs of the Milky Way, the correlation of angular momentum with environment also exists but is comparatively weak \citep[][]{Martinez_2021}. We may expect that the fainter and less massive satellites of individual galaxies have very different dynamical time scales than the more massive dEs in galaxy clusters that infall into the cluster at a higher speed. Some of the cluster dEs may experience their first in-fall such that they only recently started to become heated by their environment, while for the small local satellite galaxies many close interactions with their host may have `washed out' any correlation by now. Therefore the environment correlations we find in our sample may not be simply transferable to any environment or mass regime.

\begin{figure*}
	\centering
	\includegraphics[width=1.0\textwidth]{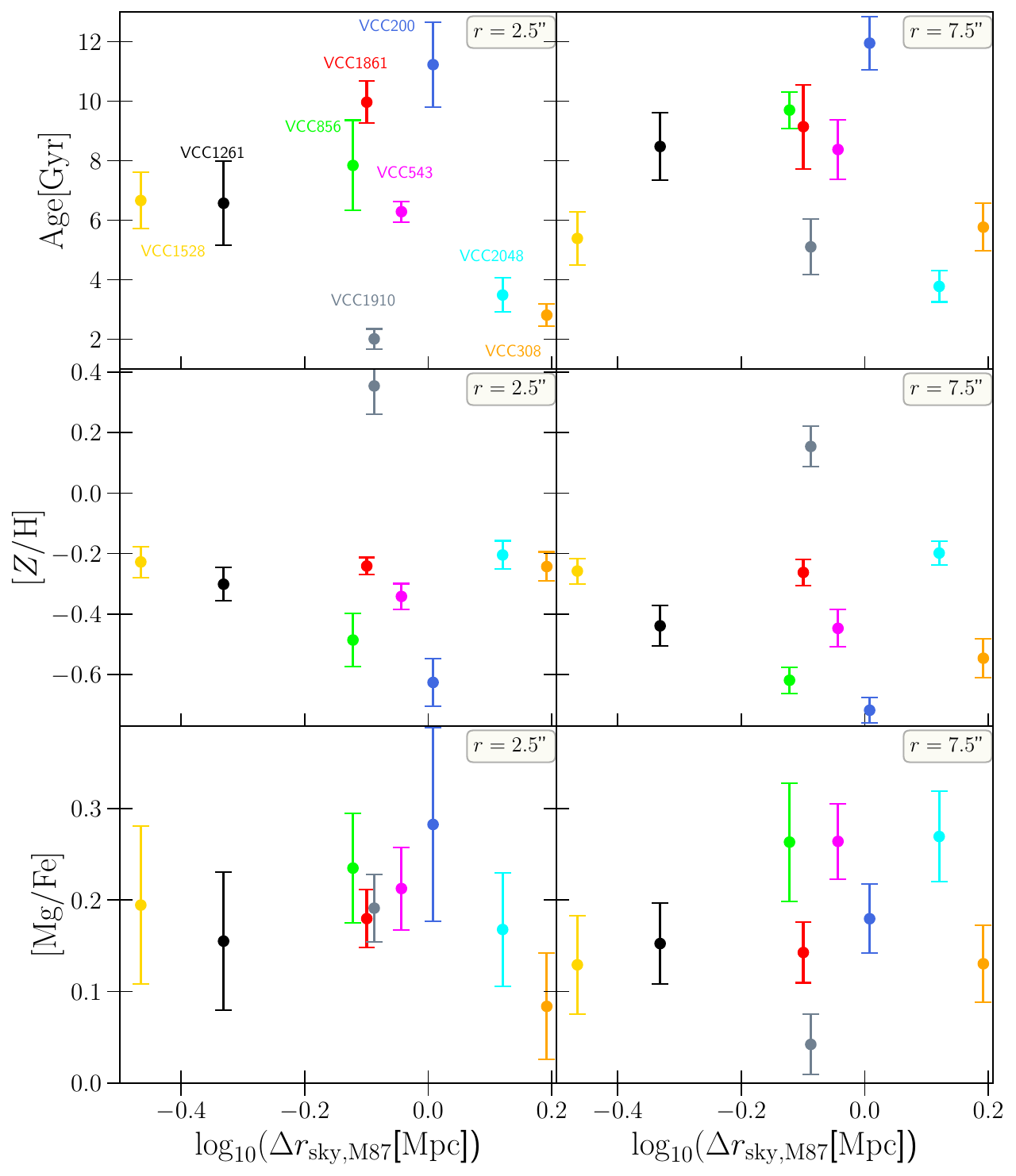}
    \caption{SSP age (\textit{Top}), metallicity (\textit{Middle}), and [Mg/Fe] abundance ratio (\textit{bottom}) extracted from the spectra at $r=2.5\arcsec$(\textit{Left}) and $r=7.5\arcsec$(\textit{Right}) versus the projected distance to M87 $\centredis$.} 
    \label{fig:pop_environment}
\end{figure*}

Contrary to the kinematics, there seems to be little to no correlation of the stellar population properties with the environment (Fig.~\ref{fig:pop_environment}). However, we do note that all dEs within the central Mpc (two-thirds of Virgo's viral radius) are older\footnote{We measure the average age of both apertures. For VCC 1910 we adopt the ages from the literature (cf. App.~\ref{append:Literature_comparison}).} than $6$ Gyr while all dEs outside this radius are considerably younger with ages of $3$ to $4$ Gyr. The two youngest galaxies (VCC 308, VCC 2048) are the galaxies furthest away from the cluster center\footnote{Note that VCC 1261 has a large line of sight distance measurement (cf. Tab.~\ref{tab:galaxy_table}). If correct VCC 1261 would be a further away from cluster center than the projected distance to M87 suggests. We have also checked whether the SSP properties are correlated with the 3D distance to M87 using the distance estimates in Tab.~\ref{tab:galaxy_table}, but the population properties remained uncorrelated with environment.} and both are located in a region with significantly lower projected number/luminosity densities than for the rest of the dEs \citep[see][]{Sybilska_2017}.  

The question whether the population properties of dEs correlate with the Virgo environment is also debated in some of the existing population studies \citep[e.g.][]{Geha_2003,Michielsen_2008,Sybilska_2017} that we discuss in App.~\ref{append:Literature_comparison}. Some find no correlation of environment with stellar populations while others find moderate correlations. In our dE sample it appears to be the case that within a projected cluster-centric distance of $\sim 1 \rm Mpc$, the dEs can have a large variety in ages and stellar populations, but at larger radii (near the virial radius of Virgo around  $\sim 1.5$ $\rm Mpc$) the dE population is dominated by much younger ages \citep[which was also noted by][]{Michielsen_2008}. This could indicate that any correlation with environment becomes quickly saturated as the dEs experience their first infall into the cluster and their stellar populations become \textit{homogeneous}. The young age of the galaxies outside the virial radius could imply that they have formed stars up until recently (see also Sec.~\ref{subsec:extended_star_formation}), while the broader distribution of ages deep within the cluster center is simply a stochastic imprint of the epoch a galaxy was being quenched at. Instead of a continuous correlation with cluster distance, one would then expect a distinct cut-off at a certain cluster radius. If this radius is indeed in the vicinity of Virgo's virial radius, our results favor a scenario where the driving quenching mechanism is a fast-acting mechanism like ram-pressure stripping. Such a process could already become efficient at the cluster's virial radius and even at the first infall a galaxy experiences. Stellar population properties found in \textit{Fornax} dEs seem to support this hypothesis more strongly than Virgo dEs, as \citet{Romero_Gomez_2023_A} found Fornax dEs exhibit no environment correlation with age and metallicity but a significant correlation with $[\alpha/Fe]$ that suggests at larger cluster-centric distances the star formation periods are more prolonged. A significant fraction of dEs may have also been quenched before arriving in the cluster due to pre-processing and, as they entered by the cluster, ram-pressure stripping triggered a short period of new star formation \citep[][]{Bidaran_2022,Romero_Gomez_2024}.

\begin{figure}
	\centering
	\includegraphics[width=1.0\columnwidth]{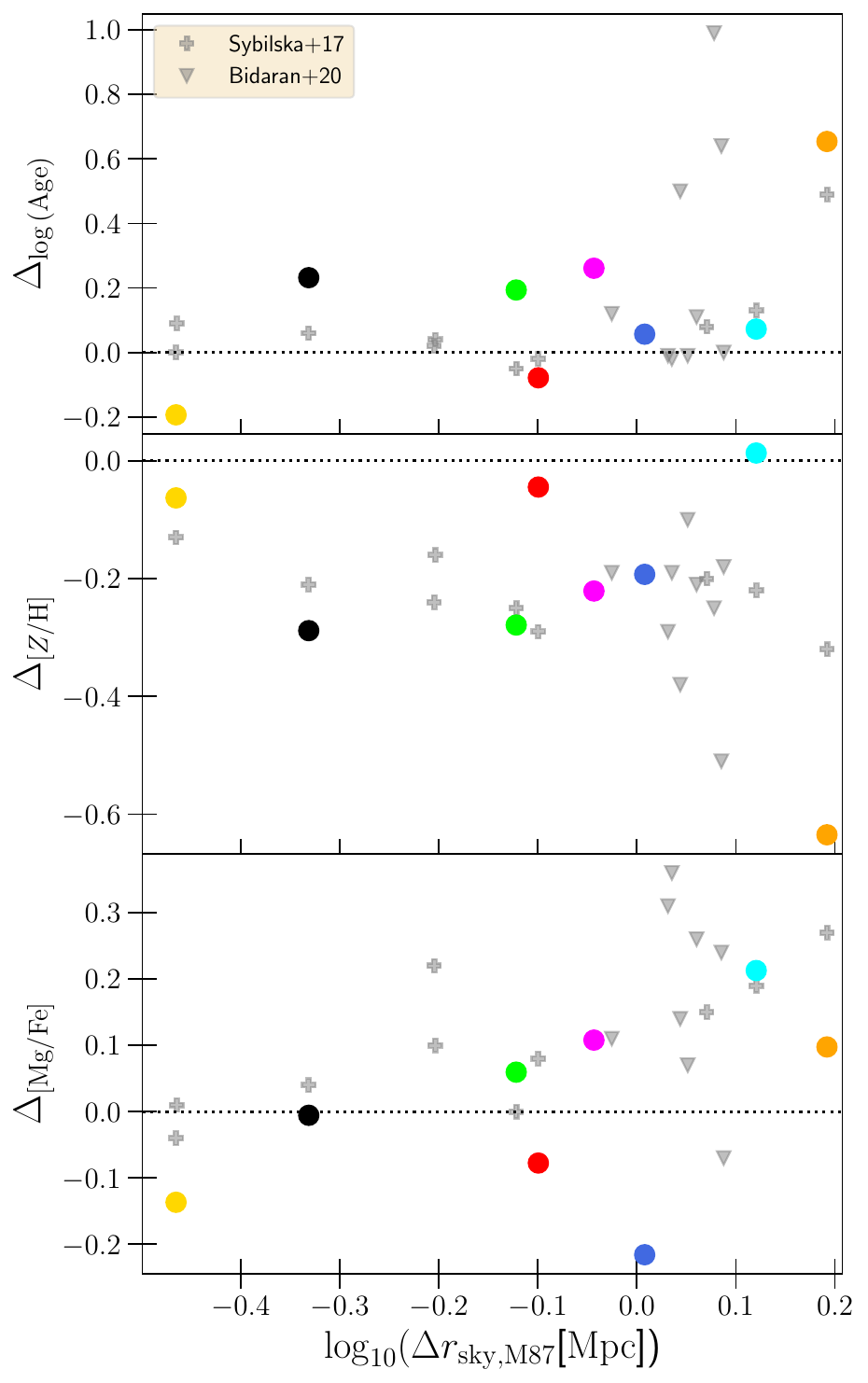}
    \caption{The log-linear population gradients versus environment for our sample together with the Virgo dEs of \citet{Sybilska_2017} and \citet{Bidaran_2023}. Both studies are based on IFU data and gradients are defined as the fitted slope of $\log \left( age \right)$, $Z/\mathrm{H}$, and [Mg/Fe] within one stellar effective radius $\log(\reff)$. The gradients for our dE sample are measured as the change between two apertures we have spectra for, i.e. $\log(r=7.5\arcsec)$ and $\log(r=2.5\arcsec)$. VCC~1910 is excluded because of its poor inner spectrum.} 
    \label{fig:pop_gradients}
\end{figure}

Environmental effects on a galaxy's stellar population will be different at different radii. Spatially resolving potential gradients is therefore important. We quantify such a gradient as the log-linear change between the two apertures of the population properties that were shown in Fig.~\ref{fig:pop_environment}. In Fig.~\ref{fig:pop_gradients} we show these gradients for our dEs as a function of their environment and compare them to gradients from other IFU studies of Virgo dEs \citep[][]{Sybilska_2017,Bidaran_2023}. Overall the age and [Mg/Fe] gradients are fairly flat for our dEs and exhibit no discernible preference to positive or negative values, but almost all dEs have moderately more metal-rich centers. Only VCC~308, which is the dE most distant from M~87 and the only one classified as having an extended blue center (Tab.~\ref{tab:galaxy_table}), and a few (but not all) of the \textit{infalling} dEs of \cite{Bidaran_2023} show more noticeable gradients. The centers of these few galaxies are younger, more metal-rich and less $\alpha$ abundant. The location of these galaxies near Virgo's virial radius suggests that as they fall into the cluster, the ram pressure exerted by the intra-cluster medium may have rejuvenated star formation in their centre \citep[][]{Bosselli_2022,Bidaran_2023}. For the galaxies that (presumably) entered the cluster a longer time ago there is no detectable correlation of the radial gradients with environment consistent with the analysis of \citet{Sybilska_2017,Bidaran_2023}. The relatively flat age and abundance gradients suggests the galaxy's main body is consistent with a nearly spatially homogeneous stellar population\footnote{This does not hold for the small blue nuclei in the centre \citep[cf.][]{Paudel_2011} which are not resolved in our study.}.

All together, the evidence for population-environment correlations in the literature are (especially in Virgo) still fairly tentative and exhibit, if any, only a moderate correlation. The inclusion of genuine field dEs in the population studies could help reveal a potential dichotomy of population properties as the global environment changes. 

The more continuous and stronger correlation of the angular momentum with environment, on the other hand, suggests that the kinematic structure evolves much more gradually than the stellar population characteristics do. These results are consistent with dEs being transformed late-type dwarfs that were \textit{quickly} quenched by their environment at the time of their first infall and, afterwards, were (more slowly) dynamically heated, losing their angular momentum in the process. In this formation scenario there exist a manifold of mechanisms that can explain this behaviour when thought of in superposition. At larger radii the quenching mechanism keeps the stellar rotation relatively intact (e.g. starvation, ram-pressure stripping) but stops the star-formation quickly. But at later stages, when the dEs are deeper in the cluster, processes like tidal disruption and harassment by more massive galaxies heat up the stellar orbits, transforming the dE further. Furthermore, if the dEs are still able to form stars during their first infall, their remaining gas disks may become misaligned with the already existing stellar distribution which could also lead to a dispersion dominated system as time progresses \citep[][]{Zeng_2024}.

However, we suspect that environmental effects are only a part of the broader picture. This is because environmental effects alone struggle to explain the dramatic increase of $\lambda$ with total mass that occurs in the ETG sequence between $10^{9}M_{\sun}$ and $10^{10}M_{\sun}$ (cf. Sec.~\ref{sec:empirics_kinematic}) because of two issues: Firstly, the LTG sequence exhibits a similar sudden $\lambda$ suppression in this mass regime (though not in $j$). Secondly, genuine field dEs that reside far outside any cluster also seem to display only low to intermediate angular momentum parameters \citep[][]{Janz_2017} and are still quiescent. Both observations suggests that, regardless of external influences, low mass galaxies in and of themselves have a low angular momentum parameter and it is possible for late-type dwarfs to be quenched even in low-density environments. Perhaps the reduced potential well makes the dwarf galaxies more susceptible to \textit{internally} induced dynamical heating and quenching, e.g. by supernovae winds or stellar bars, which causes an additional reduction of the angular momentum on top of the environmental influences.   

\section{Constraints from dynamical modeling}
\label{sec:dwarf_modelling}
In the following we present the stellar mass and kinematic structure of the dE sample with the modeling setup we described in Sec.~\ref{subsec:technique}. We plot the line of sight kinematic profiles of the best $\aicmod$ model we found for each galaxy together with the observed data in Fig.~\ref{fig:kinprofiles}. The best orbit models fit the observations well and reproduce all discernible features in the first few Gauss--Hermite moments despite the fact that we fit the entire FoV as a whole (instead of quadrants individually). This suggests that our sample galaxies do not show obvious evidence for being non-axisymmetric. In fact, it is remarkable that the kinematic moments of all dEs, fast rotators \textit{and} slow rotators, can be emulated this well by oblate axisymmetric models. In contrast, the most massive, slowly rotating ETGs often display photometric and kinematic signatures \citep[e.g.][]{Schechter_1978,Williams_1979,Binggeli_1985,Ene_2018,Neureiter_2023_b} that require \textit{triaxial} models and orbits to be emulated well. The fact that this is not necessary for any of the dEs may be another hint that they, unlike more massive ETGs, are indeed the remnants of transformed (oblate, axisymmetric) LTGs. 

\begin{figure*}
	\centering
	\includegraphics[width=1.0\textwidth]{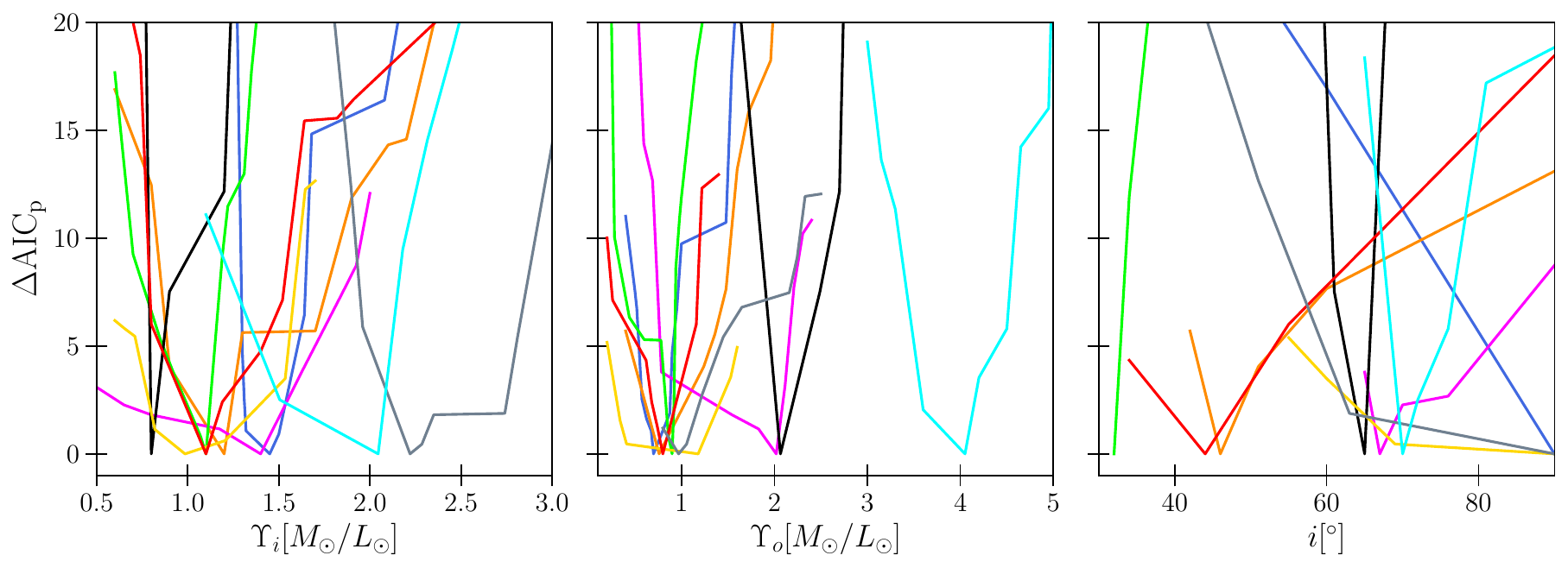}
    \caption{The $\Delta \aicmod$ envelopes for the intrinsic inner and outer mass-to-light ratios envelopes and the viewing angle of all the orbit models we probed for each dE. The stellar mass-to-light ratios are stated in their respective bands (Tab.~\ref{tab:galaxy_table}). To obtain a realistic estimate of the errors we compared the differences in the best 25 models. This is roughly equivalent to $\Delta \rm AIC \lesssim10$ criterion.} 
    \label{fig:dE_AIC_curves}
\end{figure*}

Fig.~\ref{fig:dE_AIC_curves} presents the $\aicmod$-constraints for the outer and inner mass-to-light ratios and the viewing angles we obtained from all the orbit models that were calculated. Together this 3 parameter set fully describes the 3D stellar mass distribution of the models $\rhostar=\Upsilon_{*}(\Upsilon_{i},\Upsilon_{o}) \cdot \nu(i)$.

\begin{figure}
	\centering
	\includegraphics[width=1.0\columnwidth]{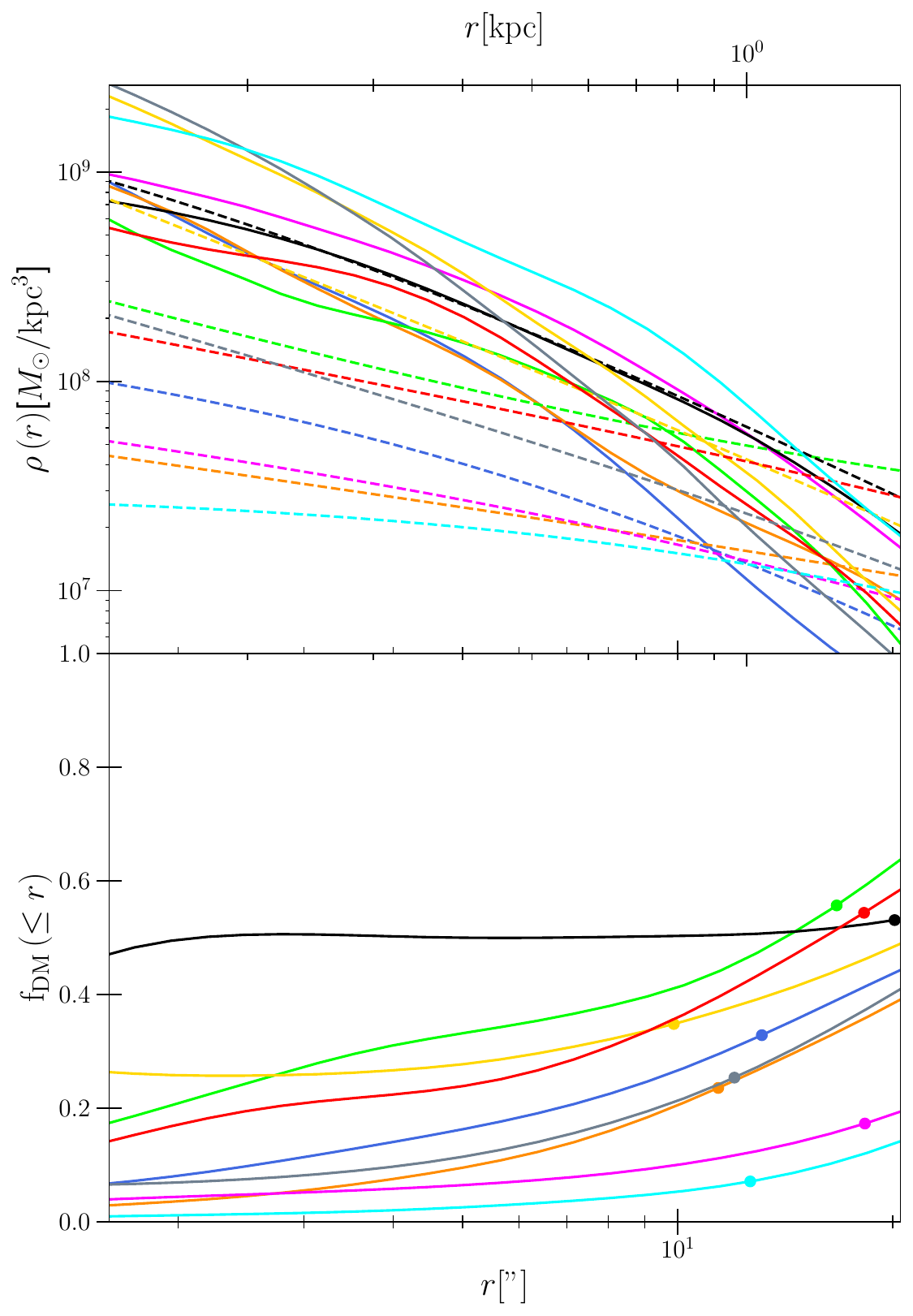}
    \caption{\textit{Top panel:} The (spherically averaged) matter densities for the dE sample that we obtained from the best $\aicmod$ model found for each dE. \textit{Solid curves:} The stellar density (including the $\Upsilon_{*}$ gradients). \textit{Dashed curves:} The corresponding dark matter halo density as modelled by the Zhao-parametrization (eq.~\ref{eq:dark_matter_model}). \textit{Bottom panel:} The cumulative Dark matter fraction $f_{\mathrm{DM}}$ within radius $r$. The dots highlight the value of $f_{\mathrm{DM}}$ at one stellar effective, the typical radial extent of the kinematic data is $1.0~\reff$.} 
    \label{fig:dE_radial_profiles}
\end{figure}

Instead of focusing only on the values of these nuisance parameters that are merely used to construct the 3D mass distribution, it is more meaningful to investigate the mass distribution they describe directly (see also argumentation in \paperrefemethods). Fig.~\ref{fig:dE_radial_profiles} illustrates the (spherically averaged) densities of both the dark and stellar mass distributions of the \textit{best} models we found for each galaxy, i.e. with $\Delta \aicmod=0$ as shown in Fig.~\ref{fig:dE_AIC_curves}. We also indicate the relative importance of both mass components by displaying the cumulative dark matter fraction $\dmfrac=\frac{M_{DM}}{M_{DM}+M_{*}}$ against the radius in Fig.~\ref{fig:dE_radial_profiles}.

Apart from VCC~1261, the baryonic stellar mass of the dEs seems to be distributed quite differently from their dark matter. The stars dominate the center, but unlike the halo their density falls off much more steeply further outside, which results in the drastic and monotonic increase in the dark matter fraction. At least within the radial regime we analyse ($\lesssim 1~\reff$), the dark matter seems to play a secondary role. We will analyse and contrast the radial density gradients and flattening of both stellar and dark matter mass in a more detailed manner in the companion paper (\paperrefeDARKMATTER). For the remainder of this work we focus on the interpretation of the recovered baryonic properties.

\subsection{Viewing angles}
\label{subsec:inclination}
We sampled 5 different inclination angles (and their corresponding deprojections) for each galaxy with the strategy described in Sec.~\ref{subsec:photometry}. The recovered inclination angles (Fig.~\ref{fig:dE_AIC_curves}) appear to be strongly constrained. VCC 856 is the only galaxy where the viewing angle reaches the lower sampling limit at $i=32\degr$. Axisymmetric deprojections of this galaxy with a significantly smaller inclination are inconsistent with the observed ellipticity profile. Therefore it seems the orbit model of VCC 856 prefers to be as close to face-on as possible, which would be in line with the faint near face-on, spiral arm signature (given the arms are aligned with the equatorial plane of the galaxy's main body spheroid). If the galaxy was closer to edge-on, we would not be able to see the pattern in the photometry. Likewise, the only other galaxy known to exhibit a hint of a visible spiral pattern, VCC 308, is also constrained to be fairly close to face-on (cf. App.~\ref{append:galdiscussion}). This consistency of the dynamically recovered viewing angles with independent photometric signatures (the spiral arms are not explicitly accounted for in the orbit models) are strengthening our confidence in the accuracy and importance of the inclination recovery which we first tested on simulations in \citet{Lipka_2021}. The average \textit{intrinsic} flattening of the stellar mass distribution which is associated with the recovered inclination angle is shown later in Fig.~\ref{fig:dE_aniso_flattening} together with the intrinsic stellar kinematic structure. We will also compare the intrinsic stellar axis ratios $q_{*}$ to the corresponding axis ratios of the dark matter component in \paperrefeDARKMATTER.

\subsection{Stellar mass-to-light ratio gradients}
\label{subsec:ml_gradients}
Dynamically derived stellar mass-to-light ratio gradients are scarce in the literature \citep[e.g.][]{Oldham_2018,Mehrgan_2024}, as mass-to-light ratios are often assumed to be spatially constant. However, even for the seemingly smooth ETGs, some studies suggest one may expect a spatial variation of $\Upsilon_{*}$. For example, for giant ETGs, several stellar population studies indicate a variation of the IMF inside the effective radius \citep[][]{Ferreras_2013,van_Dokkum_2017,Parikh_2018,Parikh_2024} which entails a corresponding variation of $\Upsilon_{*}$ there. Using the same axisymmetric modeling technique as we use, \citet{Mehrgan_2024} have detected a similar increase of $\Upsilon_\ast$ in the stellar bulges of 6 giant ETGs. All of them show a central peak in $\Upsilon_{*}$ with a gradient happening on a sub-kpc scale (more concentrated than previously anticipated).

These findings in the giant ETGs contrast with the dynamically derived \textit{stellar} mass-to-light ratio gradients of the dE sample in our study (see also Fig.~\ref{fig:ml_dyn_vs_ml_pop}). The dEs exhibit considerable diversity, with some galaxies radially increasing in $\Upsilon_{*}$, while others decrease. When calibrated in the \textit{z}-band\footnote{To calibrate the mass-to-light ratios to a common band we used a conversion factor derived from the stellar population models of \citet{Maraston_1998,Maraston_2005} assuming a Kroupa IMF and a range of metallicities and ages that are plausible for the dEs (Sec.~\ref{subsec:population_analysis}).}, the gradients of our dE sample are distributed around an average of zero.

\begin{figure}
	\centering
	\includegraphics[width=1.0\columnwidth]{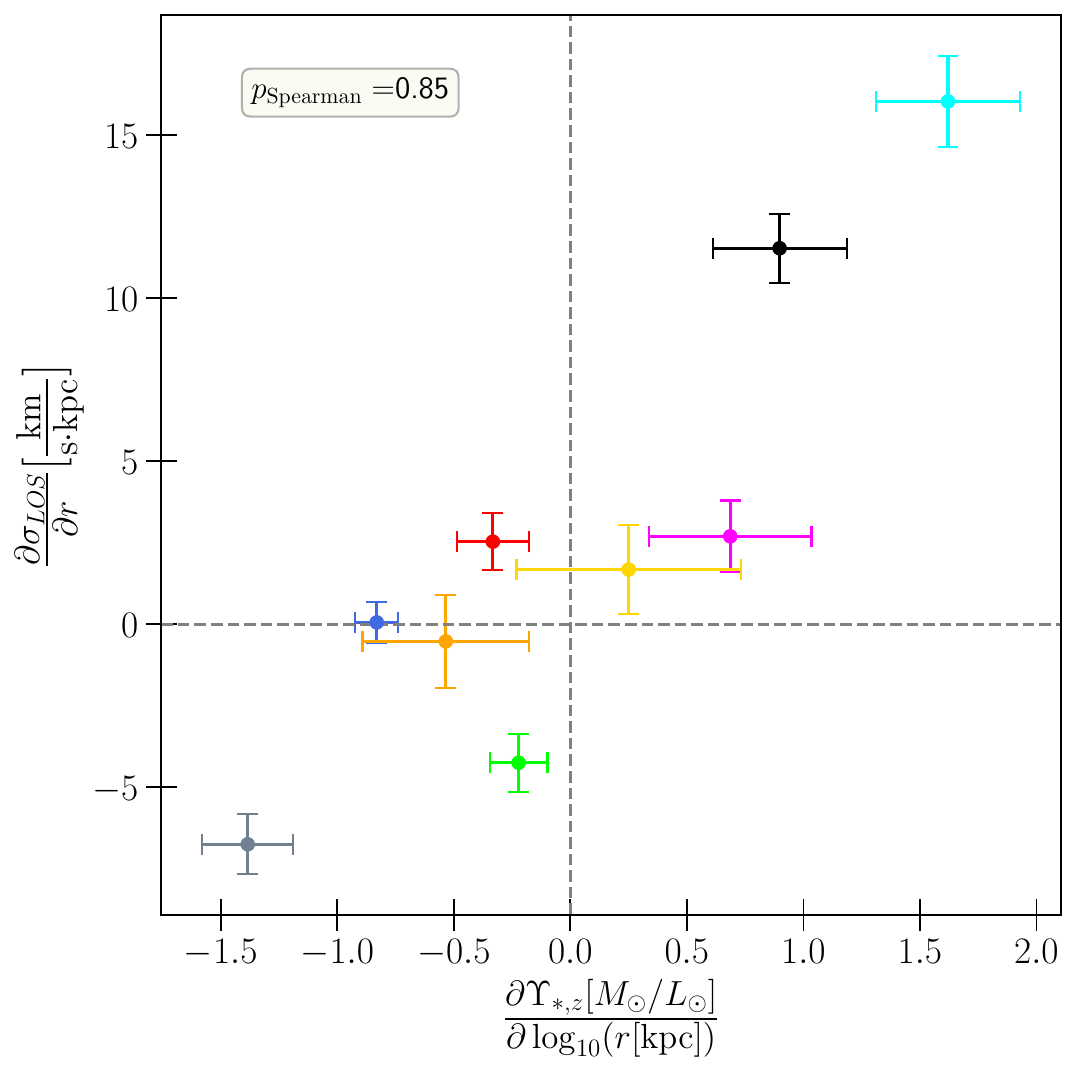}
    \caption{The observed average velocity dispersion gradients $\frac{\partial\sigma_{LOS}(r)}{\partial r}$, obtained from a linear fit to the observed radial dispersion profile, plotted against the 3D, log-linear, stellar mass-to-light ratio gradients of the dynamical models. The gradients were calibrated in the $z$-band for a consistent quantitative comparison.} 
    \label{fig:ML_GRAD-DISPERSION_GRAD}
\end{figure}

Another difference with the large ETGs is that in the dE sample \textit{all} stellar gradients are more moderate: the sample average of the \textit{absolute} gradient is $\left | \frac{\partial \Upsilon_{*,z}[M_{\sun}/L_{\sun}]}{\partial \log_{10}(r[\rm kpc])}\right | \approx 0.75 $. The most extreme gradient we find is that of VCC 2048 with $\frac{\partial \Upsilon_{*,z}[M_{\sun}/L_{\sun}]}{\partial \log_{10}(r[\rm kpc])} \approx+1.62$ (cf. Fig.~\ref{fig:ML_GRAD-DISPERSION_GRAD}). In contrast, the bulges of the giant ETGs in \citet{Mehrgan_2024} have a much larger sample average\footnote{The 3D gradients they found are stated in different bands but we approximately calibrated their values into the $z$-band for a comparison.} of $\frac{\partial \Upsilon_{*,z}[M_{\sun}/L_{\sun}]}{\partial \log_{10}(r[ \rm kpc])}\approx-4.37$. Note though this average value is largely driven by two of the giant ETGs with extreme gradients of around $-8$, while the rest of the sample is more moderate with gradients of about $-2.5$

It appears that dEs are distinct from the more massive ETGs in having much lower spatial variation with no systematic preference of rising or falling gradients. Since the stellar population properties (age and metallicity) are also approximately radially constant (Sec.~\ref{subsec:population_analysis}), the stellar component appears to be well represented by a single homogeneous stellar population, at least at the scales we investigated (i.e. between 0.1 kpc to 1 kpc). However, we do note that the study of \citet{Mehrgan_2024} was probing IMF variation in the very center of large ETGs. They choose to sample $\Upsilon_{i}$ and $\Upsilon_{o}$ at radii more concentrated in the center and not over the entire FoV like we did here. In other words, the gradients for both giant ETGs and dEs are measured between radii that are similar in physical scale (1 kpc scale) but relative to the overall galaxy size our gradients are sampled on a much more extended radial range. Perhaps one needs to probe the dwarf galaxies at much smaller scales to observe mass-to-light ratio gradients as strong as those found in giant ETGs. However, for the much smaller dEs this is currently unfeasible because our kinematic data does not resolve the very central parts well.

The inner and outer mass-to-light ratios $\Upsilon_{i}$ and $\Upsilon_{o}$ we sample with the orbit models determine the radial behaviour of the stellar mass-to-light ratio $\Upsilon_{*}(r)$ in 3D-space. In contrast, gradients recovered from a stellar population analysis (Sec.~\ref{subsec:population_analysis}) describe the stellar population properties as observed on the sky. As such, mass-to-light ratios derived from populations are actually the ratios of the projected populations on the sky. The \textit{projected} stellar mass-to-light ratios vs radius from both dynamics \textit{and} population models are shown later in Sec.~\ref{subsec:SSP_comparision}. In projection the gradients appear even more moderate than they are in 3D space.

Despite these overall only moderate mass-to-light ratio gradients detected in the dEs, we still notice interesting relationships between the dynamical $\Upsilon_{*}$ variation and other properties. For instance, the two galaxies which exhibit distinctly strong and \textit{positive} gradients, VCC 1261 and VCC 2048, also show distinctly different kinematic signatures in $\sigma$ and $h_{4}$, with them having strong central dispersion drops and negative $h_{4}$-gradients (cf. Fig.~\ref{fig:kinmap}). At the same time these two galaxies are also the galaxies with the largest measured velocity dispersion ($\sigma$$\sim$$50$ km~s$^{-1}$). The dispersion drop could be a result of recent central star formation where the young and bright stars that were being formed in the center have not been heated up yet. This lowers both the central velocity dispersion and mass-to-light ratio compared to the older, dynamically hotter main part of the galaxy. Alternatively, it could be a signature of small objects, like star clusters or minor galaxies, that were accreted and disrupted within the galaxy's extended envelope, increasing the random motions there.   

This relation between the kinematic moments and the dynamically determined $\Upsilon_{*}$-gradient seems to extend to the other galaxies: Fig.~\ref{fig:ML_GRAD-DISPERSION_GRAD} shows the intrinsic 3D mass-to-light ratio gradients of the best dynamical models we found for each galaxy versus the observed velocity dispersion gradients. The latter were determined from the slopes of a simple linear regression fit to the radial dispersions (Fig.~\ref{fig:kinprofiles}). As the Spearman correlation coefficient of $p_{\mathrm{Spearman}}=0.85$ indicates, the radial variation of the stellar mass-to-light ratio and the observed dispersion are strongly correlated. In fact if one were to exclude VCC 856 (for which the classification as a dE and its kinematic recovery are questionable) from the analysis, the evidence for this correlation becomes even stronger with $p_{\mathrm{Spearman}}=0.95$. Interestingly the relation between the dispersion and $\Upsilon_{*}$ also seems to cross the zero point, i.e. galaxies with essentially flat dispersion profiles also have flat stellar mass-to-light ratios. It seems to be the case that the local stellar mass-to-light ratio and line-of-sight dispersion increase in lockstep. Such a positive correlation with the dispersion could naturally arise through the change in potential that comes with locally changing stellar mass-to-light ratio $\Upsilon_{*}$. However, the dispersion gradients, particularly of VCC~1261 and VCC~2048, are very steep and confined to the central $5\arcsec$. The steep dispersion drops of these two galaxies towards the centre (Fig.~\ref{fig:kinprofiles}) could therefore be of different origin such as: the past accretion of smaller satellite galaxies, the presence and interaction with a bar, nucleus or embedded disk. However, none of these scenarios appears to be strongly supported by our analysis of the orbit structure that follows.

\subsection{Intrinsic kinematic structure}
\label{subsec:kin_structure}
The Schwarzschild models we employed yield constraints on the phase-space density of the stars. Therefore the 3D \textit{intrinsic} kinematic structure of the stars and their associated velocity moments can be inferred from the kinematics of the individual orbits of the model and the recovered orbital weights \citep{Thomas_2004}. 

In spherical coordinates the anisotropy parameter $\beta$ is a convenient quantity that describes by a single number the dynamical structure of the stars. It relates the second order velocity moments to one another as follows: 
\begin{equation}
\beta=1-\frac{\sigma_{\phi}^2+\sigma_{\theta}^2}{2\sigma_{r}^2}
	\label{eq:aniso_parameter}
\end{equation}
where $\sigma$ are the velocity dispersions in spherical coordinates \citep{Binney_2008}. A positive $\beta$ means the structure is radially anisotropic, i.e. the stellar velocity dispersion is larger in the radial than in the tangential direction. A negative $\beta$ on the other hand implies the opposite. 

\begin{figure*}
	\centering
	\includegraphics[width=1.0\textwidth]{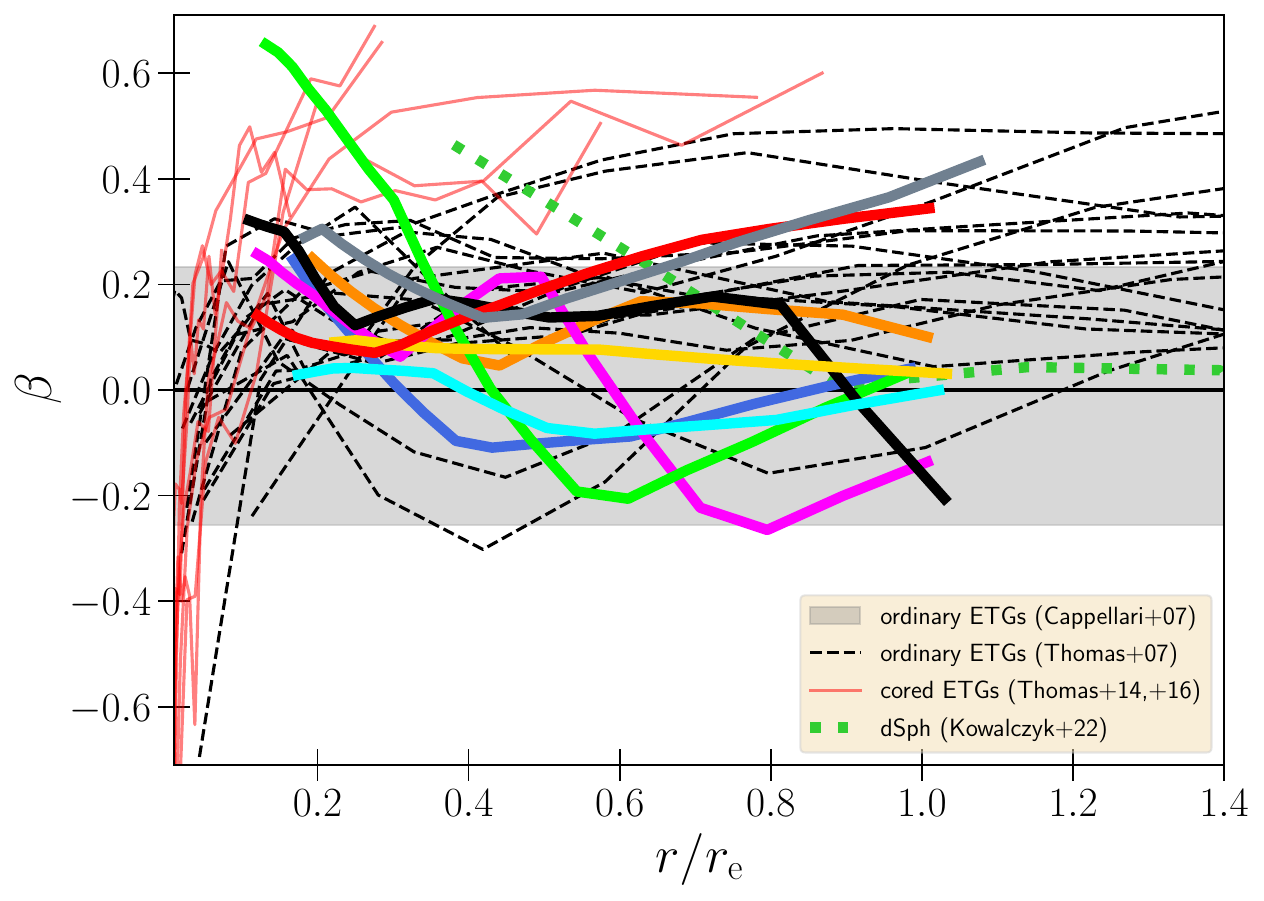}
    \caption{The spherical anisotropy parameter $\beta$ (eq.~\ref{eq:aniso_parameter}) vs radius scaled by the effective radius. The typical size of the FoV for our dE sample is about $1.0~\reff$, and outside of the FoV the anisotropy is not well constrained. The same is the case for radii smaller than the spatial resolution limit. Using our error estimation criterion (Sec.~\ref{subsec:technique}) we find a typical $1\sigma$ error of the anisotropy parameter of $\Delta \beta=0.065$ (within 1 kpc), with VCC 856 being the most uncertain at $\Delta \beta=0.11$. \textit{Red curves:} The anisotropy profiles of giant cored ETGs \citep{Thomas_2014,Thomas_2016}. Due to black-hole scouring these have formed tangentially anisotropic cores ($\lesssim0.05~\reff$) within a radially anisotropic envelope. \textit{Black dashed lines:} ETGs from the sample of \citet{Thomas_2007} which mostly consists of `ordinary' ETGs with magnitudes within $M_{B}\in [-18.8,22.6]~\mathrm{mag}$. \textit{Grey band:} The 90th percentile of the anisotropies of the 23 `ordinary' ETGs of \citet{Cappellari_2007}. Note that these values are radial averages. If the orbital structure changes from tangential to radial (as is often the case), then this averaging can make the galaxy appear closer to $\beta=0$ than it actually is. \textit{Green, dotted line:} The Fornax dSph from \citet{Kowalczyk_2022}.}
    \label{fig:dE_aniso}
\end{figure*}

Fig.~\ref{fig:dE_aniso} illustrates the spherically averaged anisotropy of the best dE models as a function of radius together with other ETG samples. Apart from VCC 856, the only genuine fast-rotator (Fig.~\ref{fig:angular_momentum}) and an outlier in many regards (App.~\ref{append:galdiscussion}), all dEs have a relatively homogeneous, isotropic orbit structure at most radii, with signs of mild radial anisotropy in the center ($<0.3~\reff$). At large radii near or outside the FoV and within scales smaller than the central spatial resolution limit, the profiles diverge more. However, this is expected as the inferred orbital motions at these scales become less reliable due to the lack of kinematic data constraints in these regions (see also \paperrefemethods). 

Despite the phenomenological diversity in the observed Gauss--Hermite moments (Fig.~\ref{fig:kinprofiles}), the intrinsic orbit structure of the galaxies seems fairly homogeneous. This lack of spatial anisotropy variations implies a more direct connection between the observed dispersion profiles (Fig.~\ref{fig:kinprofiles}) and the corresponding intrinsic \textit{mass} profiles, consistent with the strong correlation between mass-to-light ratio gradients and velocity dispersion gradients (Fig.~\ref{fig:ML_GRAD-DISPERSION_GRAD}).

An orbit structure that is essentially isotropic throughout with a mild radial anisotropy in the center is remarkable. In contrast, the $\beta$ profiles of `ordinary' ETGs can be quite heterogeneous. Power-law ETGs ($\log(M_{*}/M_{\sun})\sim 10-11$) may on average be nearly isotropic \citep[][]{Cappellari_2007} but their profiles can vary strongly with radius and are generally far from isotropic at any point \cite[e.g.][]{Thomas_2007,Thomas_2009_b,Thomas_2014,Santucci_2022}. The even more massive cored ETGs are distinct again. They are found to be tangentially biased in their center and strongly radially biased further out \citep[e.g.][]{Thomas_2014,Thomas_2016,Mehrgan_2019}, which is believed to be a result of a black hole core scouring process \citep[cf.][]{Kormendy_2009_B,Kormendy_2009,Kormendy_2013,Thomas_2014,Rantala_2018}. The scouring may even be a protracted process \citep[][]{Frigo_2021} which could also result in intermediate states being observed where the core is not yet tangentially biased \citep[e.g.][]{Neureiter_2023_b}. Conversely to both types of `ordinary' ETGs (cored and power-law) the structure of the dEs is quite isotropic and homogeneous. Furthermore the tendency to be mildly radially anisotropic in the centre appears to be unique to the dEs.

This orbit structure seems to suggest that the evolution of dEs is distinct and again we note the `trichotomy' across the stellar mass sequence from $10^{9}-10^{12}M_{\sun}$ (cf. Sec.~\ref{sec:empirics_kinematic}). However, we do caution that a comparison between the two is not straightforward because we probe different scales relative to the galaxies' sizes. The massive cored ETGs in Fig.~\ref{fig:dE_aniso} indicate that the stars typically are on more tangential orbits within about 0.5 kpc and become radial beyond $1.5 \rm kpc$. However, the physical scale involved here is the sphere-of-influence radius of the central supermassive black hole which happens to be around $0.1-1 ~\rm kpc$ for most of the massive core galaxies studied \citep[][]{Thomas_2016}. In the dEs, however, we do not resolve the sphere of influence of any black hole, i.e. we are not able to detect any scouring-related effects in our sample galaxies. 

The nearly isotropic structure (with mild radial anisotropy in the center) may be more akin to that of the less massive dwarf spheroidal galaxies (dSphs) and the dEs in the Local Group for which the anisotropy was studied. For example, \citet{De_Rijcke_2006} analyzed the velocity dispersion tensor of the 3 dE satellites of M31 based on slit kinematics and concluded all 3 galaxies are fairly isotropic within the central 0.8 kpc, the region that is most constrained by their data. \citet{Kowalczyk_2022} used Schwarzschild models for the \textit{resolved} kinematics of the Fornax dSph and found an anisotropy profile that remains fairly close to isotropy at most radii with signs of radial aniostropy in the very center. \citet{Jardel_2012,Jardel_2013_a} found $\beta$ profiles for the dSphs Fornax and Draco that are slightly radial yet close to $\beta=0$ with only a small increase in $\beta$ throughout the relevant radial range. Using a Jeans modeling technique, \citet{De_Leo_2023} found a very radially constant and close to isotropic orbit structure for the Small Magellanic Cloud after removing tidally disrupted interlopers from their analysis. The finding that the Milky Way's dSphs are fairly isotropic at all radii may be explained as a consequence of their tidal interactions with the Milky Way. N-body simulations of dSphs in a Milky Way like potential suggest that `tidal stirring' \citep[][]{Mayer_2001} could have transformed the orbit structure of initially disky dwarf galaxies to become fairly isotropic as a result of several interactions with the Milky Way \citep[cf.][]{Klimentowski_2009,Lokas_2010}. 

If this effect of isotropic transformation of dSphs via tidal interactions can be transferred to the larger scales of the more massive dEs, then the isotropic structure we find for the dEs is not necessarily their primordial state but a result of processing late-type progenitors via interactions with the cluster environment. It could also explain why VCC 856 does not conform to the homogeneous $\beta$ structure of the rest of our sample, as it has not yet been fully transformed (or is in the process of) as evidenced by the face-on spiral arms found in VCC 856. Tidal stirring would also be a natural explanation why dSphs and dEs tend to be predominantly pressure-supported and with low or moderate angular momentum (Fig.~\ref{fig:angular_momentum}) specifically in the center of Virgo (Fig.~\ref{fig:angular_momentum_vs_environment}). While the tidal stirring hypothesis may be successful in explaining the isotropic structure, the fact that our dEs inhabit fairly different environments within the Virgo cluster, and yet they all seem to have a similar isotropic structure, is peculiar. We would expect the dEs in the cluster center to have experienced on average more tidal passages, in which case we may expect a correlation of $\beta$ with the cluster environment which we do not seem to find.

Alternatively, if the isotropic structure we find is not a result of interactions with the cluster, it may stem from an \textit{internal} secular process that has driven the kinematic structure towards a specific configuration. In large ETGs the effects of these internal processes may be obscured by the more violent changes induced by mergers.

We found the \textit{spherical} anisotropy parameter $\beta$ to be almost constant with the radius and very homogeneous across all dEs in our sample. However, this impression of a completely isotropic stellar structure changes when looking at the anisotropy of the second velocity moments\footnote{In the axisymmetric models the second velocity moments equal the velocity dispersions in all but the azimuthal direction as it includes ordered motion as well.} in cylindrical coordinates defined as:
\begin{equation}
\beta_{z}=1-\frac{\left<v_{z}^2\right>}{\left<v_{r}^2\right>}
	\label{eq:beta_z}
\end{equation}
quantifying the anisotropy in the meridional plane. And in the corresponding sagittal plane:
\begin{equation}
\gamma=1-\frac{\left<v_{\phi}^2\right>}{\left<v_{r}^2\right>}
	\label{eq:gamma}
\end{equation}
  
\begin{figure*}
	\centering
	\includegraphics[width=1.0\textwidth]{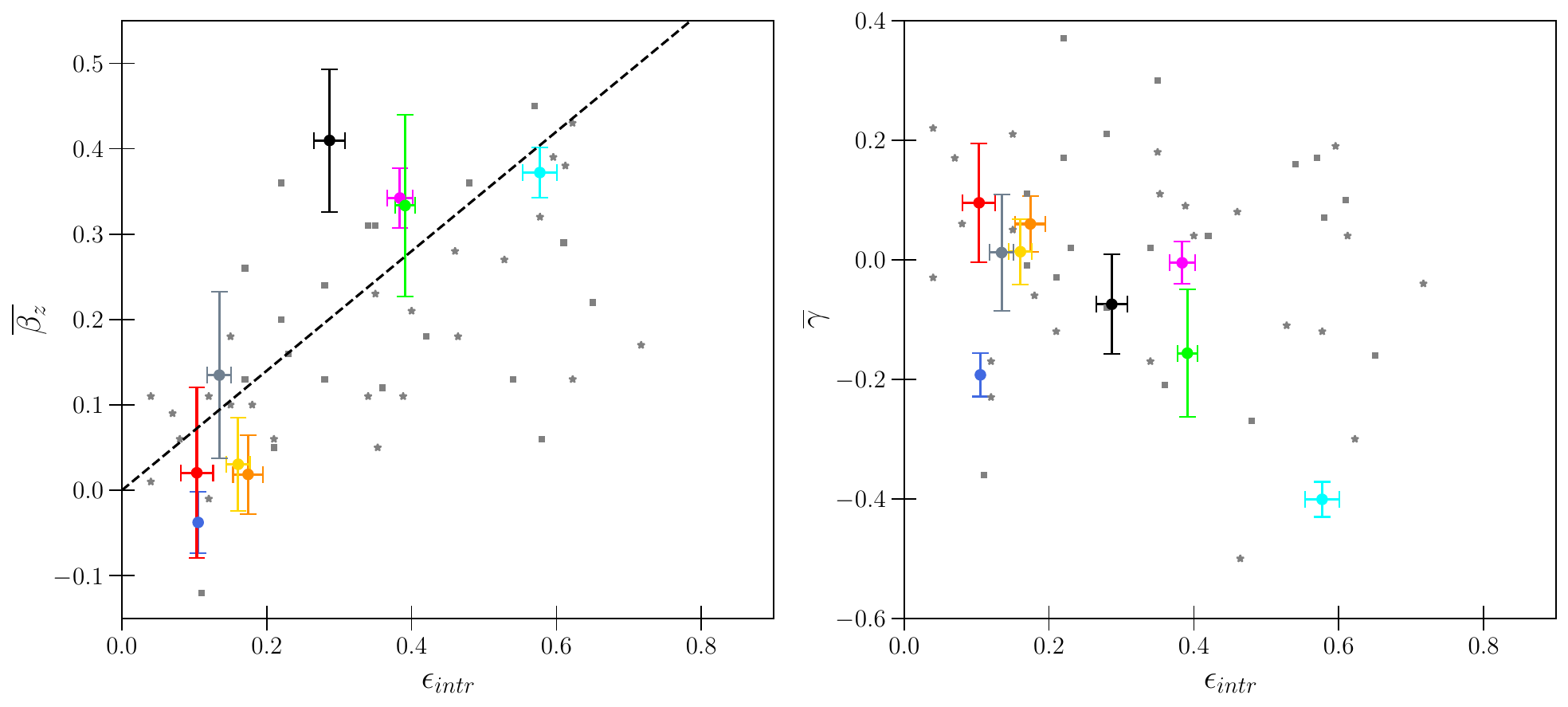}
    \caption{The average \textit{cylindrical} anisotropies $\beta_{z}$ (\textit{left}) and $\gamma$ (\textit{right}) vs the \textit{intrinsic}, average stellar ellipticity $\epsilon_{intr}$. The anisotropies were averaged within the sphere of radius $r\leq0.8 ~\rm kpc$ $(\sim 10\arcsec)$ because values outside the FoV are poorly constrained by the data. \textit{Squares:} The massive COMA ETGs of \citet{Thomas_2009_b}. \textit{Stars}: The ETGs from \citet{Cappellari_2007}. The \textit{dashed} diagonal line indicates the empirical relation $\beta_z=0.7\cdot \epsilon_{intr}$ from \citet{Cappellari_2007}, where flatter galaxies have higher anisotropy $\beta_{z}$.} 
    \label{fig:dE_aniso_flattening}
\end{figure*}

Fig.~\ref{fig:dE_aniso_flattening} shows these anisotropies (averaged within a sphere of radius $0.8 ~\rm kpc$ $\sim 10\arcsec$) versus the average \textit{intrinsic} ellipticity $\epsilon_{intr}$. For our dE sample both anisotropy parameters appear to be correlated with the intrinsic flattening. While the near spherical dEs are essentially isotropic, the velocity ellipsoid deforms as the stellar distribution becomes more flattened. The positive $\beta_z$ suggests that flattened dEs have a significantly smaller kinetic energy perpendicular to the equatorial plane, while the negative $\gamma$ suggests a relatively large azimuthal energy (only some of it comes from ordered motion). 

When compared to the angular momentum parameter of the galaxies (Fig.~\ref{fig:angular_momentum}) we find that most galaxies with a higher angular momentum parameter $\lambda_{e/2}>0.1$ also have a larger velocity dispersion in the equatorial plane ($\beta_{z} \sim 0.3$) suggesting both pressure and rotation support play a role in their flattening. There are exceptions, like VCC 1261, which is mildly flattened but has essentially no angular momentum (at least within $\reff/2$) both on the sky and also intrinsically in terms of ordered motion. Instead its flattening seems entirely supported by its relative lack of velocity dispersion in the $z$-direction. 

In Fig.~\ref{fig:dE_aniso_flattening} we also show the cylindrical anisotropies of `ordinary' ETGs \citep[][]{Cappellari_2007,Thomas_2009_b} which were obtained with axisymmetric Schwarzschild models. While the flatter dEs overlap with the `ordinary' ETGs, they trace the upper boundary in $\beta_z$ and the lower boundary in $\gamma$ of the ETG distribution. This may suggest that dEs are not heated like many of the most massive ETGs which have likely experienced mergers that would erase ordered motion more chaotically. Especially dry mergers are expected to make the orbit structure radially anisotropic, which is noticeable in a higher, positive $\gamma$ \citep[][]{Thomas_2009_b}. In comparison our dEs have low, negative $\gamma$ which makes (dry) mergers unlikely. Instead the shape of their velocity ellipsoid is still very much aligned with the orientation of a potential disky progenitor. This is compatible with the scenario that dEs stem from quenched LTG progenitors. A natural explanation for the large $\beta_{z}$ is that their kinetic energy is a relic of dissipation by a gaseous disk before the dEs were being quenched. For the flattened dEs, imprints of the disk structure remained in the orbit structure until today, while for the large fraction of near-isotropic and spherical dEs, the heating process has come to its conclusion, expunging any traces of its progenitor and its net angular momentum.

Orbit modelling of ETGs, in a similar mass range as our dEs, by \citet{Ding_2023} has suggested that the environment (cluster infall time) affects the orbits a galaxy's stars occupy. Galaxies that fell into a cluster recently have a higher fraction of orbits with high z-axis angular momentum than galaxies that entered the cluster long ago. However, contrary to the case of the angular momentum parameter $\lambda_{e/2}$ (cf. Fig.~\ref{fig:angular_momentum_vs_environment}), we do not find any strong correlation of the anisotropy and intrinsic velocity moments with the 2D (or 3D) distance to M87. Instead we find examples of flattened/anisotropic dEs but also spherical/isotropic dEs in both the cluster center and in the periphery.

All in all, dispersion support appears to be a non-negligible contributor to the dE flattening. Compared to `ordinary' ETGs the dEs have a lower amount of ordered motion (Sec.~\ref{sec:empirics_kinematic}), yet their anisotropy structure (high $\beta_z$ and low $\gamma$) that is not correlated with environment suggests their orbits are still very much aligned with the potential progenitor disk. It is not obvious why the ordered motion (i.e. angular momentum) of dEs should be diminished by the environment more effectively, while the higher kinetic energy in the equatorial plane is mostly preserved. One would expect tidal interactions that reduce the ordered motion to increase $\sigma_z$ and, hence, to reduce $\beta_{z}$. Instead, it may be more plausible that the suppressed angular momentum of dEs is not a result of environmental processing, but simply a feature of their different gravitational assembly. We revisit this scenario in \paperrefeDARKMATTERsp where the dark matter constraints allow further insights into the assembly conditions of dEs.

\section{The IMF of dwarf ellipticals}
\label{subsec:SSP_comparision}
Studies of the Milky Way and local late-type galaxies convey the picture of a \textit{universal} (Kroupa or Chabrier) IMF \citep[][]{Kroupa_2001,Kroupa_2002,Chabrier_2003,Brewer_2012}, yet the circumstances seem more complicated for early-type galaxies \citep[e.g.][]{van_Dokkum_2010}. For the most massive ETGs the results often suggest a more Salpeter-like IMF \citep[][]{Salpeter_1955} and evidence is mounting that the IMF varies with radius, age, or metallicity, and abundance ratios \citep[e.g.][]{Smith_2012,La_Barbera_2013,van_Dokkum_2017,Parikh_2018}. Determining whether the IMF of the dEs behaves more like that of the large ETGs or that of LTGs may allow us to distinguish between the formation scenarios responsible for the dEs. 

In our study we can compare two \textit{independent} measurements of the stellar mass-to-light ratios for each dE obtained from one and the same set of VIRUS-W spectra which allows us to probe the validity of an assumed IMF. The first mass-to-light ratio measurement we obtained is from the dynamical modeling (Sec.~\ref{sec:dwarf_modelling}). The second set of mass-to-light ratios stems from the stellar population analysis (Sec.~\ref{subsec:population_analysis}) where we assume a certain form for the IMF as a reference. We decided to use the Kroupa IMF as reference and calculated the present day stellar mass-to-light ratios $\Upsilon_{\rm Kroupa}$ from the SSP analysis for each of the analyzed spectra (i.e. at $r=2.5\arcsec$ and $r=7.5\arcsec$). In the following, unless stated otherwise, we specify $\Upsilon_{\rm Kroupa}$ in the same bands we used for the dynamical measurements (cf. Tab~\ref{tab:galaxy_table}). One can then compare the $\Upsilon_{\rm Kroupa}$ with the dynamically derived mass-to-light ratios and conclude whether the assumption of a Kroupa IMF is accurate or if the IMF should be more \textit{sub-Kroupa} or \textit{super-Kroupa}.

\begin{figure*}
	\centering
	\includegraphics[width=1.0\textwidth]{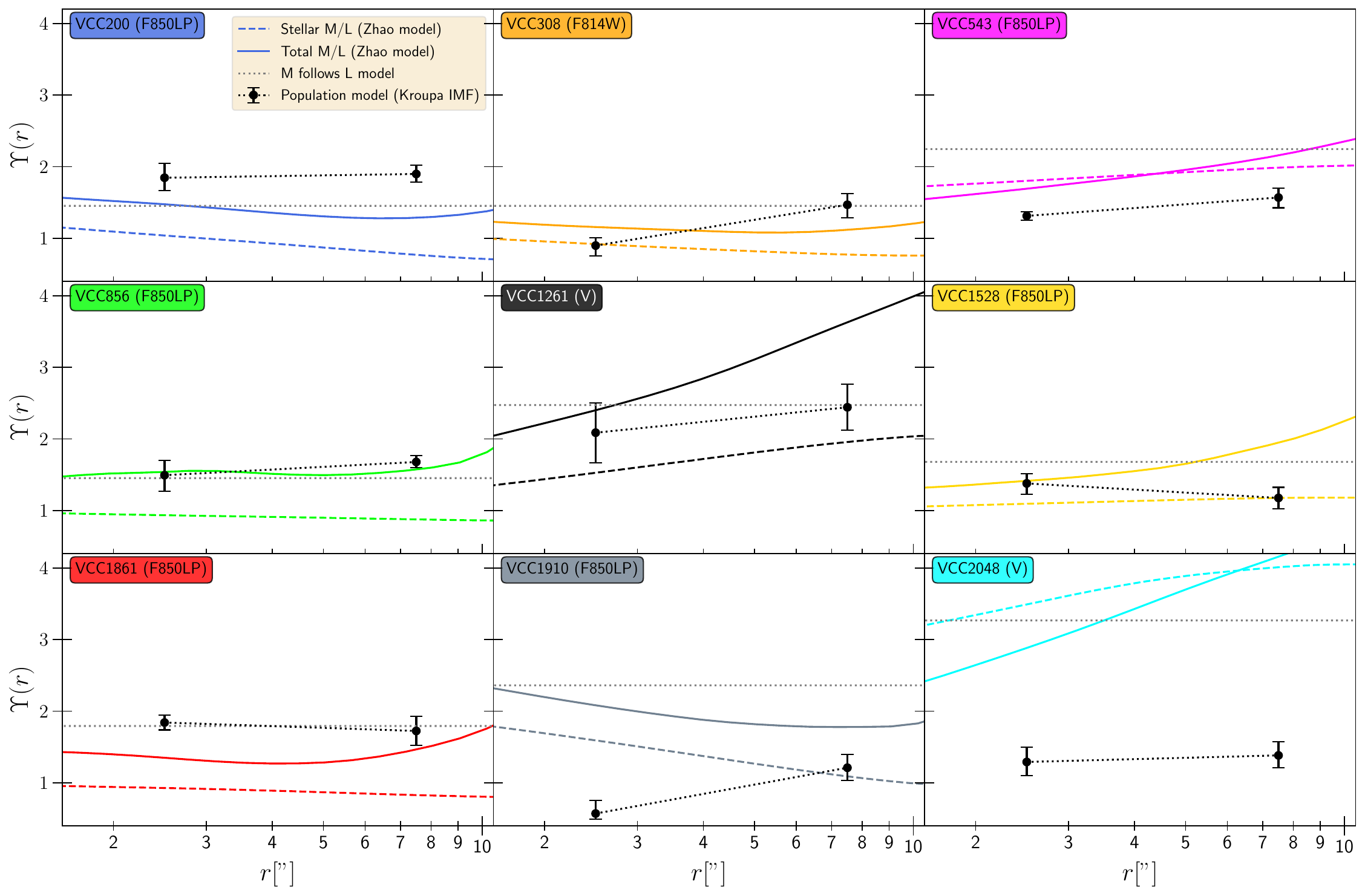}
    \caption{Comparison of the local mass-to-light ratio $\Upsilon$ as a function of radius $r[\arcsec]$ for all Virgo dEs as obtained from different models. Each panel shows one dE, with the box showing its VCC-id and the band in which the mass-to-light ratio is measured (cf. Tab.~\ref{tab:galaxy_table}). The \textit{colored} lines show the mass-to-light ratios we derived using dynamical modeling: \textit{Dashed:} The projected $\Upsilon_{*}$ of the dynamically decomposed stellar component. \textit{Solid:} The total \textit{3D} dynamical mass-to-light ratio $\Upsilon_{\mathrm{tot}}$ (i.e. including the dark matter). Note that $\Upsilon_{\mathrm{tot}}$ is not a projected quantity, as this would include all the dark matter along the LOS, i.e. even the poorly constrained DM far outside a galaxy's FoV. This can make the \textit{projected} $\Upsilon_{*}$ appear to be higher than the 3D $\Upsilon_{\mathrm{tot}}$ values at some radii, even though the latter is by definition the upper bound for the 3D stellar mass-to-light ratios. The \textit{gray, dotted horizontal} line shows the mass-to-light ratio of the best mass-follows-light model (i.e. a dynamical orbit model with no dark matter and stellar $\Upsilon$-gradient). The \textit{black dots} with errorbars show the stellar mass-to-light ratio from the population analysis (Sec.~\ref{subsec:population_analysis}) in the two annuli centered around $r=2.5\arcsec$ and $7.5\arcsec$. We also draw a connecting line between the two points to highlight a population gradient if it exists.} 
    \label{fig:ml_dyn_vs_ml_pop}
\end{figure*}

In Fig.~\ref{fig:ml_dyn_vs_ml_pop} we juxtapose the dynamical $\Upsilon_{\mathrm{dyn}}$ and the two corresponding population mass-to-light ratio $\Upsilon_{\rm Kroupa}$ which were obtained from the two spectra in the annulli centered around $r=2.5\arcsec$ and $r=7.5\arcsec$. For reference we also show the \textit{total}, \textit{3D} dynamical mass-to-light $M_{\mathrm{tot}}/L$ (i.e. including the dark matter, but excluding the SMBH) and the projected mass-to-light ratio obtained from a simple ad hoc dynamical modeling where mass follows light, i.e. a dynamical model with neither dark matter, nor a stellar gradient.    

As in other galaxy samples, we find there can be significant discrepancies between the dynamical and population mass-to-light ratios for some of the sample galaxies \citep[e.g.][]{Thomas_2011,Cappellari_2013,Posacki_2014,Mehrgan_2024}. Using the relative differences $\Delta_{*}=(\Upsilon_{\mathrm{dyn}}-\Upsilon_{\rm Kroupa})/\Upsilon_{\mathrm{mean}}$ with $\Upsilon_{\mathrm{mean}}=(\Upsilon_{\mathrm{dyn}}+\Upsilon_{\rm Kroupa})/2$ we find values up to $\Delta_{*}=\pm1$. An often used alternative to the IMF parameter $\Delta_{*}$ as defined here is the mass normalization parameter $\alpha_{\mathrm{IMF}}=\Upsilon_{\mathrm{dyn}}/\Upsilon_{\rm Kroupa}$. While the use of $\alpha_{\mathrm{IMF}}$ does not change any of the following conclusions we give values of this alternative parameter in Tab.~\ref{tab:results_table}.

\begin{figure*}
	\centering
	\includegraphics[width=1.0\textwidth]{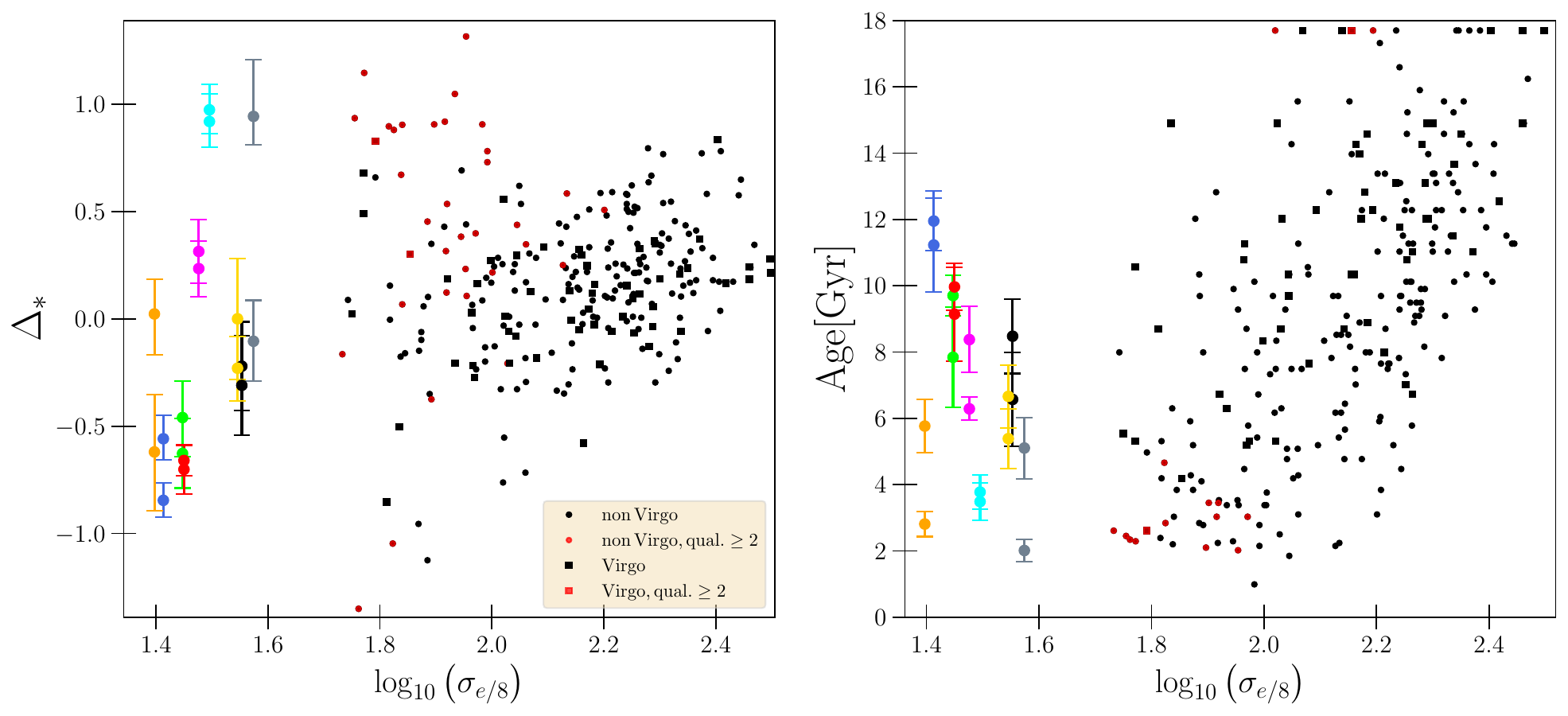}
    \caption{A comparison of the IMF parameter (\textit{left}) and the single stellar population age (\textit{right}) of dEs with those of the `ordinary' ETGs from the ATLAS$^{\rm 3D}$-survey \citep[][]{Cappellari_2011}. All galaxies are shown versus their central velocity dispersion (measured within $\reff/8$) which is essentially a proxy for their total mass. ATLAS$^{\rm 3D}$ ETGs that reside in Virgo are highlighted using squares, the others are shown as circles. We calculated the $\Delta_{*}$ for a Kroupa IMF by transforming the mass-to-light ratios from \citet{Cappellari_2013_B} which were given for a Salpeter IMF. The SSP equivalent age of the ATLAS$^{\rm 3D}$ galaxies was obtained from \citet{McDermid_2015} and is stated within a $1~\reff$ aperture. For our dEs we show the results for both apertures that we analyzed. \citet{Cappellari_2013_B} and \citet{McDermid_2015} flag their galaxies depending on the quality of the data and analysis. For details on the definition of these flags see the corresponding studies. We show galaxies flagged with the best quality in \textit{black} and `bad' galaxies that may not be as reliable in \textit{red}.} 
    \label{fig:IMF_dispersion}
\end{figure*}

Fig.~\ref{fig:IMF_dispersion} shows the IMF parameter and the SSP age for our dE sample together with the `ordinary' ETGs of the ATLAS$^{\rm 3D}$ survey \citep[][]{Cappellari_2011} scaled by the central velocity dispersion (which is a proxy for the total mass).

For giant ETGs the velocity dispersions are known to be positively correlated with the IMF parameter \citep[e.g.][]{Cappellari_2013_B,Cappellari_2016,Zhu_2024}. At the high mass end ETGs are almost all very old and their average IMF is Salpeter-like\footnote{Though this changes if spatial IMF gradients are considered \citep[][]{Mehrgan_2024}.}. But as the total galaxy mass decreases, galaxies often are younger as well and the typical IMF becomes progressively lighter, reaching Kroupa levels (i.e. $\Delta_{*}=0$). However, analogous to the break in the angular momentum parameter $\lambda$ at $\log(M_{*}/M_{\sun})\sim 10.5$ (or equivalently $\log(\sigma)\sim 2.1$), there appears a break in these ETG correlations. Above this mass threshold, $\Delta_{*}$ and the age are positively correlated with the dispersion and show relatively little scatter. Below the threshold the scatter between different galaxies increases dramatically and the IMF and age become independent of total mass and dispersion.

In this view our dEs are indistinguishable from those `ordinary' ETGs with dispersions below $\log(\sigma)\sim 2.1$. While our dEs can be both, substantially super-Kroupa or sub-Kroupa, they are within the large scatter of the `ordinary' low dispersion ETGs and they have Kroupa or Chabrier (i.e. just below Kroupa) IMF on average. A similar result was found by \citet{Tortora_2016} who compared the IMF parameter of the dEs sample of \citet{Toloba_2014} with `ordinary' ETGs. Their dEs also exhibit remarkable diversity, suggesting both super- and sub-Chabrier IMFs, with the average dE being consistent with a Chabrier IMF. This diversity they measured could have been due to the more simple, restrictive dynamical mass estimates they employed (spherical, isotropic Jeans equations), but the much more advanced Schwarzschild models employed in our work suggest a similar degree of IMF diversity.

\begin{figure*}
	\centering
	\includegraphics[width=1.0\textwidth]{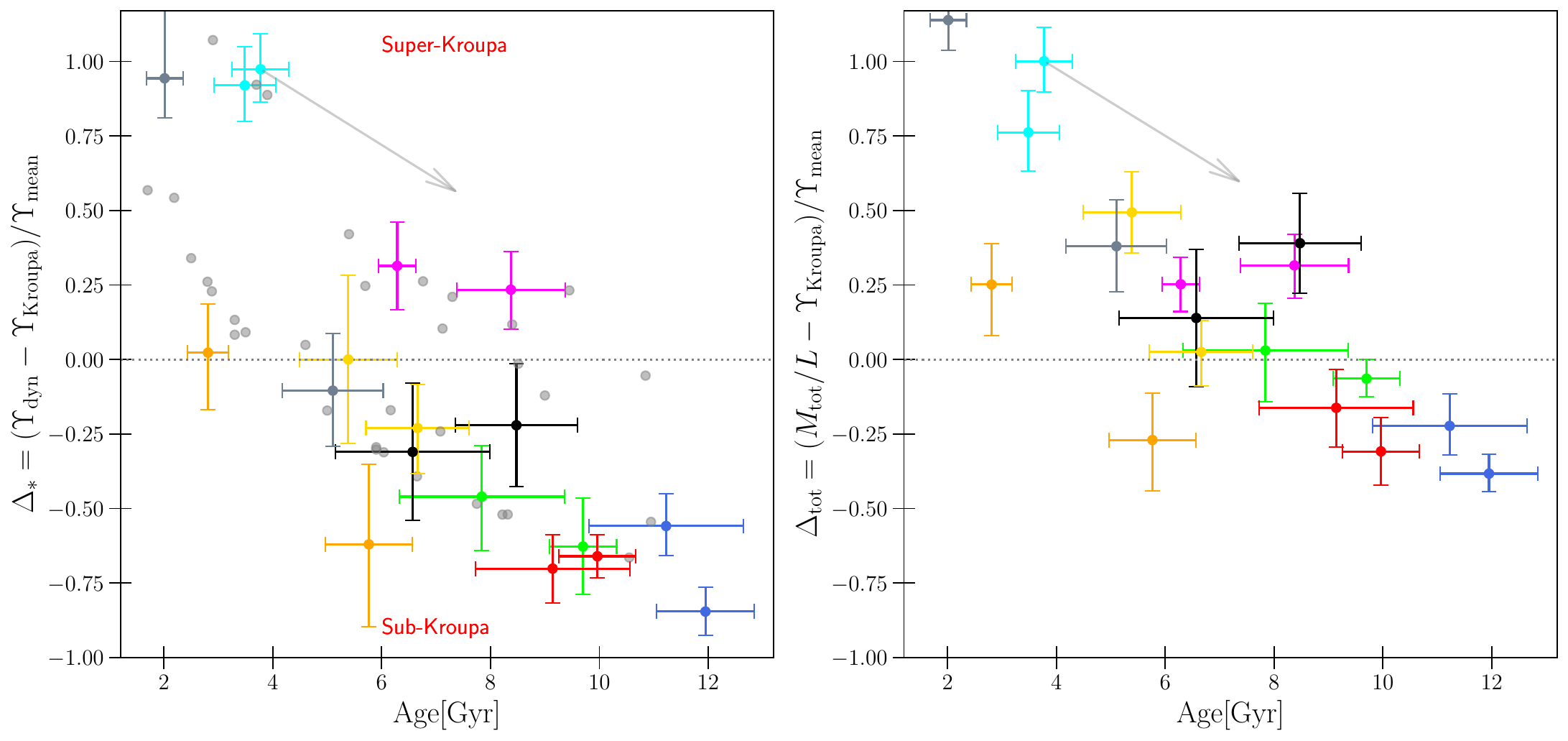}
    \caption{\textit{Left panel:} The relative differences $\Delta_{*}$ of stellar mass-to-light ratios derived from single stellar population analysis $\Upsilon_{\rm Kroupa}$ and dynamical modeling as a function of age of the SSP. \textit{Dots with error-bars:} The differences of the two mass-to-light ratios we derived from the VIRUS-W spectra binned at $r=2.5\arcsec$ and $r=7.5\arcsec$ (Sec.~\ref{subsec:population_analysis}). \textit{Gray Dots:} The differences when $\Upsilon_{\rm Kroupa}$ values are based on literature stellar population results (cf. App.~\ref{append:Literature_comparison}). The \textit{dotted, horizontal} line indicates identical population and dynamical results. Assuming the dynamical results are correct, galaxies that lie above this horizontal line would require a super-Kroupa IMF while galaxies below it a sub-Kroupa IMF. \textit{Right panel:} Similar to the left panel, but instead of comparing $\Upsilon_{\rm Kroupa}$ with the \textit{decomposed} stellar component $\Upsilon_{\mathrm{dyn}}$, the population ratio is compared to the \textit{total} dynamically mass-to-light ratio $M_{\mathrm{tot}}/L$, i.e. the dark matter component is included in the mass. Using VCC~2048 as an example, the \textit{diagonal arrow} symbolizes how an (unrealistically) large error in the SSP age would shift $\Delta_{*}$ and $\Delta_{\rm tot}$. We measure VCC~2048's population to be young, but if it were actually twice as old ($\sim 8$ Gyr), $\Upsilon_{\rm Kroupa}$ would be higher and the galaxy should actually be at the arrow's head if we measured its age correctly.} 
    \label{fig:Age_ML_Difference}
\end{figure*}

This substantial scatter of galaxies below $\log(\sigma)\sim 2.1$ poses the question whether it stems from a real non-universality of the IMF or is simply due to an increased statistical uncertainty in the dynamical or population modeling. If the IMF scatter of the low mass galaxies is a result of a non-universal IMF we may hope to identify a physical reason for changes in the IMF by looking at correlations of $\Delta_{*}$ with other properties. The strongest correlations we find are with the SSP age and metallicity. It appears that the older and the more metal-poor a stellar population is, the larger its $\Upsilon_{\rm Kroupa}$ is relative to $\Upsilon_{\mathrm{dyn}}$, which results in a smaller $\Delta_{*}$. This correlation with the stellar population properties is illustrated in the \textit{left} panel of Fig.~\ref{fig:Age_ML_Difference} which displays the differences $\Delta_{*}$ versus the SSP ages we derived from the VIRUS-W spectra. For comparison we also show the points one would get for $\Delta_{*}$ if one would assume the ages and metallicities published in the literature (cf. Fig.~\ref{fig:age_literature}). For the literature values we recalculated the mass-to-light ratios $\Upsilon_{\rm Kroupa}$ under the assumption of a Kroupa IMF. 

A similar Figure for the correlation with metallicity is shown in App.~\ref{append:IMF_metal}. Since the age of the dEs is anti-correlated with metallicity (Fig.~\ref{fig:SSP_diagnostics}) the $\Delta_{*}-[Z/\mathrm{H}]$ relation follows as a corollary. We can expect the average metallicity background for each cycle of star formation to increase with the formation epoch. This naturally changes the initial conditions for star formation over time and, possibly, the form of the IMF \citep[e.g.][]{Li_2023}. In the following sections we explore whether this correlation of the IMF parameter with age (and/or metallicity) is real and, if so, what could explain it.

\subsection{Are the dynamical models robust?}
\label{subsec:dynml_error}
The mass-to-light ratio discrepancies $\Delta_{*}$ could originate from an erroneous \textit{dynamical} stellar mass-to-light ratio estimate that is caused by a large statistical uncertainty in the dynamical decomposition of dark and stellar matter in which case the $\Upsilon_{\mathrm{dyn}}$ would not be very representative of the actual mass bound in stars. One could even assume the worst case where the mass decomposition is completely random and the $\Upsilon_{\mathrm{dyn}}$ for the different dEs randomly scatter around some average value depending on how much dark matter mass is mistakenly included. In that case we expect $\Upsilon_{\mathrm{dyn}}$ to be completely independent of age, yet the $\Delta_{*}$ would still appear to be anti-correlated with age at least qualitatively. This is because by definition the $\Upsilon_{\rm Kroupa}$ of the population models are increasing with age, which auto-correlates $\Delta_{*}$ and age. The direction of this auto-correlation is indicated by the arrow\footnote{The arrow quantifies how a change in SSP age alone would shift $\Delta_{*}$, i.e. $\Upsilon_{\mathrm{dyn}}$ and metallicity are kept the same. The shift is estimated assuming the single population models of \citet{Maraston_2005} with a Kroupa IMF.} in Fig.~\ref{fig:Age_ML_Difference}. 

However, we find the correlation between $\Delta_{*}$ and age we observe can not be driven by uncertainties in the mass \textit{decomposition} alone. This can be seen in the \textit{right} panel of Fig.~\ref{fig:Age_ML_Difference} which displays the differences $\Delta_{\mathrm{tot}}$ of $\Upsilon_{\rm Kroupa}$ relative to the \textit{total} dynamical mass-to-light ratios $M_{\mathrm{tot}}/L$. The total mass-to-light ratio $M_{\mathrm{tot}}/L$ includes the dark matter halo and, thus, reflects the total dynamical mass constraints. Consequently $\Delta_{\mathrm{tot}}$ does not depend on the merits of the mass decomposition but instead compares the dynamically (required) mass to the stellar population mass. Two issues are noticed: i) The correlation of the IMF parameter with age persists even if we include the dark matter; ii) Some of the galaxies (VCC 200, VCC 1861) have a negative $\Delta_{\mathrm{tot}}$, suggesting that the stellar population models predict a higher (local) mass than what is dynamically inferred. If correct this implies a lighter than Kroupa IMF for these two dEs or, alternatively, that there may be a systematic offset between the two modeling techniques. We will discuss the latter in App.~\ref{append:ML_check}. The fact that the anti-correlation with age persists even after including the dark matter means that independently of any mass decomposition, the total amount of dynamical mass relative to the expected stellar mass of a Kroupa IMF decreases systematically from younger to older galaxies. 

\begin{figure*}
	\centering
	\includegraphics[width=1.0\textwidth]{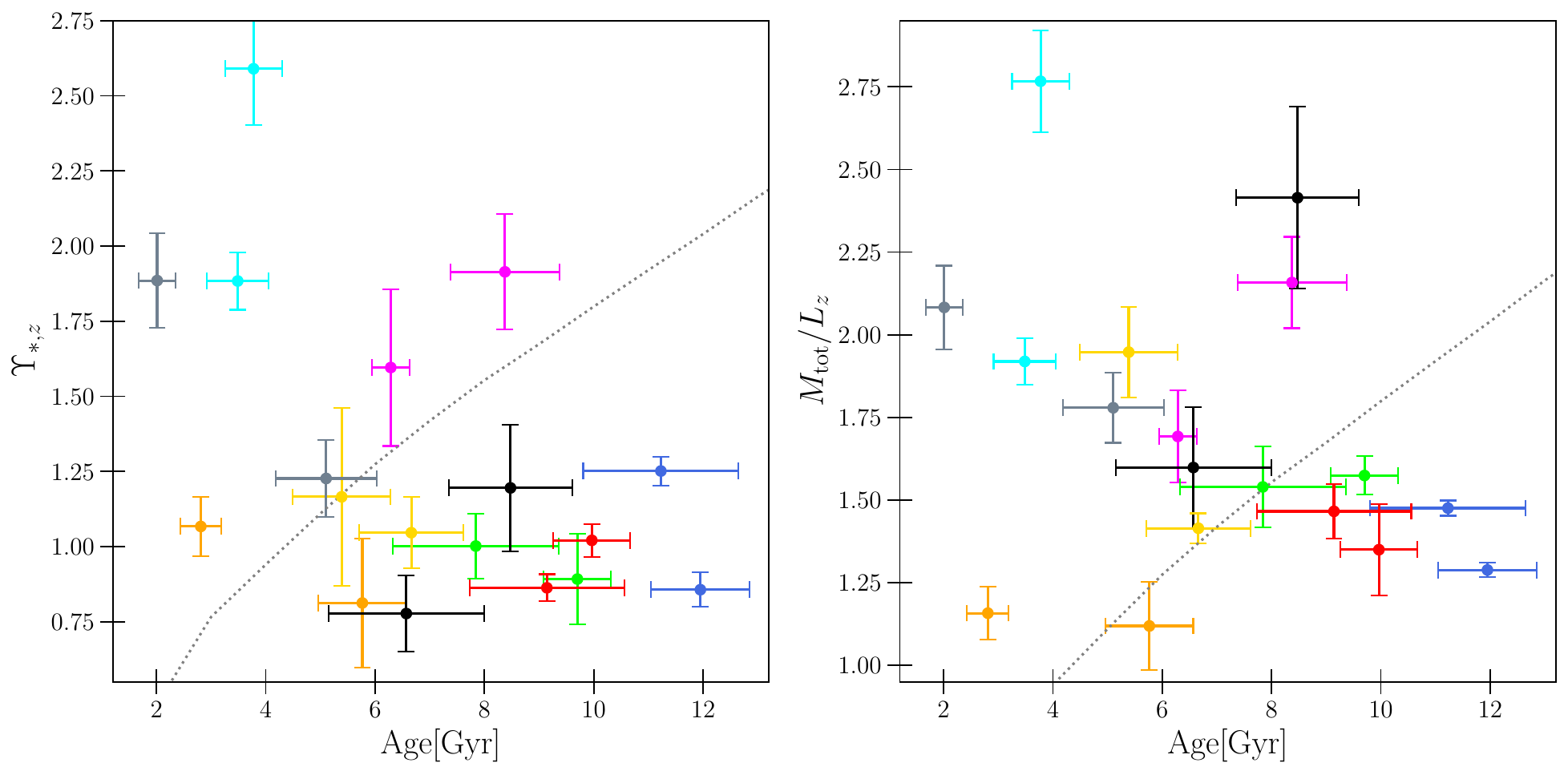}
    \caption{\textit{Left:} The dynamically recovered stellar mass-to-light ratios $\Upsilon_{*}(r)$ evaluated at radii $r=2.5\arcsec$ and $r=7.5\arcsec$ versus SSP age in the corresponding annulli. \textit{Right:} The same but for the total mass-to-light ratio (incl. dark matter). For this comparison we calibrated all mass-to-light ratios to the \textit{z}-band. The mass-to-light ratios appear to be anti-correlated with age. If the IMF is assumed to be universal across formation epochs, then this trend is counter-intuitive because one would expect mass-to-light ratio is \textit{positively} correlated with the time passed since star-formation. The dotted line indicates the expected present-day $z$-band mass-to-light ratio for a single stellar population with a universal Kroupa IMF and a metallicity of $[Z/\rm H]=-0.3$ \citep[][]{Maraston_2005} as a function of its age.} 
    \label{fig:age_vs_mldyn}
\end{figure*}

As mentioned above, $\Delta_{*}$ and population age are auto-correlated through $\Upsilon_{\rm Kroupa}$, hence one would expect to `see' an anti-correlation of $\Delta_{\mathrm{tot}}$ with age even if the dynamical mass-to-light ratio is roughly constant for all galaxies. We can remove this implicit dependence of the IMF parameter on age if we only plot the dynamically recovered mass-to-light ratio $\Upsilon_{\mathrm{dyn}}$ versus age, which is shown in Fig.~\ref{fig:age_vs_mldyn}. In both dynamical mass-to-light ratios (stars only, and total) the anti-correlation remains, although it appears to be weaker than Fig.~\ref{fig:Age_ML_Difference} indicated. In fact, the total mass-to-light ratio seems to be even more strongly anti-correlated then the stellar mass-to-light ratio.

As a sanity check, we have also computed dynamical models composed entirely of stars with a single, radially constant mass-to-light ratio optimised by fitting the kinematics. The mass profiles of these models are fixed by the light distribution. These models are worse fits with much larger $\aicmod$ values. In the sample median they are worse by $\Delta \aicmod=50$. Such fits are deemed as unacceptable models (cf. Sec.~\ref{sec:dwarf_modelling}) and due to their inflexible mass profile they often overestimate the mass of the best model at some radius (e.g. in the centre) and, in exchange, underestimate the mass at other radii (e.g. further outside). Still, the overall mass scale of these best mass-follows-light models as shown in Fig.~\ref{fig:ml_dyn_vs_ml_pop} is consistent with the mass-to-light ratios recovered with the more sophisticated dynamical models that included dark matter and stellar gradients. The anti-correlation with age remains even with the mass-follows-light models and all dynamical mass predictions seem to be fairly robust and independent of the specific choice of the dynamical model.

\subsection{Are the SSP properties robust?}
\label{subsec:pop_ml_error}
As mentioned above, the anti-correlation of the IMF parameter with age (Fig.~\ref{fig:Age_ML_Difference}) is not obvious to interpret quantitatively because $\Upsilon_{\rm Kroupa}$, and therefore also $\Delta_{*}$, are a function of the derived age. If the derived ages and metallicities are erroneous, or even completely random, one would still expect to see an anti-correlation because larger SSP ages always imply larger $\Upsilon_{\rm Kroupa}$ and vice versa. This effect is again quantified by the arrow shown in Fig.~\ref{fig:Age_ML_Difference}. If the population of a galaxy that we measure to be young is actually twice as old, then $\Delta_{*}$ and $\Delta_{\mathrm{tot}}$ would be reduced significantly. While such an unrealistically large error in age quantitatively is still not sufficient to bring all galaxies to a single universal level of $\Delta_{*}$, it makes it difficult to judge the strength of the correlation by eye.  

This is why a comparison of two independent measurements (like Fig.~\ref{fig:age_vs_mldyn}) is essential to quantify the strength or existence of such a correlation. Assuming the conditions are comparable for all dEs, e.g. differences in halo formation, environment, etc. are not a major factor in the measured $\Upsilon_{*}$ differences, then one may expect the dynamical mass-to-light ratios to be a function of the stellar population age. When a galaxy has stopped forming new stars and passively ages, we expect its $\Upsilon_*$ and $M_{\mathrm{tot}}/L_{z}$ to gradually increase\footnote{Of course younger galaxies could still have dust/gas left which could boost the mass-to-light ratio. However, none of the dEs in our sample exhibit any signs of dust, and they all have a fairly spatially constant color \citep[][]{Ferrarese_2006} apart from the negligibly small central nuclei (cf. App.~\ref{append:galdiscussion}). If there still are small undetectable differences in the dust content, we do not expect them to be able to explain the large variation in mass-to-light ratios across age we find (Fig.~\ref{fig:age_vs_mldyn}).}. If we assume a universal IMF we can quantify the expected change of the mass-to-light ratio from this aging process by evolving a typical stellar population. This is illustrated for a Kroupa IMF by the dotted line in Fig.~\ref{fig:age_vs_mldyn}. In terms of strength this expected mass-to-light ratio change per Gyr is roughly comparable to the trend suggested by the measured $M_{\mathrm{tot}}/L_{z}$ vs age measurements, but its sign is reversed. If we assume a universal IMF and robust dynamical measurements, this implies the actual age trend is roughly opposite to what is measured with the SSP models.   

Even assuming the worst case for the SSP models, which would be that they are entirely unconstrained and the derived ages are completely random, it seems strange that the dynamical and population measurement conspired to produce exactly the opposite to what one would expect from the mass-to-light ratio vs age behaviour. For the anti-correlation of the IMF parameter to be meaningless and solely be a result of errors in the SSP properties it would not be sufficient to randomly re-draw ages. Instead it would require the dEs we measured to be very old ($\sim 12$ $\mathrm{Gyr}$) to actually be very young ($\sim 4$ $\mathrm{Gyr}$) and vice versa.   

\begin{figure*}
	\centering
	\includegraphics[width=1.0\textwidth]{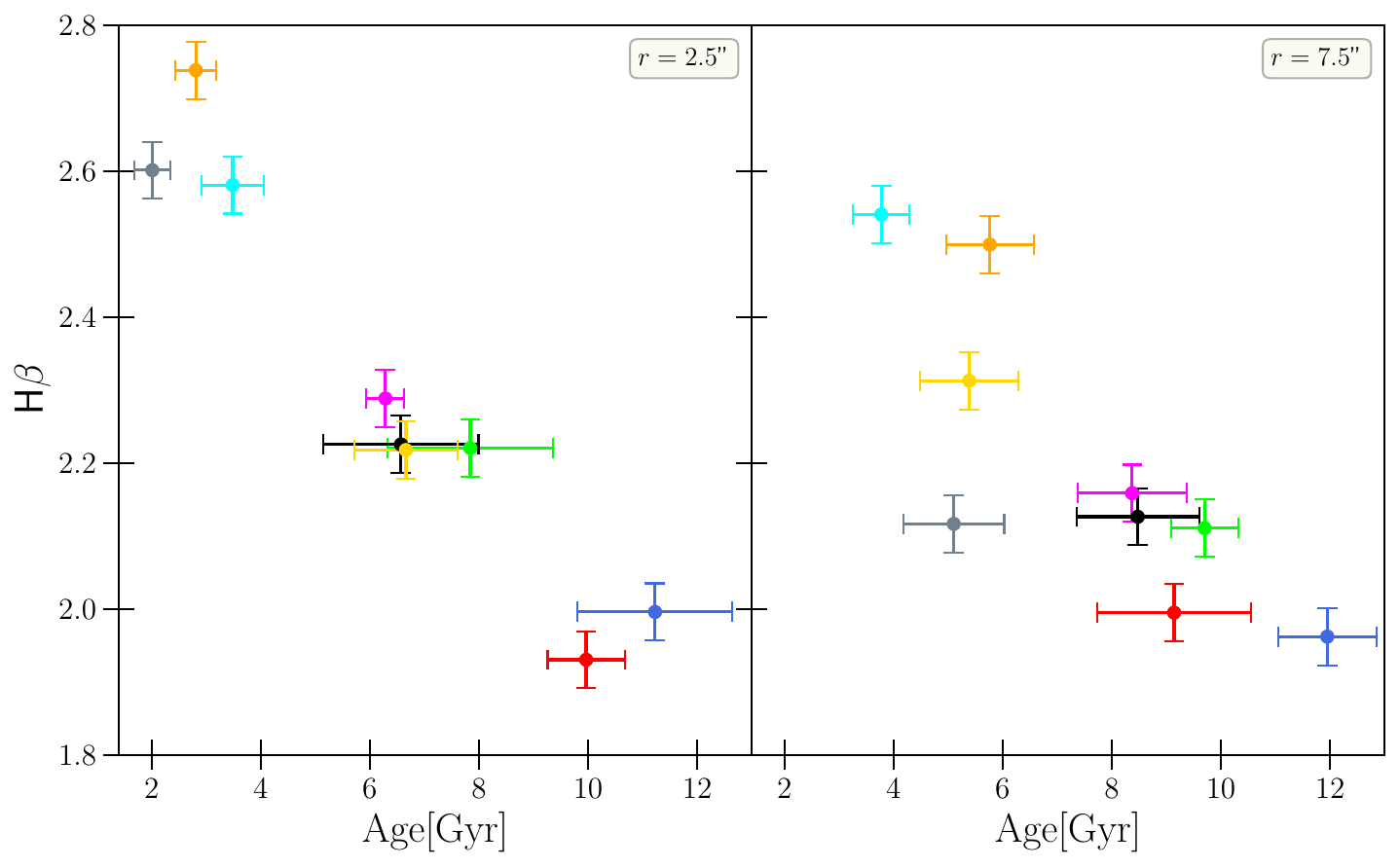}
    \caption{The H$\beta$ index vs. the SSP age derived from the two spectra at $r=2.5\arcsec$ \textit{(left)} and $r=2.5\arcsec$ \textit{(left)}. The measurement of the strong age-indicator H$\beta$ appears to be a strong predictor of the age, which suggests the age we derived is a direct result of the measured spectra and not a random modeling result.} 
    \label{fig:hbeta}
\end{figure*}

While such a coincidence can not be excluded, given such a small sample, we also have no reason to believe that the SSP properties are entirely random either. The stellar population ages are a direct consequence of the measured H$\beta$ indices (the strongest age indicator in the VIRUS-W spectra) which is shown in Fig.~\ref{fig:hbeta}. As expected, the H$\beta$ indices are anti-correlated with age and have realistic values. This suggests that there is no substantial issue within the SSP modeling itself.

Of course an error could already happen at the level of the index measurement. For example, unmasked emission lines can partially fill up the absorption lines, distorting the corresponding Lick indices, or similarly, problems in the spectral continuum determination or flux calibration could affect the indices. However, we did not find any evidence of emission lines in the spectra. Furthermore the comparison with previously published SSP results shows the same relative trends in age and metallicity we found (Fig.~\ref{fig:age_literature}), e.g. if we measure one galaxy to be younger and more metal rich than another, then this is also detected in the other results, even if the absolute differences in age and metallicity may be slightly offset.

In conclusion, for the anti-correlation of the IMF parameter to be solely a result of an uncertainty in the derived SSP ages, a systematic error at the level of the Lick indices present throughout the literature studies would be necessary that is strong enough to be able to change and reverse the derived trends with SSP properties. Or, alternatively, both SSP models and dynamical models have very large uncertainties but coincidentally produced the anti-correlation. A coincidence is not necessarily excluded given the small sample size. To tackle this, a larger number of independent and robust measurements of the \textit{dynamical} mass-to-light ratios will be needed. However, as of now not many dynamical constraints have been published and the existing ones rely on rough virial mass estimates or mass-follows-light models. These rough estimates are often further impacted by the low spectral resolution of the data which biases the inferred velocity dispersions (App.~\ref{append:Literature_comparison}).  

\subsection{Star formation history}
\label{subsec:extended_star_formation}
An assumption that comes along with the SSP modeling is that all stars have formed in a \textit{single} rapid burst of star formation. However, this may not be reflective of the actual star-formation history (SFH) the dEs have experienced. Two alternatives (or a mixture of both) are conceivable: i) The dEs have a bursty history, i.e. the galaxies are constituted of multiple distinct stellar populations that formed violently in multiple, separate bursts. ii) The dEs had a slow but prolonged star formation over several Gyr before they were quenched. 

A bursty SFH can happen if internal or external processes (e.g. mergers, ram-pressure-stripping, re-accretion, ...) are able to continuously trigger and stop multiple star-formation bursts. A plausible scenario is that after a violent initial burst, the resulting supernova feedback expels most of the gas temporally (but not indefinitely) into the intracluster medium before the galaxy re-accretes the gas again, triggering a secondary star burst \citep[][]{Seo_2023}. Such a process would depend on both environment and total inital mass. For example, a high mass dE in a low-density environment will be able to hold onto more of its gas without expelling it indefinitely and, thus, is able to trigger multiple bursts.

A slow and gradual SFH on the other hand can happen if the galaxy has a continuous supply of gas (e.g. through wet minor mergers or further gas accretion). Again high-mass dEs that inhabited low-density environments in their past and, thus, gravitationally dominated their local environment are likely candidates for such a prolonged SFH. The galaxies would have continuously formed stars during their LTG progenitor phase only stopping when quenched and transformed to a dE (e.g. as they experience their first infall into the cluster).

In either case (bursty or continuous) the extended SFH could lead to the anti-correlation of the dynamically measured mass-to-light ratios with age (Fig.~\ref{fig:age_vs_mldyn}). Supposing all dEs have formed in the same epoch (e.g. 12 Gyrs ago) but depending on their total mass and environment, some of them (the ones we find to be younger and metal-rich) were able to have an extended SFH. Then the apparently young and metal-rich galaxies could have significant amounts of older stellar populations with a higher mass-to-light ratio, which may be overshined by the youngest component from the last star-formation period. Moreover, they could host far more stellar remnants (from long passed SF periods) then the SSP models would predict. In that case the mass bound in stars would be higher than the luminosity suggests and the actual mass-to-light ratio would be higher than that of the most recently formed sub-population which might dominate the SSP model. 

Several of our findings support this SFH explanation for the $\Delta_{*}$-age anti-correlation. Firstly, the younger galaxies appear to have almost solar-like metallicity (Fig.~\ref{fig:age_literature}), suggesting the youngest population has been chemically enriched by the past SFH. Secondly, the two youngest galaxies are also found in the lowest density environments near the cluster's virial radius (Fig.~\ref{fig:pop_environment}), which may imply they have only recently been quenched. Thirdly and lastly, the dEs generally have low, near solar-like [Mg/Fe] ratios when compared to the higher ratios of more massive ETGs\footnote{A lower [Mg/Fe] is an indication of a more extended SFH because Mg (as an $\alpha$ element) is made predominately by supernova of type II whereas Fe builds up more gradually via type Ia.}, which implies the SFH of dEs were more gradual. This is not new and numerous studies of several abundance ratios have found that dEs have a chemical composition that indicate a prolonged SFH \citep[e.g.][]{Geha_2003,Michielsen_2008,Sen_2018,Romero_Gomez_2023_A}.  

\begin{figure*}
	\centering
	\includegraphics[width=1.0\textwidth]{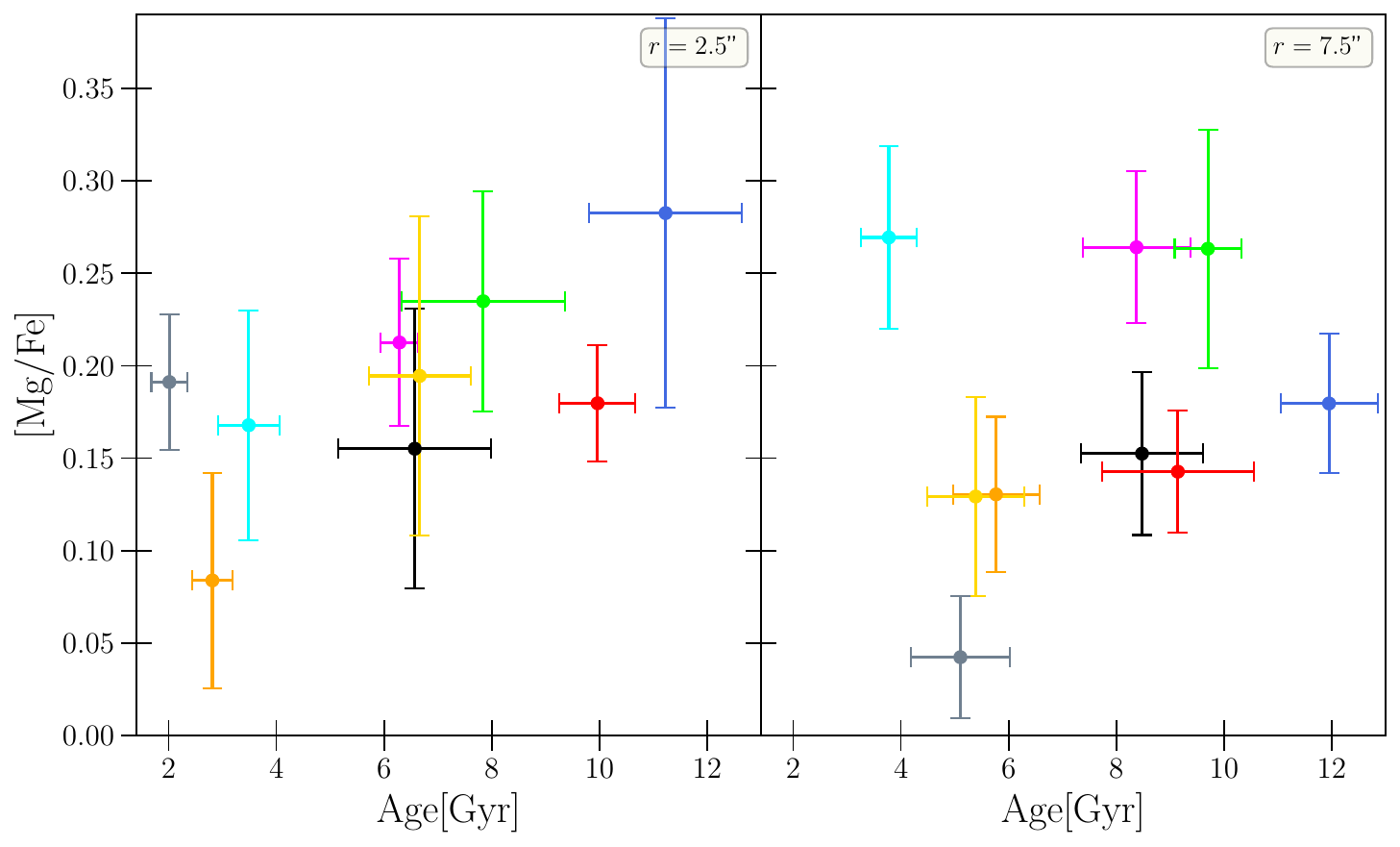}
    \caption{The abundance ratio [Mg/Fe] (a proxy for extended star formation) vs the single population age. Especially in their center ($r=2.5\arcsec$), dEs that are young also have lower solar-like [Mg/Fe], which suggests they had a more extended star formation history than older ones which (presumably) were quenched earlier.} 
    \label{fig:age-abund}
\end{figure*}

If the SSP ages we find are more representative of the epoch of last star-formation (i.e. of the time they were being quenched) and less so of the galaxy's time of formation, then we may expect that the [Mg/Fe] ratios are correlated with the SSP ages. Indeed we find signs of such a correlation as illustrated in Fig.~\ref{fig:age-abund}. Especially in the center ($r=2.5\arcsec$), the abundance ratios are strongly correlated with age (more so than with environment, see Fig.~\ref{fig:pop_environment}). 

If the dEs indeed have experienced a complex star-formation history over several Gyr, then one may be able to unravel the $\Delta$-age anti-correlation by modeling them with multiple populations (if one can get the degeneracies under control). As mentioned in App.~\ref{append:Literature_comparison}, \citet{Rys_2015} have tested the simplest form of extended star formation (2-burst models) on some of our sample galaxies. Unfortunately, for most of those dEs they often had trouble finding a solution for the older population, in which case they fixed its age to 12 Gyr to find a solution. Still, qualitatively, their results are fairly supportive of a complex, extended SFH. For example, while 70 to 80\% of the light of VCC 2048 (SSP age of $\sim$ 3-4 Gyr) comes from a young population (2 to 3 Gyr), about half the mass is currently found in the second older 12 Gyr population. While we can expect the IMF parameter of the dEs with old SSP ages ($\sim$12 Gyr) are robust (the \textit{single} population assumption is correct), this is less clear for the dEs with younger SSP ages. In a scenario like the above for VCC~2048 where half the mass is bound in an older `hidden' population, a single-population-model would result in an underestimation of $\Upsilon_{\mathrm{Kroupa}}$ and therefore an increase of $\Delta_{*}$. As such the observed $\Delta_{*}$-age anti-correlation could be an artifact originating from not accounting for the extended SFHs of dEs with younger ages. However, quantitatively it seems implausible that the SSP assumption alone could be responsible for the variation in the IMF parameter: to bring the very high $\Delta_{*}$ of the youngest dEs (e.g. VCC~2048) down to the sub-Kroupa level of the oldest (e.g. VCC~200) their $\Upsilon_{\rm Kroupa}$ would have to be up to 5.77 times higher than what we estimated with the SSP models. Nonetheless we may expect that the assumption of a single population could artificially increase the significance of the $\Delta_{*}$--age correlation. A robust modeling of the (potentially) complex SFH of dEs will be needed in the future.

It is likely that galaxies in the mass regime of our dE sample are particularly prone to display a complex extended SFH. More massive galaxies quench quickly because of feedback from active galactic nuclei. Less massive galaxies also quench quickly because they are much more vulnerable to rapid quenching by their environment or supernova feedback. However, in the intermediate regime of the dEs there may be a `sweetspot' at which a galaxy is just about massive enough to hold onto its gas reservoir for a prolonged period without ejecting it due to violent internal feedback. For example, while for the less massive galaxies gas could be removed indefinitely and rapidly in a \textit{single} quenching event, the slightly more massive dEs could be able to reaccrete some the ejected gas and rejuvenate their star formation. The results of \citet{Romero_Gomez_2023_B} indeed suggest that such a maximum-SFH sweetspot is somewhere between $10^8$ and $10^9 M_{\sun}$. They analysed the SFHs of 3 different galaxy samples (Local Group dSphs, Fornax dEs, massive ETGs) and found that the dEs formed their stellar mass more slowly than the galaxies in the other two samples. Surveys that study the mass regime between $10^{6}-10^{10}M_{\sun}$ using a \textit{single} cohesive sample/analysis may be able to confirm the existence and location of this sweetspot in the future.

\subsection{IMF variation with formation time}
\label{subsec:IMF_variation}
If an extended SFH is not sufficient to explain the correlations of $\Delta_{*}$ with age and metallicity quantitatively and the mass-to-light ratios from dynamical and SSP modeling are indeed robust, the alternative physical explanation is that the assumption of a universal Kroupa IMF is wrong and instead the true IMF changes over time and metal content. To be consistent with the dynamical masses, the galaxies with $\Delta_{*}>0$ would require a \textit{Super}-Kroupa IMF, while galaxies with $\Delta_{*}<0$ require a \textit{Sub}-Kroupa IMF. In that case the correlation suggests that the IMF changed from Sub-Kroupa IMF in the early Universe to a Super-Kroupa IMF in the present, while crossing the intermediate Kroupa IMF at around 4-8 Gyr ago. Our results are not the first indication that masses derived from SSP-modeling with an assumed universal IMF maybe be overestimated for older populations. For example, the anti-correlation observed in ETGs between central dark matter fraction and SSP age may well be explained by a lighter IMF of older populations \citep[cf.][]{Napolitano_2010,Tortora_2014}. This is also not the first time a variation of the IMF with time and metallicity content is proposed \citep[e.g.][]{Dokkum_2008,Li_2023}, with the theory being that at early formation epochs the IMF is more `bottom-light', or analogously more `top-heavy', than present-day IMFs. Such IMFs may be caused by the on average higher temperatures of the star forming clouds in the early Universe (e.g. due to CMB heating) or more effective stellar feedback in the early, low metallicity environments \citep[][]{Larson_2005,Chon_2022,Chon_2024}. 

Regardless, the large change in $\Delta_{*}$ we observe across our sample galaxies does require a significant change of the IMF during the lifetime of the Universe. At this point the variation (or universality) of the IMF with time is still a hotly debated issue \citep[for a review see][]{Bastian_2010}. While our dE sample does suggest a potential variation of the IMF across formation epoch, the evidence across the literature is still conflicting depending on the IMF measurement probe that is being used. For example, while our dwarf models suggest a \textit{negative} correlation of the IMF parameter with age, observations of `relic' galaxies (which are assumed to be probes of conditions in the early Universe) suggest the opposite trend: a \textit{positive} correlation with age \citep[][]{Navarro_2023}. \citep[][]{Parikh_2018} who studied the IMF of `ordinary' ETGs with masses $\log(M_{*}/M_{\sun})\sim 10-11$ find a similar, but stronger, metallicity trend as we do but their trend with age is again the opposite. If on the other hand we analyse the IMF parameters and SSP ages of the ATLAS$^{\rm 3D}$-survey \citep[][]{Cappellari_2011} we find no strong correlation with age in either direction. Only the few young ATLAS$^{\rm 3D}$ galaxies with $\sigma<100$ km~s$^{-1}$ tentatively show signs of $\Delta_{*}$--age anti-correlation. 

In conclusion, while some of the above discussed potential causes for the $\Delta_{*}$-age correlation we found appear more plausible than others, we believe a superposition of the discussed issues could be the most likely explanation. To confirm the veracity of this correlation, a larger sample will have to be investigated that ideally includes a larger variety in total mass and environments (e.g. field galaxies). At the current level of accuracy we do not believe we can employ the SSP results to complement or help improve the dynamical models of our dE sample, e.g. by using them as additional constraints to help decompose baryonic and dark matter (\paperrefeDARKMATTER). Neither do we deem the SSP results to be robust enough to corroborate the veracity of the dynamical mass decomposition. Future simultaneous population and dynamical analysis applied to the same data sets, such as done in this study, may be very conducive to test current modeling assumptions and confirm or deny our findings regarding the IMF and/or a prolonged SFH. 

A problem with investigating a potential non-universality of the IMF due to age (or metallicity) is that the single stellar population estimates are most sensitive to the epoch when a galaxy has \textit{stopped} forming stars than to the epoch when the galaxy has actually formed. Complementary to more sophisticated population models which may be able to account for complex SFH one may also look for probes of the galaxy formation epoch that are independent of stellar activity. Studying galaxies' dark matter halo densities (\paperrefeDARKMATTER) could be such a way to gauge the epoch of gravitational assembly independently and to discriminate whether the dEs have assembled in different epochs (implying a non-universality of the IMF with age) or all at a similar, early epoch (implying varying degrees of SFH and the IMF trend with age to be an artifact).

\section{Summary and Conclusions}
\label{sec:conclusions}
We have presented new kinematic data for a sample of 9 Virgo-Cluster dEs obtained with the high-resolution IFU-spectrograph VIRUS-W. It is the first 2D kinematic study of these objects with such a high spectral resolution which allows the recovery of velocity dispersions reliably down to $\sigma \sim 15$ km~s$^{-1}$ and out to approximately $1~\reff$. Previous observations with lower spectral resolution tend to yield dispersions that were biased high by up to $20-50\%$ (App.~\ref{append:Literature_comparison}). We also provide the first spatially resolved measurements of the higher order non-gaussian moments of the LOSVDs for these galaxies. The dEs exhibit diverse but systematic kinematic signatures in all Gauss--Hermite moments up to $h_{4}$. We find dEs with strong central dispersion drops as well as flat and slightly decreasing dispersion profiles. Many of the galaxies follow the same $v-h_{3}$ anti-correlation known from more massive galaxies. Some of the dEs also has a central peak in $h_{4}\sim+0.1$, whereas the profile decreases radially to $\sim 0$ which is often associated with a central dispersion drop.

\textbf{Mild dynamical mass-to-light ratio gradients.} We have used the spatially resolved LOSVDs to construct orbit-superposition models which allowed us to dynamically constrain their 3D intrinsic mass and kinematic structure. This study is the first attempt to dynamically constrain stellar mass-to-light ratio \textit{gradients} on the scales of dEs. The gradients we recover are generally low to moderate but we find a strong positive correlation of the stellar mass-to-light ratio gradients with the observed line-of-sight velocity dispersion profiles. Galaxies that feature a radially increasing $\sigma$ also increase in their stellar mass-to-light ratio with radius and vice versa, while dEs with a flat dispersion profile are also flat in their mass-to-light ratio. Averaged over the whole sample, the gradients are distributed around zero. 

\textbf{SSP gradients.} We also binned the spectra in two annulii to perform a single stellar population analysis with the goal to derive age, metallicity and abundance ratios. Our SSP results are in broad agreement with most of the existing literature, but the scatter and measurement error are considerable. Unlike the case for the LOSVDs, we have no reason to believe that our SSP results are significantly more reliable than the already existing measurements. We find little to no correlation between SSP results and cluster environment. Combining the SSP analysis with the dynamical results we do \textit{not} find any strong evidence of a spatial variation of the IMF as found for the bulges of massive ETGs \citep[][]{Mehrgan_2024,Parikh_2024}. On average dEs are consistent with a Kroupa-like IMF. Taken together, both SSP gradients and dynamical models paint a spatially homogeneous picture of the luminous matter. Apart from some outliers for which the interaction with the intra-cluster medium has presumably rejuvenated recently some central star formation, the stellar population properties and stellar mass-to-light ratios of dEs change at most only moderately. This suggests that the bulk of a dE's stars (within the apertures we investigated) has formed in parallel from the same IMF and dEs were quenched all at once.  

\textbf{Anti-correlation between M/L and age.} While the stellar changes within a single galaxy might be small, we find a much larger heterogeneity across the different sample dEs. Dwarf ellipticals display a greater diversity in age compared to the generally much older giant ETGs, with dEs SSP ages ranging from 2 to 12 Gyr. We find that their mass-to-light ratios are anti-correlated with this single population age, i.e. the younger the galaxy is, the more the dynamical mass-to-light ratio exceeds what is expected for a Kroupa IMF. This either suggests a dependence of the IMF on formation epoch, or alternatively, a more complex, prolonged star formation history. The former scenario suggests the Virgo dEs are constantly being produced in different formation epochs and environments, while the latter suggests all dEs have formed at roughly the same time early in the Universe, but the ones found to be young have had prolonged, complex star formation history whereas older galaxies were quenched early on. Compared to `ordinary' ETGs, little is known about the IMF of dwarf ellipticals yet, but our results suggest that future IMF studies will require more sophisticated population models than what was used for most `ordinary' ETG studies. To establish conclusively whether the IMF varies with age or not will require sophisticated extended SFH modeling. 

\textbf{Suppressed angular momentum.} In terms of their \textit{projected} angular momentum, the majority of dEs are either classified as intermediate rotators or slow rotators. The result stands in opposition to the angular momentum parameter of more massive ETGs. A plenitude of large samples, which study ETGs in the high mass ranges $\sim10^{10}-10^{12}M_{\sun}$, have identified a transition in the kinematic structure of ETGs around $10^{11}M_{\sun}$ \citep[e.g.][]{Emsellem_2011,Jin_2020,Santucci_2023} where angular momentum parameter and orbit structure change dramatically. For the most massive ETGs, the angular momentum parameter is low, more stars occupy more radial orbits, and they form tangentially anisotropic cores due to black hole scouring. In contrast `intermediate-mass' ETGs ($\leq 10^{11}M_{\sun}$) can often have more tangential orbit contributions and higher angular momentum. However, the results for the dEs suggest a \textit{second} point of change in the kinematic structure that occurs in the ETG sequence at even lower masses around $M_{*}\sim 10^{9.5}M_{\sun}$, at which point the amount of ordered motion is reduced again. This trichotomy in the ETG mass sequence appears to be traced by a corresponding behaviour in the total specific angular momentum $j=J/M$, such that dEs have significantly less angular momentum than expected from (dark matter only) structure formation models. The correlation of the angular momentum of dEs with their environment suggests that external influences play a role in this reduction. The galaxies become more susceptible to tidal perturbations and ram-pressure stripping as their total mass decreases. However, even the dEs in our sample that are at the cluster's virial radius appear to be only moderately rotating, which could suggest that the environment can not be the sole driver of the momentum reduction. \textit{Internally} induced heating processes could also play a significant role, and similar to the external processes, their impact can be expected to increase as the potential well of the galaxies becomes shallower. Together both external and internal influences may play an important role in shaping the orbital structure of dEs, but instead the suppressed angular momentum could also be more so a result of their distinct gravitational assembly rather than the result of those secular processes. This will be explored further in \paperrefeDARKMATTER. To investigate these different processes, future studies that include genuine field dEs far outside any cluster or group may be helpful to break this degeneracy.  

\textbf{A 3D orbit structure aligned with its star-forming progenitor.}
The above `trichotomy' in the angular momentum of ETGs is also echoed in the \textit{intrinsic} orbital structure. The dEs behave distinctly from both intermediate mass ETGs and the most massive ETGs. In spherical coordinates the intrinsic anisotropy parameter $\beta$ is close to isotropic or mildly radial. In contrast, more massive ETGs typically exhibit an anisotropy structure that varies more strongly with radius and is far from isotropy. Intermediate mass ETGs exhibit a range of different orbital structures and can be both significantly radial as well as tangential. The more massive cored ETGs are more homogeneous to each other, but have tangentially anisotropic cores due to black hole scouring. If we analyse the dEs in cylindrical coordinates, we find that intrinsically flattened dEs have an increased kinetic energy in their equatorial plane despite their aforementioned suppressed angular momentum. Flattened dEs in our sample tend to have a higher $\sigma_{\phi}$ in the equatorial plane rather than enhanced radial motions, which contributes significantly to their flattening by the (low) angular momentum. This orbit structure makes (dry) mergers very unlikely, and it may be interpreted as a (partially) intact relic of their star-forming progenitors. The low energy perpendicular to the disk again poses the question of how efficient external/internal heating mechanism really are, and whether the low angular momentum is more so an expression of dE assembly (\paperrefeDARKMATTER). 

All in all, dEs appear to be \textit{spatially} homogeneous in their stellar structure within the scales we investigated. Any stellar sub-structure must be too weak to be detectable on the macro level of the dynamical and population models. Still, detailed photometry studies \citep[e.g.][]{Barazza_2002} suggest dEs exhibit, even if only faint, substructure (see also discussion App.~\ref{append:galdiscussion}). In our sample galaxies most notably are the unresolved blue nuclei which seem to be more or less present in any dE, even if initially classified as a non-nucleated dE. Furthermore, some dEs may have weak embedded disks as suggested by faint face-on spiral arms or disky isophotes. Still, currently we do not find compelling evidence for the need of 2-component model structure (e.g. disk, bulge, bars,...) on a macro level. Perhaps future studies with a higher resolving power and $S/N$ may be able to use the detailed photometric substructure in their analysis.

The homogeneous \textit{spatial} structure we observe in the dEs contrasts with significant heterogeneity across the different sample galaxies. The dE population seems to be very different in their stellar mass and kinematic structure when compared to the rest of the ETG sequence. These differences may be explained by the very different evolutionary channel the dEs follow when compared to the more massive ETGs: For the latter a history of mergers and continued accretion of material has played the dominating role in shaping the orbit and mass-to-light ratio structure. Whereas for the dEs internal feedback process and external influences by the environment may have changed them significantly. This may have also left an imprint in the dark matter halos and shape of the mass distribution which we will investigate in \paperrefeDARKMATTER.

\begin{acknowledgments}
\section*{ACKNOWLEDGEMENTS}
We thank the anonymous referee for comments and suggestions that helped improving the paper. This work is based on observations obtained with the Harlan J. Smith Telescope at the McDonald Observatory, Texas. Computing has been carried out on the COBRA and RAVEN HPC systems at the Max Planck Computing and Data Facility (MPCDF), Germany. We also made frequent use of the NASA/IPAC Extragalactic Database (NED), operated by the Jet Propulsion Laboratory and the California Institute of Technology, NASA’s Astrophysics Data System bibliographic services, and the HyperLeda database \citep{Paturel_2003}. 
\end{acknowledgments}

\vspace{5mm}
\facilities{Smith, HST}
\software{astropy \citep{astropy_2022}}

\appendix

\section{Example for a typical LOSVD recovery}
\label{append:kindata}
\begin{figure*}
	\centering
	\includegraphics[width=1.0\textwidth]{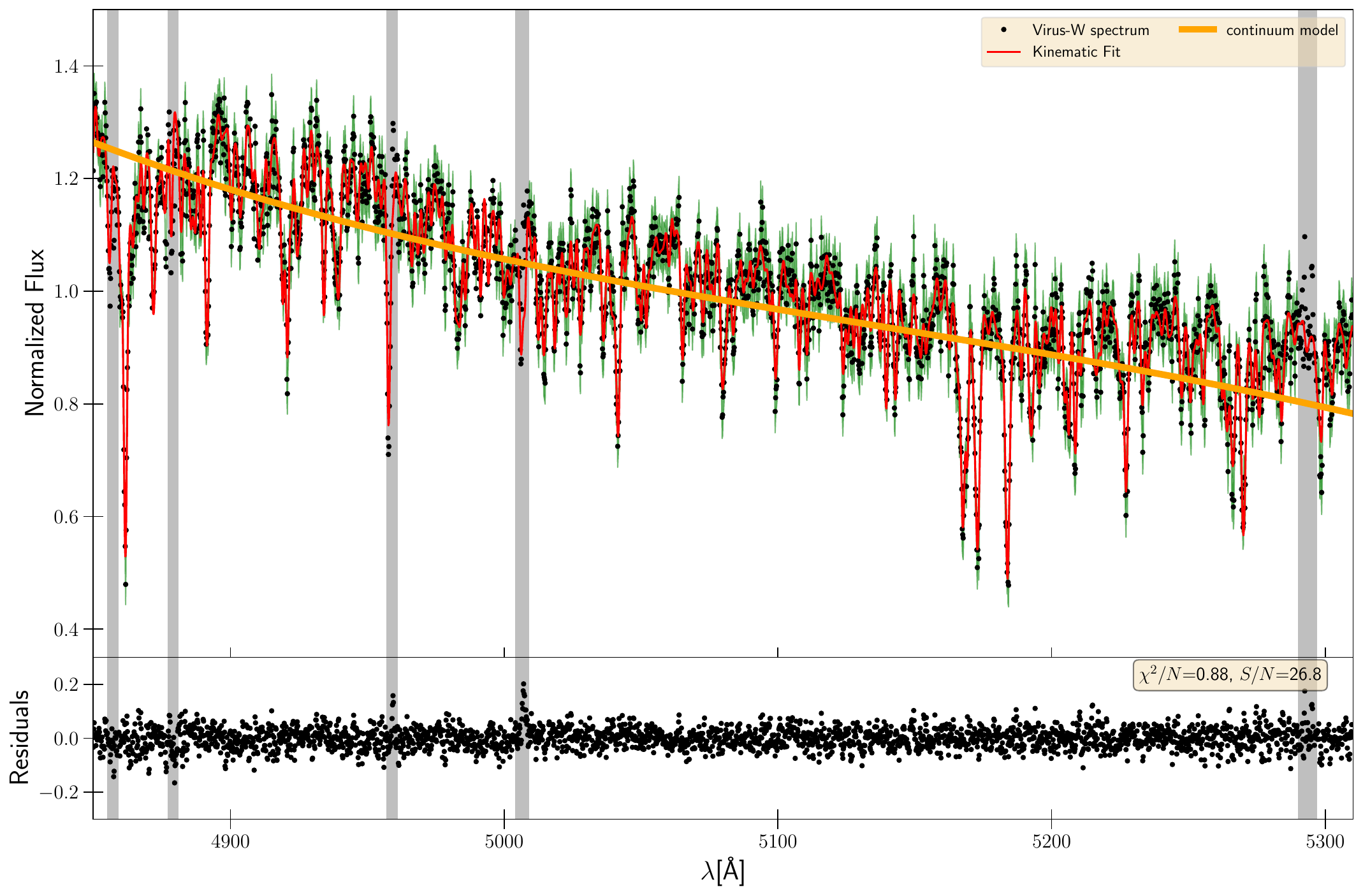}
    \caption{\textit{Top:} Example of a typical Voronoi binned spectrum (one of the bins of VCC~200 within the central 5\arcsec). \textit{Black:} The observed VIRUS-W spectrum. The \textit{green} band around the data indicates the $1\sigma$ flux uncertainty. \textit{Red:} The corresponding Fit model. \textit{Orange:} The continuum modelled by a 2nd order multiplicative polynomial. The \textit{gray} regions indicate the parts which were masked before the fit. \textit{Bottom panel:} The Residuals between Fit and Data.}
    \label{fig:spectrum_example}
\end{figure*}

\begin{figure}
	\centering
	\includegraphics[width=1.0\columnwidth]{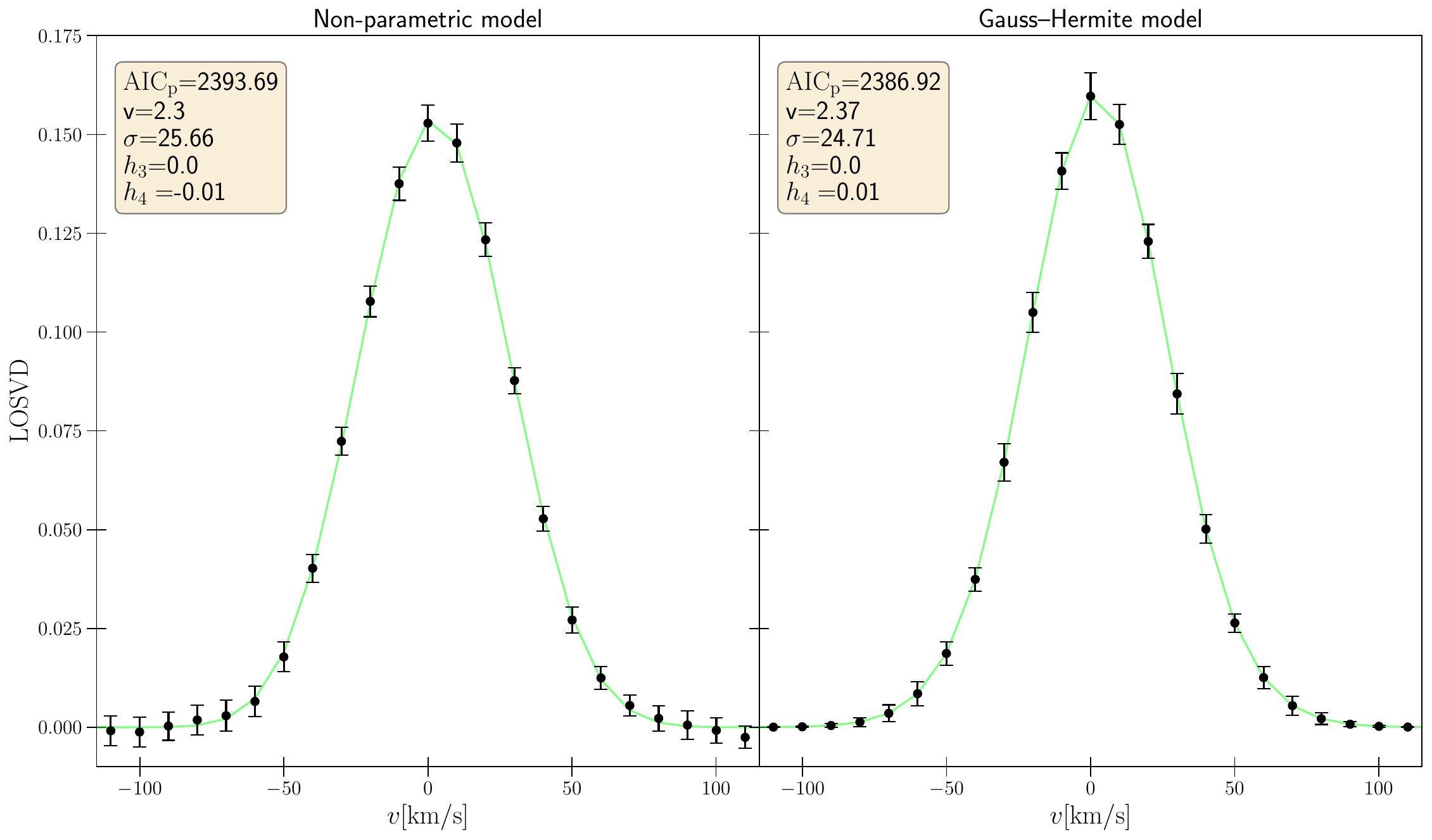}
    \caption{Comparison of an optimized LOSVD model parametrized by a Gauss--Hermite series (\textit{left}) and a non-parametric description (\textit{right}) for the same Voronoi Bin of VCC 200 ($S/N=26$). Both are broadly consistent with one another. In the case of this specific Voronoi Bin, the Gauss--Hermite representation is slightly preferred by $\aicmod$. The error suppression by a Gauss--Hermite parametrization at high velocities is substantial which is why we readjusted the errors as described in Sec.~\ref{subsec:spectroscopy}.}
    \label{fig:LOSVD_comparison}
\end{figure}

Fig.~\ref{fig:spectrum_example} shows an example for a VIRUS-W spectrum fit used to extract the kinematical and stellar population properties as is outlined in Sec.~\ref{subsec:spectroscopy}. A comparison of the LOSVD recovery with a Gauss--Hermite model vs a non-parametric model description of the LOSVD is shown in Fig.~\ref{fig:LOSVD_comparison}. In this case the Gauss--Hermite model is preferred as it achieves a slightly smaller $\aicmod$. Still, generally both LOSVD model choices are fairly consistent with each other for a given Voronoi bin.

\section{Comparison to existing kinematic and stellar population studies}
\label{append:Literature_comparison}
\begin{figure*}
	\centering
	\includegraphics[width=1.0\textwidth]{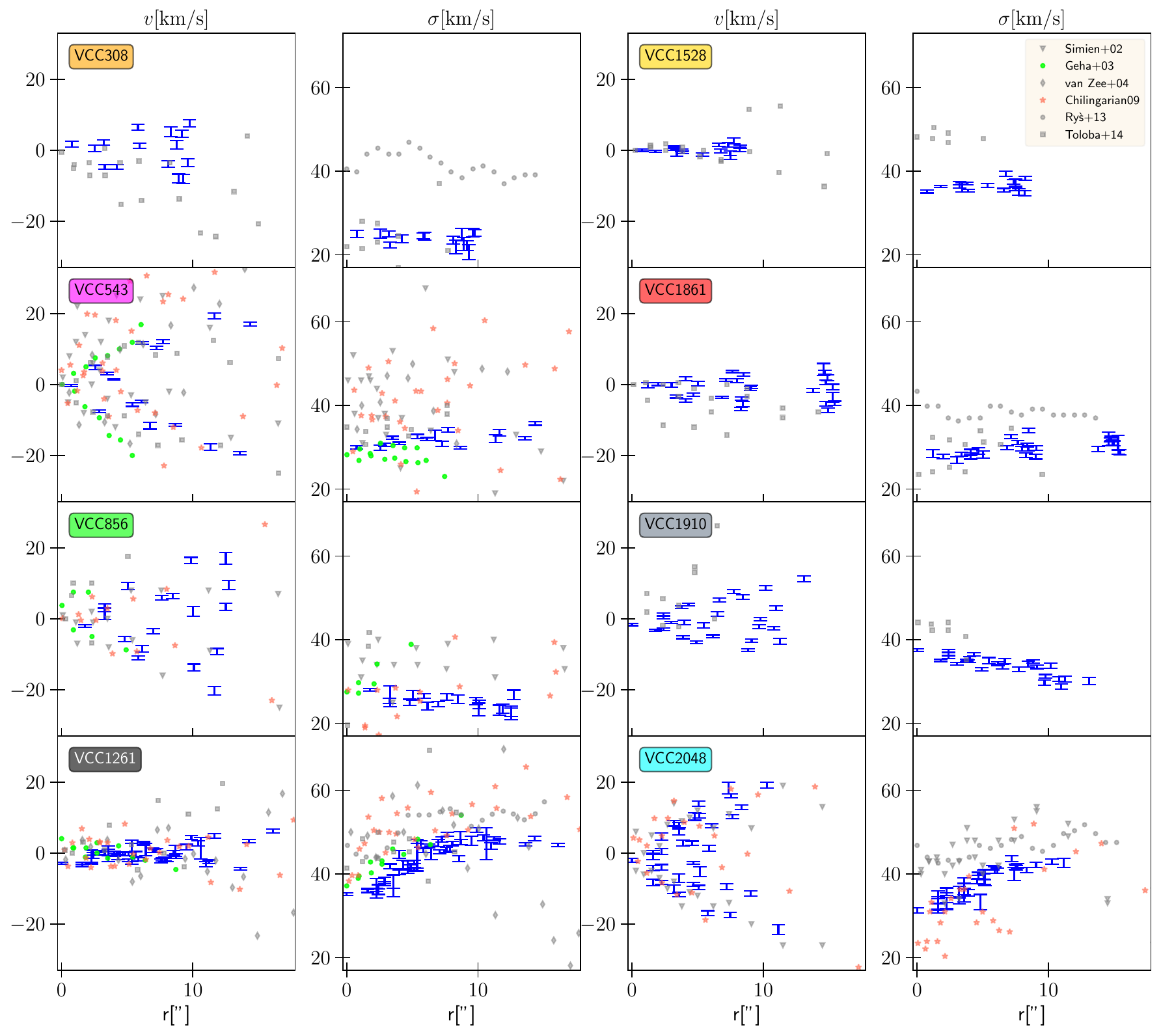}
    \caption{Literature comparison for the mean (line-of-sight) velocity and dispersion. In this paper we differentiate between sample galaxies using the color-coding indicated by the background color of the label here. There is no published data available for VCC 200 so we excluded it from this plot. The color-coding that is used for VCC 200 in all other Figures is \textit{dark blue}. To avoid overcrowding we do not show errorbars for the literature. For the studies where we did not find any raw data in a table format we reverse engineered the approximate values from their Figures. A few literature data points are not displayed here because they lie outside the plot boundaries. \textit{Blue with errorbars}: Our kinematic data (see also Fig.~\ref{fig:kinprofiles}). \textit{Gray triangles}: Data from \citet{Simien_2002}. \textit{Green dots}: Data from \citet{Geha_2003}. \textit{Gray Diamonds}: Data from \citet{van_Zee_2004}. \textit{Red stars}: Data from \citet{Chilingarian_2009}. \textit{Gray dots}: Dispersion from the IFU study of \citet{Rys_2013}. Since they have $\sim 100$ Voronoi bins we only plot the dispersion averaged within ellipticals bins as shown in Fig.5 of \citet{Rys_2013}. This reduces their high bin-to-bin scatter in the dispersion significantly, yet the higher dispersion remains. Their mean velocities were not displayed in the same manner which is why they are not shown here. \textit{Gray squares}: Data from \citet{Toloba_2014}.}
    \label{fig:kinematic_literature}
\end{figure*}

\begin{figure*}
	\centering
	\includegraphics[width=1.0\textwidth]{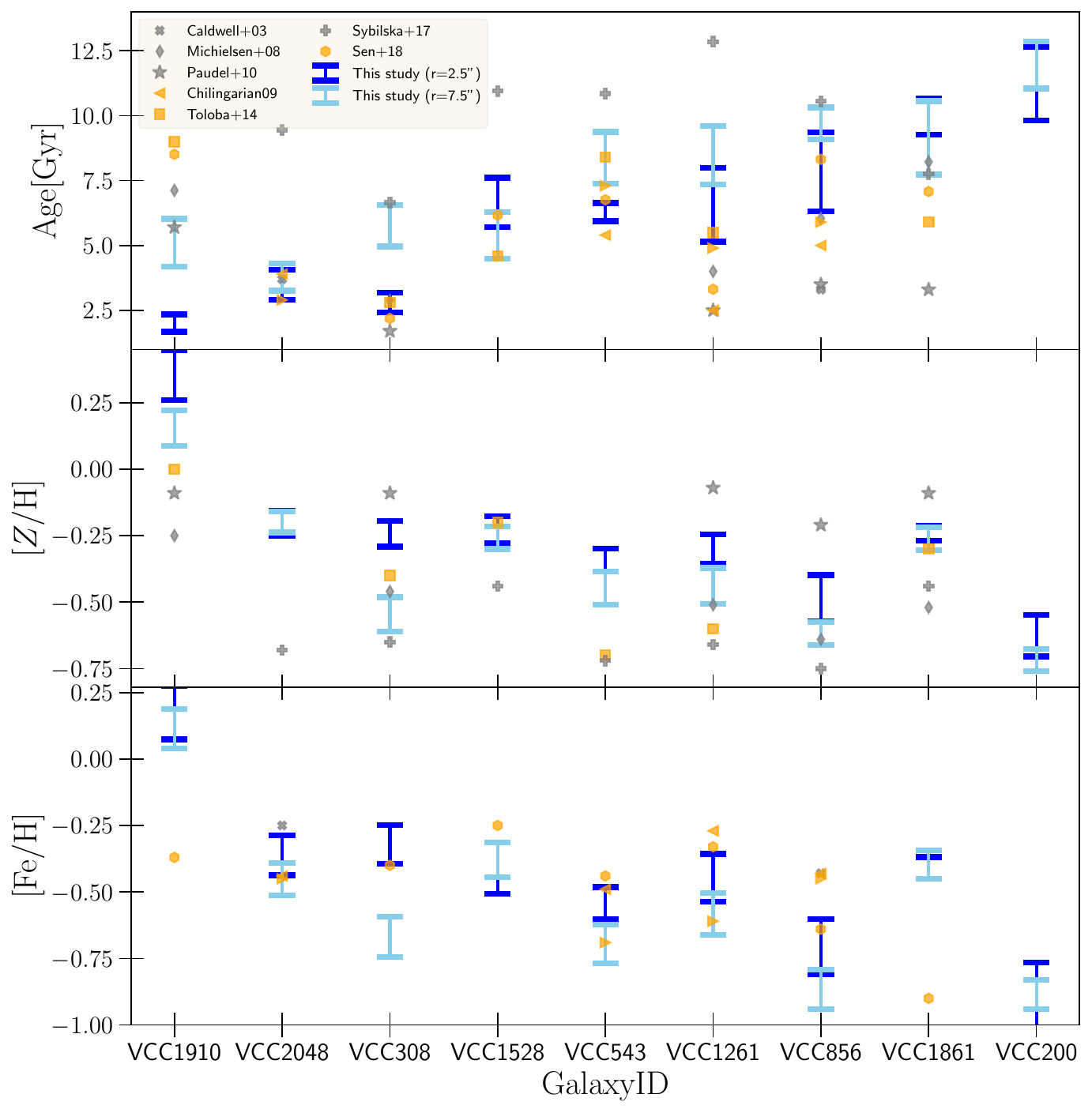}
    \caption{Literature comparison of single stellar population properties. We ordered the galaxies from left to right by the average age we derived in the two apertures. (\textit{Top:} Age in Gyr, \textit{Middle:} Metallicity $[Z/\rm H]$, \textit{Bottom:} Iron abundance [Fe/H] for the studies that trace metallicity in terms of [Fe/H]. We use $[Z/\rm H] = [\rm Fe/H] + 0.94\cdot[\alpha/ \rm Fe]$ \citep{Thomas_2003} to convert our metallicity $[Z/\rm H]$ to [Fe/H]. The \textit{blue} errorbars show our results for the central spectrum at $r=2.5\arcsec$ (\textit{dark blue}) and the outer one at $r=7.5\arcsec$ (\textit{light blue}). Studies with a spectral resolution of $R\gtrsim2000$ are highlighted in \textit{orange}, error bars from the literature are not shown to avoid overcrowding (they are generally larger than our errors). Our results for the central spectrum of VCC 1910 are not trustworthy (see Sec.~\ref{subsec:population_analysis}). For the values of \citet{Chilingarian_2009} we plotted the values for the circum-nuclear region and the outer region and indicate the region by the direction of the arrow. We excluded the results of \citet{Rys_2013,Rys_2015} from this plot as they are models composed of two different stellar population models, although we note that they employed the same data as \citet{Sybilska_2017}. We are not aware of any existing analysis for VCC 200.}
    \label{fig:age_literature}
\end{figure*}

In the last two decades, almost all of the galaxies in our sample were the subject of kinematic and stellar population studies already. In the following we discuss and compare our kinematic and population results with the studies we are aware of, discuss the significance of our results and highlight the need for a high resolving power. The published studies we compare our results to usually include other dEs as well, but we will focus only on those galaxies that are also part of our own sample. We plot the age, metallicity, rotation velocity and velocity dispersion from previous measurements together with our results in Fig.~\ref{fig:kinematic_literature} and Fig.~\ref{fig:age_literature} (for those studies where we were able to extract the respective data in a consistent way).

As part of a larger sample of early-type galaxies with various sizes, the kinematics of VCC 543, VCC 856, and VCC 2048 were studied by \citet{Simien_2002} using a long-slit spectrograph, which achieved a resolving power of $R=5050$, or an instrumental dispersion that is just below, or at least at a similar level to, the minimum velocity dispersion we measured for these galaxies. They measured the velocity and dispersion of the spectra out to $\sim$ 20-25\arcsec, finding velocity curves consistent with ours, although the scatter is considerable. The dispersions of VCC 543 and VCC 856, however, show significant, qualitative and quantitative disagreements: they scatter a lot, show no clear radial gradient and are generally much higher than the values we find for VCC 543 and VCC 856. For VCC 2048 the dispersion profile agrees qualitatively with ours showing a dispersion drop in the center (albeit shallower). However, again the dispersions are biased high on average. The discrepancies are likely caused by their larger instrumental dispersion which is close to the velocity dispersions of these galaxies. An additional factor could be that \citet{Simien_2002} did not include higher order deviations from Gaussian LOSVDs in their fits which could artificially broaden the LOSVDs. 

The spectral resolution was even lower in the studies of \citet{Caldwell_2003},  \citet{van_Zee_2004}, \citet{Michielsen_2008} and \citet{Paudel_2010} ($R$$\sim$$1500$, $R$$\sim$$2200$, $\sigma_{\mathrm{instr}}$$\sim$$170$ km~s$^{-1}$ and $\sigma_{\mathrm{instr}}$$\sim$$280$ km~s$^{-1}$, respectively). Accordingly, when stated, the respective velocity dispersions measured in these studies are again higher than ours. The dispersions of \citet{van_Zee_2004} are all offset by more than 10 km~s$^{-1}$ while those of \citet{Caldwell_2003} are even outside the plot range of Fig.~\ref{fig:kinematic_literature}. \citet{Michielsen_2008} and \citet{Paudel_2010} only stated stellar population properties and no kinematics. \citet{Caldwell_2003} discussed both population properties and kinematics. The age and metallicity \citet{Caldwell_2003} determined for VCC 2048 agree with our results within the one sigma error while their age for VCC 856 is significantly lower. However, they did not find old ages for any of the low $\sigma$-galaxies in their sample. The SSP results of \citet{Michielsen_2008} are based on a slightly higher resolution and seem to agree better with our results. On the other hand \citet{Paudel_2010} found systematically younger and more metal rich populations than we (and other studies) do.

\citet{Chilingarian_2009} re-analyzed the data of \citet{Simien_2002} and \citet{van_Zee_2004} using a full spectral fitting technique instead of Lick indices. The dispersions of VCC 543, VCC 856, VCC 1261 are relatively consistent with those older studies in that their scatter is high and the average dispersion is significantly larger than what we find. Only VCC 2048 does not appear to systematically offset high, and in fact it is the only example where a significant portion of the literature dispersions fall \textit{below} our values. While about half of the spatial bins of VCC 2048 fall close to the dispersions we derived, the other half scatters to the low side. The fact that some of these bins have such a low dispersion is peculiar and seems inconsistent with the dispersions from \citet{Simien_2002} which were derived from the same raw data. The population properties \citet{Chilingarian_2009} derived were spatially resolved by separately analysing the co-added spectra in \textit{two} different radial bins along the slit: A circum-nuclear bin (excluding the central blue nuclei) and an outer bin (near or just within $\sim$$1~\reff$). In Fig.~\ref{fig:age_literature} we plot both, the values of the circum-nuclear and the outer region. The population properties agree well, and in fact, even the radial behaviour is (at least qualitatively) consistent with our findings.   

\citet{Toloba_2014} analysed the rotation, dispersion and population properties of a large dE sample using optical long-slit spectroscopy from 3 different telescopes with the spectral resolution ranging from $R$$\sim$$1900$ to $3300$. As is the case with most previous studies, the velocity curves are fairly consistent with our findings. However, unlike many other low spectral resolution studies, they also found quantitatively consistent dispersions for some (but not all) of the low-$\sigma$ galaxies (specifically VCC 308, VCC 543, VCC 1861). This could be because they aimed for a very high $S/N$ for the purposes of a better dispersion recovery. High $S/N$ can counteract low resolution effects as shown in the recovery simulations of \citet{Toloba_2011} and \citet{Eftekhari_2022}. We plot the ages and metallicity they derived for the summed spectra within $1~\reff$ in Fig.~\ref{fig:age_literature} (some of their spectra include more Lick indices than ours, such as the age-sensitive H$\alpha$ line; e.g. for VCC 308, VCC 543, VCC 1528, VCC 1861 they include this index). We excluded VCC 856 from our plot as the age they derived is at the boundary of their grid ($14.1 \, \mathrm{Gyr}$). \citet{Sen_2018} reanalyzed the data of \citet{Toloba_2014} with a special focus on different abundance ratios in dEs. For this purpose they derived ages and metallicities for the very central spectra ($r<\reff/8$) from a combination of 23 Lick indices. Overall their results are in good agreement with \citet{Toloba_2014} and with ours, not only in age but also in metallicity (Fig.~\ref{fig:age_literature}).
 
\citet{Geha_2003} is the only comparison study with better data in terms of resolution ($R$$\sim$$13000$) and spectral coverage (3900\r{A}-11000\r{A}) (highlighted in green in Fig.~\ref{fig:kinematic_literature}). Their dispersions for 
VCC 543, VCC 856, and VCC 1261 are quantitatively much more consistent with ours. We suspect that minor differences could be caused by the fact that only Gaussian LOSVDs are fitted and because they only used K-type stellar templates. 

With the advent of IFUs, 2D kinematic studies of dwarf galaxies have become more prevalent. In a series of papers \citep{Rys_2013,Rys_2015} were the first to analyze the kinematics and stellar population of the galaxies in our sample using an IFU: SAURON in its low resolution mode ($R$$\sim$$1300$ within the spectral range of $4760$\r{A}-$5300$\r{A}). The velocity maps are fairly consistent in that VCC 1261 and VCC1861 are essentially non-rotators, while VCC 308 and VCC 2048 show intermediate to strong rotation respectively. Their dispersions on the other hand scatter significantly between neighbouring Voronoi bins with differences of up to 100 km~s$^{-1}$. Therefore in Fig.~\ref{fig:kinematic_literature} we only show their radially averaged dispersions which reduces the bin-to-bin scatter, but still the dispersions seem systematically high.  They did not find dispersions below $\sim 40$ km~s$^{-1}$. We excluded their ages and metallicities from Fig.~\ref{fig:age_literature} because they used models with 2 distinct stellar populations, i.e. they are not directly comparable to single population models. Still, such two-burst models are the first step to modeling a possibly more complex star formation history. In this context, we will discuss their results in Sec.~\ref{subsec:extended_star_formation}.

\citet{Sybilska_2017} re-analyze the data of \citet{Rys_2013} using the same observation and reduction setup but adding more dEs (including VCC 543, VCC 856, VCC 1528). Among other quantities, they derived global dispersions of the \textit{integrated} spectra within $1~\reff$. Similar to the original results of \citet{Rys_2013}, the dispersions are higher than the average dispersions we measured, especially so for the lower-$\sigma$ galaxies of our sample. Unlike \citet{Rys_2013,Rys_2015}, they derived population parameters using Lick indices with a \textit{single} population instead of a two burst model, which makes the comparison with our population parameters more meaningful. Compared to many previous studies their results for the integrated spectra within $1~\reff$ yielded very high ages and lower metallicities. In fact, many of the dEs that are also in our sample have no upper constraint for the age as the associated errors reach the edge of their parameter grid (14 Gyr).

The kinematic comparison with the existing literature suggests LOSVDs from lower resolution spectra seem to be artificially broadened especially when the galaxy's actual dispersion is lower or at the level of the instrumental dispersion. In some cases, low resolution effects may be mitigated by a very high signal-to-noise ratio \citep{Toloba_2011} but this requires that the true noise level in the data is estimated accurately in the first place. Furthermore a too low signal-to-noise of the binned spectra leads to large bin-to-bin scatter which in turn blurs spatial any systematic spatial signatures like the $\sigma$-drops present in VCC 1261 and VCC 2048. 

We conclude that in the case of the dEs where the dispersion can be as low as 20 km~s$^{-1}$, a resolution $R\gtrsim 8000$ with a $S/N\gtrsim15$ is required to obtain the width of the LOSVDs in an unbiased manner. For our Virgo galaxies this study is the first to use high-resolution, high $S/N$, spatially resolved spectra. This allows an unbiased dispersion recovery all the while covering the full 2D kinematic information. It reveals that many of the galaxies have significantly lower dispersions than previous studies suggested. Given the resolution of VIRUS-W we should be able to measure dispersions down to as low as $\sim 15$ km~s$^{-1}$, yet the homogeneous dispersion profiles of some galaxies (VCC~200 and VCC~308) could be suggestive that the limit may be reached already at $\sim 20$ km~s$^{-1}$. We tested this two-fold. Firstly by correctly recovering a (gaussian) toy LOSVD with $\sigma=15$ km~s$^{-1}$ from a Monte-Carlo mock spectrum of a single stellar population with the same resolution and $S/N$ as the VCC~200 observations. And secondly by adding the \textit{real} spectra of VCC~200 and VCC~308 into larger bins, thus, doubling the $S/N$ per bin. Fitting these spectra did not change the dispersions as $\sigma$ remained consistent within $\lesssim 0.5$ km~s$^{-1}$ with the lower $S/N$ spectra. From these two tests we conclude the dispersions of these two galaxies are robust and the resolution limit is not yet reached. 

Concerning stellar population properties, the results seem broadly in agreement with those of the literature as long as we disregard our results for VCC 1910 (see Sec.~\ref{subsec:population_analysis}). The metallicity values of the dEs are all sub-solar $[Z/\rm H]\in [-0.75,-0.2]$ and span a large range in ages from 3 to 12 Gyr. Both age and metallicity appear to be slightly correlated. Some of this correlation could still be an artifact of the age--metallicity degeneracy, however, the independent measurements of the two spectra at $2.5\arcsec$ and $7.5\arcsec$ seem broadly consistent for a given galaxy suggesting that they are robust and that the dEs have a radially homogeneous stellar population. Only VCC 308 seems to be significantly younger and metal-rich in its center. This is expected because VCC 308 is the only galaxy in our sample with an extended blue center and as such classified as a dE(bc), while the other dEs do not show color gradients (outside their blue nuclei which are not well resolved; see App.~\ref{append:galdiscussion}). We note, however, that many of the literature results have large 1$\sigma$ error bars and that \citet{Paudel_2010}, for example, found systematically younger and metal-rich galaxies, whereas \citet{Sybilska_2017} find the opposite: systematically older and more metal-poor populations. This may suggest that the age--metallicity degeneracy is not always broken sufficiently. While the population studies with an intermediate spectral resolution (highlighted in yellow in Fig.~\ref{fig:age_literature}) tend to agree better with our results than those with lower resolution, there is no strong systematic trend, e.g. that low resolution spectra would yield population properties that are biased in a specific direction. We suspect the signal-to-noise of the spectra plays the more important role if one wants to recover unbiased stellar population properties. 

\section{Do population models overestimate the total dynamical mass?}
\label{append:ML_check}
For a few of the galaxies in Fig.~\ref{fig:Age_ML_Difference} the IMF parameter $\Delta_{\mathrm{tot}}$ is slightly negative which suggests that (locally) the total mass predicted dynamically is \textit{lower} than the population models imply. Assuming both mass-to-light ratio estimates are accurate this implies that the IMF of these galaxies is lighter than Kroupa even if we assume they have no dark matter. While this is not per se an issue this could hint at a problem in either the population or dynamical estimates of the galaxies with negative $\Delta_{\mathrm{tot}}$. In the following we will discuss possible issues.

Firstly, the negative $\Delta_{\mathrm{tot}}$ does not necessarily imply that the total, cumulative with radius, dynamical mass is in fact lower than the SSP predictions. The dynamical models we employ have much more flexibility to spatially vary their $M_{\mathrm{tot}}/L$-profiles thanks to the 5 halo parameters (eq.~\ref{eq:dark_matter_model}) and the log-linear stellar mass-to-light ratios. In contrast, the SSP models are fixed to the light distribution at the integrated spectra around the two probed radii $r=2.5\arcsec$ and $r=7.5\arcsec$. Consequently, the dynamical models are able to change the mass-to-light ratios at all other radii much more easily which could (occasionally) produce negative $\Delta_{\mathrm{tot}}$ at the two radii evaluated by the inflexible SSP models. To test this we have also dynamically modelled simple mass-follows-light models that have no dark matter component and only a single global mass-to-light ratio sampled in $\sim0.2$ steps, and as such these dynamical models are much more comparable in their flexibility to the SSP models. Overall the best mass-follows-light models, shown in Fig.~\ref{fig:ml_dyn_vs_ml_pop}, are consistent with the mass-to-light ratios recovered with the more flexible dynamical models that included dark matter and stellar gradients. For most of the galaxies (VCC 308, VCC 1861) that have negative $\Delta_{\mathrm{tot}}$, this explanation lifts some of the tension regarding the negative $\Delta_{\mathrm{tot}}$. For the remaining galaxy (VCC 200) with $\Delta_{\mathrm{tot}}<0$, the total mass recovery of the Schwarzschild models could be biased low, or the mass-to-light ratio obtained from the stellar population analysis could be biased high. 

None of our application of the dynamical modeling on simulations \citep[e.g.][]{Lipka_2021,Neureiter_2023_a} have suggested that the \textit{total} dynamical mass could be systematically \textit{underestimated}. If anything it would be easier to bias dynamical masses higher instead of lower. For example, spectra with too low of a resolution would overestimate the observed velocity dispersion (cf. App.~\ref{append:Literature_comparison}) and, as a consequence, also the inferred dynamical mass. Of course the dynamical models assume the systems are in dynamical equilibrium, which could affect our mass inference if they are not. However, given the old age and photometric structure of VCC 200 \citep[][]{Ferrarese_2006} we have no particularly strong reason to believe this galaxy is more out of equilibrium than the other dEs in the sample. Instead the very old age of VCC 200 (12 Gyr) could be slightly overestimated, which could bias its SSP mass-to-light ratio $\Upsilon_{\rm Kroupa}$ higher, resulting in a negative $\Delta_{\mathrm{tot}}$. Unfortunately, VCC 200 is also the only galaxy in our sample for which no reference SSP results exist in the literature so we cannot confirm its age. However, considering its low H$\beta$ index ($\lesssim 2.0$) the galaxy is very likely older than 9 Gyr. In conclusion, we suspect the age of VCC 200 is either slightly overestimated by 1 or 2 Gyr and consistent with Kroupa IMF or, alternatively, the age is correct and the galaxy has a IMF slightly lighter than Kroupa.

\section{IMF--metallicity coupling}
\label{append:IMF_metal}
\begin{figure}
	\centering
	\includegraphics[width=1.0\columnwidth]{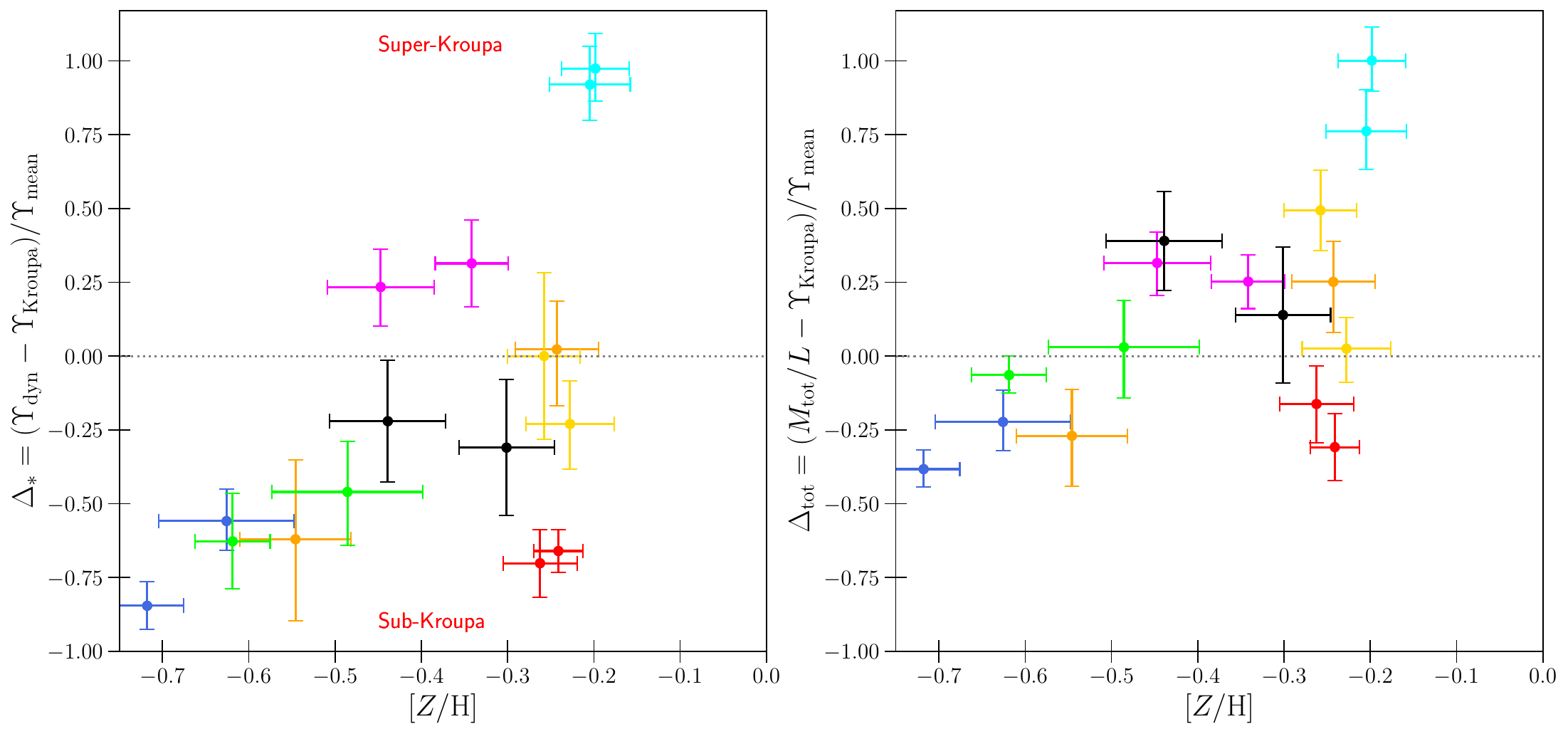}
    \caption{Analogous to Fig.~\ref{fig:Age_ML_Difference} but for the metallicity instead of the age. VCC~1910 is not in the plot range due to its ( likely wrong ) super-solar metallicity. Due to the anti-correlation of age and metallicity (Fig.~\ref{fig:SSP_diagnostics}) of our dEs the positive correlation of the IMF parameter and the metallicity is a corollary of the IMF and age relation.}
    \label{fig:IMF_metal}
\end{figure}

Fig.~\ref{fig:IMF_metal} shows the relation of IMF parameter with the metallicity. Apart from VCC~1861 the IMF parameter is positively correlated with $[Z/\mathrm{H}]$. If the $\Delta_{*}$ we derived are robust, it could be that either age or metallicity (or both together) are the physical reason behind the variety seen in the IMF of dEs. For ordinary ETGs strong, positive correlation of the IMF parameter and the metallicity have been noticed before \citep[e.g.][]{Martin_Navarro_2015,Parikh_2018,Li_2023}.

\section{Phenomenological discussion of each galaxy - Colors, Substructure, environment, kinematic signatures}
\label{append:galdiscussion}
In this section, we discuss each dE holistically based solely on the photometric and kinematic data we derived and what we found in the existing literature. A detailed description of the \textit{HST} photometry, isophotes and colors can be found in \citet{Ferrarese_2006} for all galaxies in our sample, except for VCC 308. In the following, unless stated otherwise, all colors are stated in $g-z$ bands.

\textbf{VCC 200:}
Classified as a dE2(N), the ellipticity structure is actually fairly round for large parts of the galaxy, only experiencing a double peak up to $\epsilon$$\sim$$0.2$ at $0.5\arcsec$ and $5\arcsec$ \citep{Ferrarese_2006}. This signature may stem from a ring structure inhabiting an otherwise rounder galaxy. Boxiness/diskiness parameters scatter but are otherwise consistent with 0. The $g-z$ color is constant with radius at $1.2-1.3$ mag. Only in the center, it has a slightly bluer ($1.1$ mag) and bright nucleus \citep[cf.][]{Hamraz_2019}. We find the galaxy exhibits intermediate rotation ($\leq 10$ km~s$^{-1}$) and a constant, low dispersion ($\sim 25$ km~s$^{-1}$) within the VIRUS-W FoV. While one may discern a hint of $v$-$h_3$ anticorrelation in the maps, even higher Gauss--Hermite moments scatter significantly (but stay within $\pm0.1$). If the heliocentric velocity and distance to us and M87 are to be believed, VCC 200 is at the backside of the cluster far away from the center and is moving towards it. This stands in contrast to its very old SSP age (Sec.~\ref{subsec:population_analysis}). Within our sample, VCC 200 seems to be an outlier in regard to its stellar population, with it being very old ($>10\rm Gyr$) and metal-poor compared to the rest of our results. Unfortunately, it is also the only galaxy for which we are not aware of any existing kinematic or stellar population studies. 

\textbf{VCC 308:}
VCC 308 is the only galaxy in our sample classified as a dE(bc) \citep[cf.][]{Lisker_2007}, meaning it is significantly bluer in the center, showing an extended radial color gradient becoming redder further outside (i.e. not just a distinct blue nucleus). However, in the case of VCC 308 this gradient is relatively small in terms of radial extent and magnitude \citep[cf.][]{Lisker_2006_b}. One could debate whether it is really that distinct to the rest of our sample galaxies. \citet{Lisker_2006_a} also find the photometry shows weak signs of spiral arms that are seen face-on. The kinematics of VCC 308 we measured is similar to that of VCC 200. It has intermediate rotation while having the lowest dispersion values ($20-25$ km~s$^{-1}$) in our sample (still significantly higher than the resolution of VIRUS-W). Higher Gauss--Hermite moments are noisy, but we identify a hint of a $v-h_{3}$ anti-correlation and a positive $h_{4}$ within the vast majority of the Voronoi bins. VCC 308 is far away from the center in a very low projected density environment \citep[][]{Sybilska_2017}.

\textbf{VCC 543:}
VCC 543 appears to be fairly elongated, being classified as a dE5. While the designation by \citet{Lisker_2007} suggests no nucleus, we and \citet{Hamraz_2019,Ferrarese_2006} find a detectable, round, and slightly bluer nucleus in the galaxies center. Other than that, the galaxy is quite regular, showing no preference for either diskiness nor boxiness or any detectable substructures. The galaxy shows a strong, linearly rising, velocity signal that is anti-correlated with $h_{3}$. Within 10\arcsec, $h_{4}$ appears to experience a radial drop off, however, further outside the signal becomes quite noisy, showing no clear trend. The dispersion increases with radius, showing no signs of plateauing within the FoV. The galaxy is in the foreground of the cluster, moving slightly away from it, possibly because it has previously passed the cluster center.

\textbf{VCC 856:}
At first glance the galaxy appears to be a typical, fairly round dE ($\epsilon$$\sim$$0.1$) with a \textit{g}-\textit{z}-color of $\sim1.2~ \rm mag$ and a bluer ($\sim 1.0 ~\rm mag$) nucleus that dominates its center. However, as first noted by \citet{Jerjen_2000}, one can see a faint signature of face-on spiral arms in the galaxy. Perhaps we see the galaxy during its transformation from a dwarf spiral into a dE. Despite the \textit{presumably} face-on disk, the galaxy displays clears signs of rotation around its axis, suggesting at least some degree of inclination. The dispersion is generally low, with a \textit{decreasing} dispersion. It stands out significantly from the rest of the dE sample by exhibiting a strongly \textit{rising} $h_{4}$ profile with comparatively larger error bars. The scatter in $h_3$ is large, possibly due to the spiral arms. 

\textbf{VCC 1261:}
VCC 1261 is the brightest galaxy in our sample. Despite its significant flattening ($\epsilon$$\sim$$0.4$) the galaxy's isophotes are regular ellipticals ($a_{4} \sim 0$) showing no signs of an embedded disk. In its center it hosts a rounder, bluer ($1.0 ~\rm mag$ compared to $1.2 ~\rm mag$) nucleus. Within $\sim 10\arcsec$ the kinematics show little rotation but a strongly rising dispersion reaching $\sigma=50$ km~s$^{-1}$. Accompanied by this dispersion rise is a significant drop in $h_{4}$ down to $0$ at $10\arcsec$, suggesting a strong $\sigma$-$h_{4}$-correlation. In the outermost radial bins the galaxy behaves quite differently with the dispersion and $h_{4}$ plateauing, while the rotation abruptly becomes more significant. The orbit of VCC 1261 within Virgo may be odd, as it seems to be far behind the cluster yet is still moving away from Virgo relatively rapidly. 

The fact that VCC 1261 appears to be a very flattened galaxy, yet having essentially no rotation in our and most other kinematic studies \citep[][]{Geha_2003,van_Zee_2004,Chilingarian_2009,Rys_2013,Toloba_2014,Toloba_2015,Sybilska_2017,Sen_2018}, seems to be in conflict with the hypothesis that that dEs are remnants of more disky, gas-rich, late-type dwarfs. However, in our and other studies we observe a rise in velocity beyond $12\arcsec$ which could suggest that we simply do not probe the galaxy at large enough radii to see a more pronounced rotation signal. The same could apply to the other non-rotator in our sample, VCC 1528. Both non-rotators are, perhaps not coincidentally, the two dEs nearest to the cluster center (Tab.~\ref{tab:galaxy_table}). Initially their mass and extent could have been larger than the other dEs in our sample, but their increased likelihood of interactions via harassment and ram-pressure stripping could have made the two dEs significantly fainter at larger radii, effectively `shrinking' them to the regime of our sample. This appears fairly plausible because we will find that these two galaxies have a significantly more massive dark matter halo than the other dEs (cf. \paperrefeDARKMATTER). This scenario is also suggested by \citet{Beasley_2009} who studied the kinematics of VCC 1261's globular clusters (GC), which, unlike studies based on the galaxy's integrated light, allows constraints on $v$ and $\sigma$ much farther out at several effective radii. However, this also comes with fairly large uncertainties and the assumption that the galaxy's stars and GCs are closely kinematically associated. The GC motions they found suggest significant rotation ($v/\sigma>1$) at larger radii. Future IFU studies with deeper kinematics may resolve whether the velocity of the stars is indeed rising further beyond $1~\reff$ for the two galaxies. 

It could also be that VCC~1261 is an outlier. The E7/S0 galaxy NGC~4550 for example is known to host two co-spatial, \textit{counter-rotating} disks of equal mass which results in a net streaming motion that is extremely low, yet the systems is very flat \citep[][]{Rubin_1992,Rix_1992,Emsellem_2007}. A sign that such a counter-rotating disk could be embedded within the central $\sim 10\arcsec$ where $v=0$ is the congruent decrease in $\sigma$ and increase in $h_{4}$ towards the centre (Fig.~\ref{fig:kinprofiles}). However, from its orbit structure obtained from the dynamical modeling (Fig.~\ref{fig:dE_aniso} and Fig.~\ref{fig:dE_aniso_flattening}) we do not find strong evidence in favor of this scenario.

\textbf{VCC 1528:}
Despite being classified as non-nucleated dE in \citet{Lisker_2007}, we and \citet{Ferrarese_2006} find the galaxy to host a resolved, blue nucleus. Apart from this nucleus the galaxy is fairly red ($\sim 1.35 ~\rm mag$) near its center, progressively becoming bluer ($\sim 1.2  ~\rm mag$) at larger radii. At large radii ($>10\arcsec$) the galaxy appears fairly round ($\epsilon \sim 0.1$). However, it is significantly more elliptical in the center with $\epsilon \sim 0.25$. The shapes of the ellipses are very regular. Within its small FoV, VCC 1528 is a non-rotator with a virtually flat (or very slightly rising) dispersion profile. Higher Gauss--Hermite moments scatter, especially near the edge of the FoV, but are overall consistent with (and close to) zero. Since the galaxy becomes flatter and bluer outside the FoV, a change in kinematics may be expected. VCC 1528 is the dE closest to Virgo's 3D center. 

\textbf{VCC 1861:}
Classified as dE0, the galaxy is very round and, as such, has no well constrained position angle. The color of the galaxy ($\sim$ $1.3 ~\rm mag$) and the bluer, very bright nucleus are typical for our sample. The isophote ellipses are regular and we find no substructures in the galaxy. VCC 1861 shows intermediate rotation, and a slightly rising dispersion profile within $\sim 10\arcsec$, plateauing at larger radii. The LOSVDs are roughly symmetric with only a tentative sign of a $v$-$h_{3}$ anti-correlation in the kinematic map. Similar to VCC 1261 and VCC 2048, the galaxy has strongly peaked LOSVDs in the center ($h_4 \sim 0.1$) which steadily become more Gaussian($h_{4}$$\sim$$0$) with increasing radius. VCC 1861 may be associated with a locally denser sub-clump of galaxies formed around the large elliptical galaxy M60. 

\textbf{VCC 1910:}
VCC 1910 is the reddest galaxy in our sample with $g-z\approx1.4 ~\rm mag$, but has a typical blue nucleus with $g-z\approx1.0 ~\rm mag$. Akin to VCC 200, the ellipticity has two distinct and extended peaks at $0.8\arcsec$ and $6\arcsec$ reaching $\epsilon$$\sim$$0.2$. VCC 1910 is an intermediate rotator which, together with VCC 856, stands out in that the dispersion peaks in the center and then steadily drops of with radius. In the maps one can again see signs of $v$-$h_{3}$ anti-correlation, but overall higher moments scatter significantly and are consistent with zero. VCC 1910 might also be associated with the M60 sub-clump, but its net velocity suggests it moves relative quickly towards us.

\textbf{VCC 2048:}  
Together with VCC 543 and VCC 1261 the galaxy is on the bluer end of our sample with a color of $1.2 ~\rm mag$ and a blue nucleus at $1.0 ~\rm mag$. And, again together with VCC 543 and VCC 1261, it is significantly elongated ($\epsilon$ $\sim$$0.5$ to $0.6$). It also hosts a large disk that causes the isophotes to be disky over a large radial range. For a more in-depth study of this galaxy's photometry and a bulge-disk decomposition, we refer the reader to \citet{Kormendy_2012}. VCC 2048 shows very strong rotation around its minor axis and, very similar to the other flattened galaxy VCC 1261, has a strongly rising dispersion profile that plateaus at $\sim 8\arcsec$ without signs of a drop-off within the FoV. The velocity and skewness $h_{3}$ are clearly anti-correlated, and $h_{4}$ has a strong peak in the center which starts to drop off with increasing radius, even becoming slightly negative in the outermost bins. Similar to VCC 308, VCC 2048 is located at a very low projected cluster density \citep[][]{Sybilska_2017}.   
\section{Angular momentum versus stellar mass and $\lambda_{e}$}
\label{append:Lambda}
\begin{figure}
	\centering
	\includegraphics[width=1.0\columnwidth]{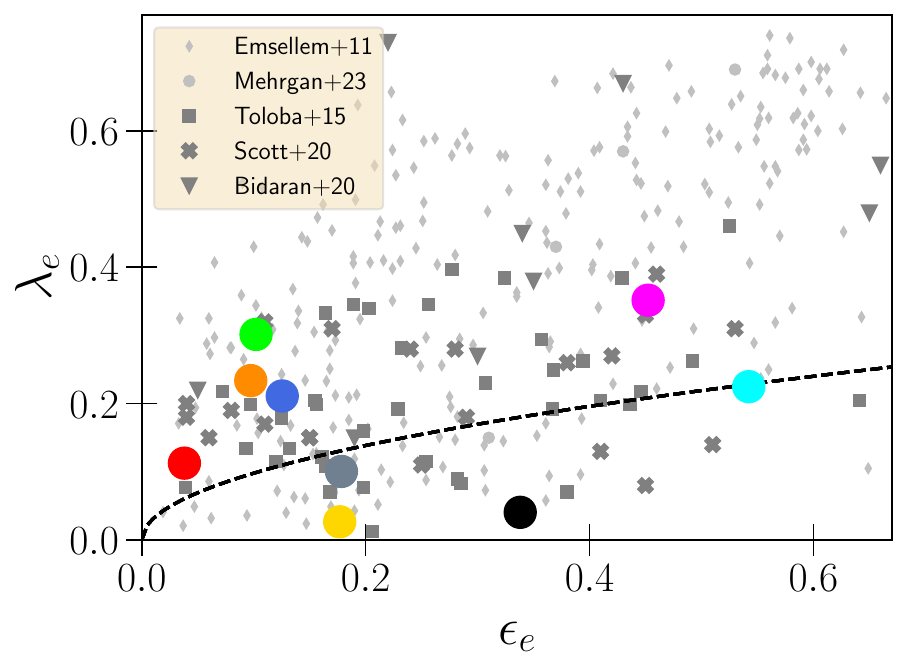}
    \caption{Alternative version of Fig.~\ref{fig:angular_momentum} analysed within 1.0 instead of 0.5 stellar effective radii. For some galaxies in the VIRUS-W sample and in the literature this requires an extrapolation because the kinematic maps do not extend far enough. The typical distribution of dEs does not change much with the increase in aperture apart from two of the infalling dEs of \citet{Bidaran_2020} becoming outliers.}
    \label{fig:angular_momentum_1REFF}
\end{figure}

\begin{figure*}
	\centering
	\includegraphics[width=1.0\textwidth]{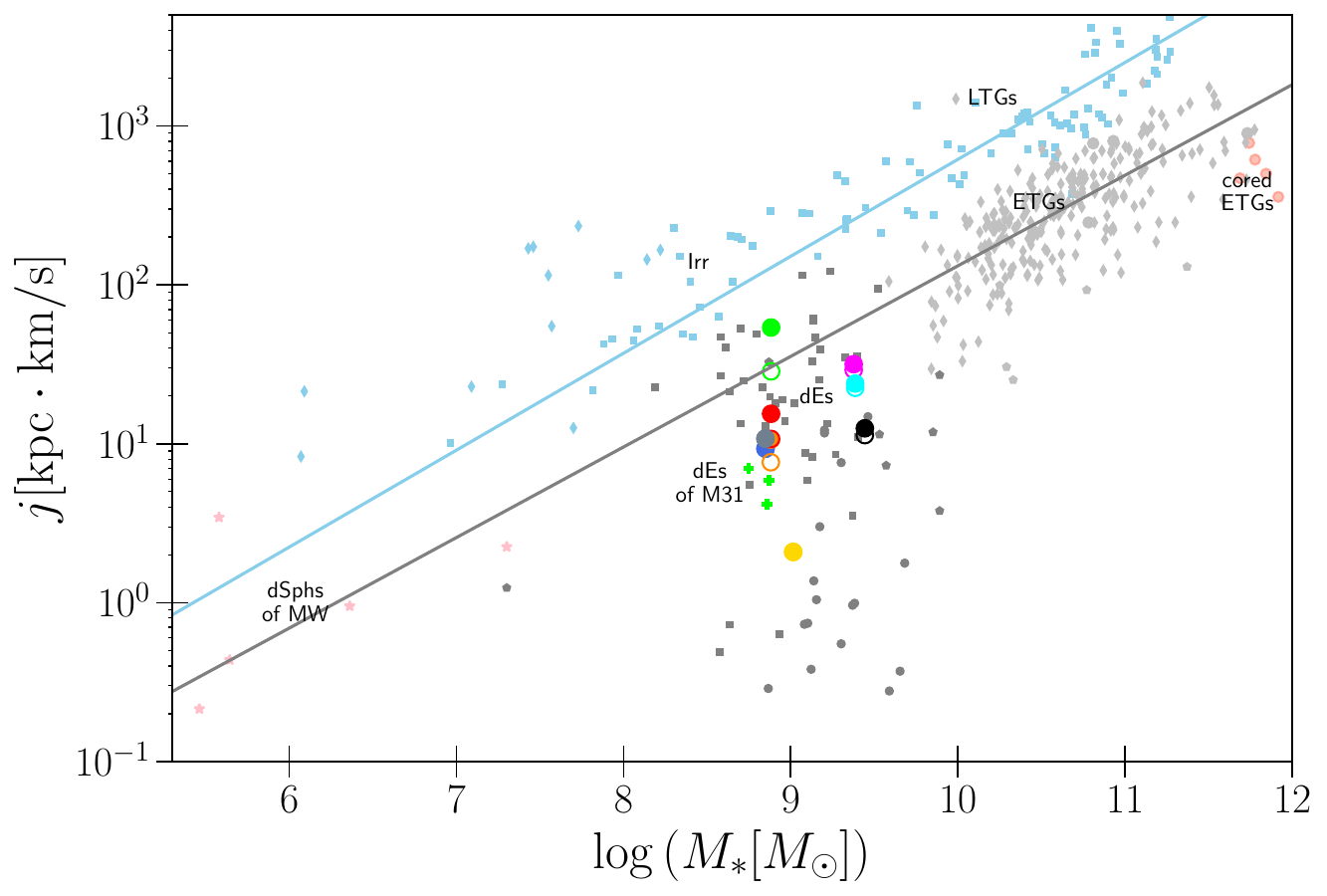}
    \caption{As Fig.~\ref{fig:specific_momentum} but versus the total stellar mass $M_{*}$. \textit{Gray diagonal:} Standard $j-M_{*}$ scaling relation from \citet{Pulsoni_2023} for fast rotators, i.e. this is \textit{not} a fit to any of the galaxies shown here. \textit{Blue diagonal:} Standard scaling relation for massive LTGs \citep[][]{Teodoro_2023}.}
    \label{fig:specific_momentum_vs_mass}
\end{figure*}

Fig.~\ref{fig:angular_momentum_1REFF} and Fig.~\ref{fig:specific_momentum_vs_mass} are alternative versions to Fig.~\ref{fig:angular_momentum} and Fig.~\ref{fig:specific_momentum}. The former shows the angular momentum parameter within $1~\reff$ and the latter the specific angular momentum $j$ against stellar mass instead of $B$-band magnitude. In some of the studies shown in Fig.~\ref{fig:specific_momentum_vs_mass} the stellar masses were not stated, and we proceeded as follows. For the \citet{Toloba_2015} dEs we use the stellar masses obtained from \citet{Tortora_2019}. For the \citet{Martinez_2021} dSphs we adopted stellar masses from \citet{Hayashi_2020}. For the Local Group dEs \citep[][]{Geha_2006,Geha_2010} we used values from \citet{Mateo_1998}. For the `ordinary' ETGs of \citet{Bender_1990} we estimate a mass from the $B$-band magnitudes and an assumed mass-to-light ratio of $5$; for their dwarfs we again used the values from \citet{Mateo_1998} and \citet{Hayashi_2020}. For the dEs of \citet{Geha_2003} we convert from magnitudes and assume the median mass-to-light ratio that was given for a sub-sample of these dEs \citep[see][]{Geha_2002}. To estimate the stellar masses of \citet{Emsellem_2011} we use eq.~28 of \citet{Cappellari_2013} and the values stated therein

\bibliography{literature}{}

\begin{thebibliography}{}
\expandafter\ifx\csname natexlab\endcsname\relax\def\natexlab#1{#1}\fi
\providecommand{\url}[1]{\href{#1}{#1}}
\providecommand{\dodoi}[1]{doi:~\href{http://doi.org/#1}{\nolinkurl{#1}}}
\providecommand{\doeprint}[1]{\href{http://ascl.net/#1}{\nolinkurl{http://ascl.net/#1}}}
\providecommand{\doarXiv}[1]{\href{https://arxiv.org/abs/#1}{\nolinkurl{https://arxiv.org/abs/#1}}}

\bibitem[{{Ahn} {et~al.}(2012){Ahn}, {Alexandroff}, {Allende Prieto}, {Anderson}, {Anderton}, {Andrews}, {Aubourg}, {Bailey}, {Balbinot}, {Barnes}, {Bautista}, {Beers}, {Beifiori}, {Berlind}, {Bhardwaj}, {Bizyaev}, {Blake}, {Blanton}, {Blomqvist}, {Bochanski}, {Bolton}, {Borde}, {Bovy}, {Brandt}, {Brinkmann}, {Brown}, {Brownstein}, {Bundy}, {Busca}, {Carithers}, {Carnero}, {Carr}, {Casetti-Dinescu}, {Chen}, {Chiappini}, {Comparat}, {Connolly}, {Crepp}, {Cristiani}, {Croft}, {Cuesta}, {da Costa}, {Davenport}, {Dawson}, {de Putter}, {De Lee}, {Delubac}, {Dhital}, {Ealet}, {Ebelke}, {Edmondson}, {Eisenstein}, {Escoffier}, {Esposito}, {Evans}, {Fan}, {Femen{\'\i}a Castell{\'a}}, {Fern{\'a}ndez Alvar}, {Ferreira}, {Filiz Ak}, {Finley}, {Fleming}, {Font-Ribera}, {Frinchaboy}, {Garc{\'\i}a-Hern{\'a}ndez}, {Garc{\'\i}a P{\'e}rez}, {Ge}, {G{\'e}nova-Santos}, {Gillespie}, {Girardi}, {Gonz{\'a}lez Hern{\'a}ndez}, {Grebel}, {Gunn}, {Guo}, {Haggard}, {Hamilton}, {Harris}, {Hawley}, {Hearty}, {Ho}, {Hogg}, {Holtzman},
  {Honscheid}, {Huehnerhoff}, {Ivans}, {Ivezi{\'c}}, {Jacobson}, {Jiang}, {Johansson}, {Johnson}, {Kauffmann}, {Kirkby}, {Kirkpatrick}, {Klaene}, {Knapp}, {Kneib}, {Le Goff}, {Leauthaud}, {Lee}, {Lee}, {Long}, {Loomis}, {Lucatello}, {Lundgren}, {Lupton}, {Ma}, {Ma}, {MacDonald}, {Mack}, {Mahadevan}, {Maia}, {Majewski}, {Makler}, {Malanushenko}, {Malanushenko}, {Manchado}, {Mandelbaum}, {Manera}, {Maraston}, {Margala}, {Martell}, {McBride}, {McGreer}, {McMahon}, {M{\'e}nard}, {Meszaros}, {Miralda-Escud{\'e}}, {Montero-Dorta}, {Montesano}, {Morrison}, {Muna}, {Munn}, {Murayama}, {Myers}, {Neto}, {Nguyen}, {Nichol}, {Nidever}, {Noterdaeme}, {Nuza}, {Ogando}, {Olmstead}, {Oravetz}, {Owen}, {Padmanabhan}, {Palanque-Delabrouille}, {Pan}, {Parejko}, {Parihar}, {P{\^a}ris}, {Pattarakijwanich}, {Pepper}, {Percival}, {P{\'e}rez-Fournon}, {P{\'e}rez-R{\`a}fols}, {Petitjean}, {Pforr}, {Pieri}, {Pinsonneault}, {Porto de Mello}, {Prada}, {Price-Whelan}, {Raddick}, {Rebolo}, {Rich}, {Richards}, {Robin}, {Rocha-Pinto},
  {Rockosi}, {Roe}, {Ross}, {Ross}, {Rossi}, {Rubi{\~n}o-Martin}, {Samushia}, {Sanchez Almeida}, {S{\'a}nchez}, {Santiago}, {Sayres}, {Schlegel}, {Schlesinger}, {Schmidt}, {Schneider}, {Schultheis}, {Schwope}, {Sc{\'o}ccola}, {Seljak}, {Sheldon}, {Shen}, {Shu}, {Simmerer}, {Simmons}, {Skibba}, {Skrutskie}, {Slosar}, {Sobreira}, {Sobeck}, {Stassun}, {Steele}, {Steinmetz}, {Strauss}, {Streblyanska}, {Suzuki}, {Swanson}, {Tal}, {Thakar}, {Thomas}, {Thompson}, {Tinker}, {Tojeiro}, {Tremonti}, {Vargas Maga{\~n}a}, {Verde}, {Viel}, {Vikas}, {Vogt}, {Wake}, {Wang}, {Weaver}, {Weinberg}, {Weiner}, {West}, {White}, {Wilson}, {Wisniewski}, {Wood-Vasey}, {Yanny}, {Y{\`e}che}, {York}, {Zamora}, {Zasowski}, {Zehavi}, {Zhao}, {Zheng}, {Zhu}, \& {Zinn}}]{Ahn_2012}
{Ahn}, C.~P., {Alexandroff}, R., {Allende Prieto}, C., {et~al.} 2012, \apjs, 203, 21, \dodoi{10.1088/0067-0049/203/2/21}

\bibitem[{{Astropy Collaboration} {et~al.}(2022){Astropy Collaboration}, {Price-Whelan}, {Lim}, {Earl}, {Starkman}, {Bradley}, {Shupe}, {Patil}, {Corrales}, {Brasseur}, {N{\"o}the}, {Donath}, {Tollerud}, {Morris}, {Ginsburg}, {Vaher}, {Weaver}, {Tocknell}, {Jamieson}, {van Kerkwijk}, {Robitaille}, {Merry}, {Bachetti}, {G{\"u}nther}, {Aldcroft}, {Alvarado-Montes}, {Archibald}, {B{\'o}di}, {Bapat}, {Barentsen}, {Baz{\'a}n}, {Biswas}, {Boquien}, {Burke}, {Cara}, {Cara}, {Conroy}, {Conseil}, {Craig}, {Cross}, {Cruz}, {D'Eugenio}, {Dencheva}, {Devillepoix}, {Dietrich}, {Eigenbrot}, {Erben}, {Ferreira}, {Foreman-Mackey}, {Fox}, {Freij}, {Garg}, {Geda}, {Glattly}, {Gondhalekar}, {Gordon}, {Grant}, {Greenfield}, {Groener}, {Guest}, {Gurovich}, {Handberg}, {Hart}, {Hatfield-Dodds}, {Homeier}, {Hosseinzadeh}, {Jenness}, {Jones}, {Joseph}, {Kalmbach}, {Karamehmetoglu}, {Ka{\l}uszy{\'n}ski}, {Kelley}, {Kern}, {Kerzendorf}, {Koch}, {Kulumani}, {Lee}, {Ly}, {Ma}, {MacBride}, {Maljaars}, {Muna}, {Murphy}, {Norman},
  {O'Steen}, {Oman}, {Pacifici}, {Pascual}, {Pascual-Granado}, {Patil}, {Perren}, {Pickering}, {Rastogi}, {Roulston}, {Ryan}, {Rykoff}, {Sabater}, {Sakurikar}, {Salgado}, {Sanghi}, {Saunders}, {Savchenko}, {Schwardt}, {Seifert-Eckert}, {Shih}, {Jain}, {Shukla}, {Sick}, {Simpson}, {Singanamalla}, {Singer}, {Singhal}, {Sinha}, {Sip{\H{o}}cz}, {Spitler}, {Stansby}, {Streicher}, {{\v{S}}umak}, {Swinbank}, {Taranu}, {Tewary}, {Tremblay}, {de Val-Borro}, {Van Kooten}, {Vasovi{\'c}}, {Verma}, {de Miranda Cardoso}, {Williams}, {Wilson}, {Winkel}, {Wood-Vasey}, {Xue}, {Yoachim}, {Zhang}, {Zonca}, \& {Astropy Project Contributors}}]{astropy_2022}
{Astropy Collaboration}, {Price-Whelan}, A.~M., {Lim}, P.~L., {et~al.} 2022, \apj, 935, 167, \dodoi{10.3847/1538-4357/ac7c74}

\bibitem[{Audet \& Dennis(2006)}]{Audet_2006}
Audet, C., \& Dennis, J.~E. 2006, SIAM Journal on Optimization, 17, 188, \dodoi{10.1137/040603371}

\bibitem[{{Barazza} {et~al.}(2002){Barazza}, {Binggeli}, \& {Jerjen}}]{Barazza_2002}
{Barazza}, F.~D., {Binggeli}, B., \& {Jerjen}, H. 2002, \aap, 391, 823, \dodoi{10.1051/0004-6361:20020875}

\bibitem[{{Barnes} \& {Hernquist}(1992)}]{Barnes_1992}
{Barnes}, J.~E., \& {Hernquist}, L. 1992, \nat, 360, 715, \dodoi{10.1038/360715a0}

\bibitem[{{Bastian} {et~al.}(2010){Bastian}, {Covey}, \& {Meyer}}]{Bastian_2010}
{Bastian}, N., {Covey}, K.~R., \& {Meyer}, M.~R. 2010, \araa, 48, 339, \dodoi{10.1146/annurev-astro-082708-101642}

\bibitem[{{Beasley} {et~al.}(2009){Beasley}, {Cenarro}, {Strader}, \& {Brodie}}]{Beasley_2009}
{Beasley}, M.~A., {Cenarro}, A.~J., {Strader}, J., \& {Brodie}, J.~P. 2009, \aj, 137, 5146, \dodoi{10.1088/0004-6256/137/6/5146}

\bibitem[{{Bender}(1988)}]{Bender_1988_B}
{Bender}, R. 1988, \aap, 193, L7

\bibitem[{{Bender} {et~al.}(1992){Bender}, {Burstein}, \& {Faber}}]{Bender_1992}
{Bender}, R., {Burstein}, D., \& {Faber}, S.~M. 1992, \apj, 399, 462, \dodoi{10.1086/171940}

\bibitem[{{Bender} \& {Nieto}(1990)}]{Bender_1990}
{Bender}, R., \& {Nieto}, J.~L. 1990, \aap, 239, 97

\bibitem[{{Bender} {et~al.}(1991){Bender}, {Paquet}, \& {Nieto}}]{Bender_1991}
{Bender}, R., {Paquet}, A., \& {Nieto}, J.~L. 1991, \aap, 246, 349

\bibitem[{{Bender} {et~al.}(1994){Bender}, {Saglia}, \& {Gerhard}}]{Bender_1994}
{Bender}, R., {Saglia}, R.~P., \& {Gerhard}, O.~E. 1994, \mnras, 269, 785, \dodoi{10.1093/mnras/269.3.785}

\bibitem[{{Bender} {et~al.}(1989){Bender}, {Surma}, {Doebereiner}, {Moellenhoff}, \& {Madejsky}}]{Bender_1989}
{Bender}, R., {Surma}, P., {Doebereiner}, S., {Moellenhoff}, C., \& {Madejsky}, R. 1989, \aap, 217, 35

\bibitem[{{Bidaran} {et~al.}(2020){Bidaran}, {Pasquali}, {Lisker}, {Coccato}, {Falc{\'o}n-Barroso}, {van de Ven}, {Peletier}, {Emsellem}, {Grebel}, {La Barbera}, {Janz}, {Sybilska}, {Vijayaraghavan}, {Gallagher}, \& {Gadotti}}]{Bidaran_2020}
{Bidaran}, B., {Pasquali}, A., {Lisker}, T., {et~al.} 2020, \mnras, 497, 1904, \dodoi{10.1093/mnras/staa2097}

\bibitem[{{Bidaran} {et~al.}(2022){Bidaran}, {La Barbera}, {Pasquali}, {Peletier}, {van de Ven}, {Grebel}, {Falc{\'o}n-Barroso}, {Sybilska}, {Gadotti}, \& {Coccato}}]{Bidaran_2022}
{Bidaran}, B., {La Barbera}, F., {Pasquali}, A., {et~al.} 2022, \mnras, 515, 4622, \dodoi{10.1093/mnras/stac2005}

\bibitem[{{Bidaran} {et~al.}(2023){Bidaran}, {La Barbera}, {Pasquali}, {van de Ven}, {Peletier}, {Falc{\'o}n-Barroso}, {Gadotti}, {Sybilska}, \& {Grebel}}]{Bidaran_2023}
---. 2023, \mnras, 525, 4329, \dodoi{10.1093/mnras/stad2546}

\bibitem[{{Binggeli} {et~al.}(1985){Binggeli}, {Sandage}, \& {Tammann}}]{Binggeli_1985}
{Binggeli}, B., {Sandage}, A., \& {Tammann}, G.~A. 1985, \aj, 90, 1681, \dodoi{10.1086/113874}

\bibitem[{{Binggeli} {et~al.}(1987){Binggeli}, {Tammann}, \& {Sandage}}]{Binggeli_1987}
{Binggeli}, B., {Tammann}, G.~A., \& {Sandage}, A. 1987, \aj, 94, 251, \dodoi{10.1086/114467}

\bibitem[{{Binney} \& {Tremaine}(2008)}]{Binney_2008}
{Binney}, J., \& {Tremaine}, S. 2008, {Galactic Dynamics: Second Edition} ({Princeton University Press})

\bibitem[{{Blakeslee} {et~al.}(2009){Blakeslee}, {Jord{\'a}n}, {Mei}, {C{\^o}t{\'e}}, {Ferrarese}, {Infante}, {Peng}, {Tonry}, \& {West}}]{Blakeslee_2009}
{Blakeslee}, J.~P., {Jord{\'a}n}, A., {Mei}, S., {et~al.} 2009, \apj, 694, 556, \dodoi{10.1088/0004-637X/694/1/556}

\bibitem[{{Boch} \& {Fernique}(2014)}]{Boch_2014}
{Boch}, T., \& {Fernique}, P. 2014, in Astronomical Society of the Pacific Conference Series, Vol. 485, Astronomical Data Analysis Software and Systems XXIII, ed. N.~{Manset} \& P.~{Forshay}, 277

\bibitem[{{Boselli} {et~al.}(2022){Boselli}, {Fossati}, \& {Sun}}]{Bosselli_2022}
{Boselli}, A., {Fossati}, M., \& {Sun}, M. 2022, \aapr, 30, 3, \dodoi{10.1007/s00159-022-00140-3}

\bibitem[{{Brewer} {et~al.}(2012){Brewer}, {Dutton}, {Treu}, {Auger}, {Marshall}, {Barnab{\`e}}, {Bolton}, {Koo}, \& {Koopmans}}]{Brewer_2012}
{Brewer}, B.~J., {Dutton}, A.~A., {Treu}, T., {et~al.} 2012, \mnras, 422, 3574, \dodoi{10.1111/j.1365-2966.2012.20870.x}

\bibitem[{Burnham \& Anderson(2002)}]{burnham_2002}
Burnham, K., \& Anderson, D. 2002, Model selection and multimodel inference: a practical information-theoretic approach (Springer Verlag)

\bibitem[{{Butler} {et~al.}(2017){Butler}, {Obreschkow}, \& {Oh}}]{Butler_2017}
{Butler}, K.~M., {Obreschkow}, D., \& {Oh}, S.-H. 2017, \apjl, 834, L4, \dodoi{10.3847/2041-8213/834/1/L4}

\bibitem[{{Caldwell} {et~al.}(2003){Caldwell}, {Rose}, \& {Concannon}}]{Caldwell_2003}
{Caldwell}, N., {Rose}, J.~A., \& {Concannon}, K.~D. 2003, \aj, 125, 2891, \dodoi{10.1086/375308}

\bibitem[{{Cappellari}(2016)}]{Cappellari_2016}
{Cappellari}, M. 2016, \araa, 54, 597, \dodoi{10.1146/annurev-astro-082214-122432}

\bibitem[{{Cappellari} \& {Copin}(2003)}]{Cappellari_2003}
{Cappellari}, M., \& {Copin}, Y. 2003, \mnras, 342, 345, \dodoi{10.1046/j.1365-8711.2003.06541.x}

\bibitem[{{Cappellari} {et~al.}(2007){Cappellari}, {Emsellem}, {Bacon}, {Bureau}, {Davies}, {de Zeeuw}, {Falc{\'o}n-Barroso}, {Krajnovi{\'c}}, {Kuntschner}, {McDermid}, {Peletier}, {Sarzi}, {van den Bosch}, \& {van de Ven}}]{Cappellari_2007}
{Cappellari}, M., {Emsellem}, E., {Bacon}, R., {et~al.} 2007, \mnras, 379, 418, \dodoi{10.1111/j.1365-2966.2007.11963.x}

\bibitem[{{Cappellari} {et~al.}(2011){Cappellari}, {Emsellem}, {Krajnovi{\'c}}, {McDermid}, {Scott}, {Verdoes Kleijn}, {Young}, {Alatalo}, {Bacon}, {Blitz}, {Bois}, {Bournaud}, {Bureau}, {Davies}, {Davis}, {de Zeeuw}, {Duc}, {Khochfar}, {Kuntschner}, {Lablanche}, {Morganti}, {Naab}, {Oosterloo}, {Sarzi}, {Serra}, \& {Weijmans}}]{Cappellari_2011}
{Cappellari}, M., {Emsellem}, E., {Krajnovi{\'c}}, D., {et~al.} 2011, \mnras, 413, 813, \dodoi{10.1111/j.1365-2966.2010.18174.x}

\bibitem[{{Cappellari} {et~al.}(2013{\natexlab{a}}){Cappellari}, {Scott}, {Alatalo}, {Blitz}, {Bois}, {Bournaud}, {Bureau}, {Crocker}, {Davies}, {Davis}, {de Zeeuw}, {Duc}, {Emsellem}, {Khochfar}, {Krajnovi{\'c}}, {Kuntschner}, {McDermid}, {Morganti}, {Naab}, {Oosterloo}, {Sarzi}, {Serra}, {Weijmans}, \& {Young}}]{Cappellari_2013}
{Cappellari}, M., {Scott}, N., {Alatalo}, K., {et~al.} 2013{\natexlab{a}}, \mnras, 432, 1709, \dodoi{10.1093/mnras/stt562}

\bibitem[{{Cappellari} {et~al.}(2013{\natexlab{b}}){Cappellari}, {McDermid}, {Alatalo}, {Blitz}, {Bois}, {Bournaud}, {Bureau}, {Crocker}, {Davies}, {Davis}, {de Zeeuw}, {Duc}, {Emsellem}, {Khochfar}, {Krajnovi{\'c}}, {Kuntschner}, {Morganti}, {Naab}, {Oosterloo}, {Sarzi}, {Scott}, {Serra}, {Weijmans}, \& {Young}}]{Cappellari_2013_B}
{Cappellari}, M., {McDermid}, R.~M., {Alatalo}, K., {et~al.} 2013{\natexlab{b}}, \mnras, 432, 1862, \dodoi{10.1093/mnras/stt644}

\bibitem[{{Chabrier}(2003)}]{Chabrier_2003}
{Chabrier}, G. 2003, \pasp, 115, 763, \dodoi{10.1086/376392}

\bibitem[{Chilingarian(2009)}]{Chilingarian_2009}
Chilingarian, I.~V. 2009, Monthly Notices of the Royal Astronomical Society, 394, 1229, \dodoi{10.1111/j.1365-2966.2009.14450.x}

\bibitem[{{Chon} {et~al.}(2024){Chon}, {Hosokawa}, {Omukai}, \& {Schneider}}]{Chon_2024}
{Chon}, S., {Hosokawa}, T., {Omukai}, K., \& {Schneider}, R. 2024, \mnras, 530, 2453, \dodoi{10.1093/mnras/stae1027}

\bibitem[{{Chon} {et~al.}(2022){Chon}, {Ono}, {Omukai}, \& {Schneider}}]{Chon_2022}
{Chon}, S., {Ono}, H., {Omukai}, K., \& {Schneider}, R. 2022, \mnras, 514, 4639, \dodoi{10.1093/mnras/stac1549}

\bibitem[{{Chowdhury} \& {Chengalur}(2017)}]{Chowdhury_2017}
{Chowdhury}, A., \& {Chengalur}, J.~N. 2017, \mnras, 467, 3856, \dodoi{10.1093/mnras/stx355}

\bibitem[{{{\c{S}}en} {et~al.}(2018){{\c{S}}en}, {Peletier}, {Boselli}, {den Brok}, {Falc{\'o}n-Barroso}, {Hensler}, {Janz}, {Laurikainen}, {Lisker}, {Mentz}, {Paudel}, {Salo}, {Sybilska}, {Toloba}, {van de Ven}, {Vazdekis}, \& {Yesilyaprak}}]{Sen_2018}
{{\c{S}}en}, {\c{S}}., {Peletier}, R.~F., {Boselli}, A., {et~al.} 2018, \mnras, 475, 3453, \dodoi{10.1093/mnras/stx3254}

\bibitem[{{De Leo} {et~al.}(2023){De Leo}, {Read}, {Noel}, {Erkal}, {Massana}, \& {Carrera}}]{De_Leo_2023}
{De Leo}, M., {Read}, J.~I., {Noel}, N. E.~D., {et~al.} 2023, arXiv e-prints, arXiv:2303.08838, \dodoi{10.48550/arXiv.2303.08838}

\bibitem[{{de los Reyes} {et~al.}(2023){de los Reyes}, {Kirby}, {Zhuang}, {Steidel}, {Chen}, \& {Wheeler}}]{de_los_Reyes_2023}
{de los Reyes}, M. A.~C., {Kirby}, E.~N., {Zhuang}, Z., {et~al.} 2023, \apj, 951, 52, \dodoi{10.3847/1538-4357/acd189}

\bibitem[{{de Nicola} {et~al.}(2020){de Nicola}, {Saglia}, {Thomas}, {Dehnen}, \& {Bender}}]{deNicola_2020}
{de Nicola}, S., {Saglia}, R.~P., {Thomas}, J., {Dehnen}, W., \& {Bender}, R. 2020, \mnras, 496, 3076, \dodoi{10.1093/mnras/staa1703}

\bibitem[{{De Rijcke} {et~al.}(2006){De Rijcke}, {Prugniel}, {Simien}, \& {Dejonghe}}]{De_Rijcke_2006}
{De Rijcke}, S., {Prugniel}, P., {Simien}, F., \& {Dejonghe}, H. 2006, \mnras, 369, 1321, \dodoi{10.1111/j.1365-2966.2006.10377.x}

\bibitem[{{Dekel} \& {Birnboim}(2006)}]{Dekel_2006}
{Dekel}, A., \& {Birnboim}, Y. 2006, \mnras, 368, 2, \dodoi{10.1111/j.1365-2966.2006.10145.x}

\bibitem[{{Di Teodoro} {et~al.}(2023){Di Teodoro}, {Posti}, {Fall}, {Ogle}, {Jarrett}, {Appleton}, {Cluver}, {Haynes}, \& {Lisenfeld}}]{Teodoro_2023}
{Di Teodoro}, E.~M., {Posti}, L., {Fall}, S.~M., {et~al.} 2023, \mnras, 518, 6340, \dodoi{10.1093/mnras/stac3424}

\bibitem[{{Ding} {et~al.}(2023){Ding}, {Zhu}, {van de Ven}, {Coccato}, {Corsini}, {Costantin}, {Fahrion}, {Falc{\'o}n-Barroso}, {Gadotti}, {Iodice}, {Lyubenova}, {Mart{\'\i}n-Navarro}, {McDermid}, {Pinna}, \& {Sarzi}}]{Ding_2023}
{Ding}, Y., {Zhu}, L., {van de Ven}, G., {et~al.} 2023, \aap, 672, A84, \dodoi{10.1051/0004-6361/202244558}

\bibitem[{{Dressler}(1980)}]{Dressler_1980}
{Dressler}, A. 1980, \apj, 236, 351, \dodoi{10.1086/157753}

\bibitem[{{Efstathiou} \& {Jones}(1979)}]{Efstathiou_1979}
{Efstathiou}, G., \& {Jones}, B.~J.~T. 1979, \mnras, 186, 133, \dodoi{10.1093/mnras/186.2.133}

\bibitem[{{Eftekhari} {et~al.}(2022){Eftekhari}, {Peletier}, {Scott}, {Mieske}, {Bland-Hawthorn}, {Bryant}, {Cantiello}, {Croom}, {Drinkwater}, {Falc{\'o}n-Barroso}, {Hilker}, {Iodice}, {Napolitano}, {Spavone}, {Valentijn}, {van de Ven}, \& {Venhola}}]{Eftekhari_2022}
{Eftekhari}, F.~S., {Peletier}, R.~F., {Scott}, N., {et~al.} 2022, \mnras, 517, 4714, \dodoi{10.1093/mnras/stac2606}

\bibitem[{{Emsellem} {et~al.}(2007){Emsellem}, {Cappellari}, {Krajnovi{\'c}}, {van de Ven}, {Bacon}, {Bureau}, {Davies}, {de Zeeuw}, {Falc{\'o}n-Barroso}, {Kuntschner}, {McDermid}, {Peletier}, \& {Sarzi}}]{Emsellem_2007}
{Emsellem}, E., {Cappellari}, M., {Krajnovi{\'c}}, D., {et~al.} 2007, \mnras, 379, 401, \dodoi{10.1111/j.1365-2966.2007.11752.x}

\bibitem[{{Emsellem} {et~al.}(2011){Emsellem}, {Cappellari}, {Krajnovi{\'c}}, {Alatalo}, {Blitz}, {Bois}, {Bournaud}, {Bureau}, {Davies}, {Davis}, {de Zeeuw}, {Khochfar}, {Kuntschner}, {Lablanche}, {McDermid}, {Morganti}, {Naab}, {Oosterloo}, {Sarzi}, {Scott}, {Serra}, {van de Ven}, {Weijmans}, \& {Young}}]{Emsellem_2011}
---. 2011, \mnras, 414, 888, \dodoi{10.1111/j.1365-2966.2011.18496.x}

\bibitem[{{Ene} {et~al.}(2018){Ene}, {Ma}, {Veale}, {Greene}, {Thomas}, {Blakeslee}, {Foster}, {Walsh}, {Ito}, \& {Goulding}}]{Ene_2018}
{Ene}, I., {Ma}, C.-P., {Veale}, M., {et~al.} 2018, \mnras, 479, 2810, \dodoi{10.1093/mnras/sty1649}

\bibitem[{{Fabricius} {et~al.}(2008){Fabricius}, {Barnes}, {Bender}, {Drory}, {Grupp}, {Hill}, {Hopp}, \& {MacQueen}}]{Fabricius_2008}
{Fabricius}, M.~H., {Barnes}, S., {Bender}, R., {et~al.} 2008, in Society of Photo-Optical Instrumentation Engineers (SPIE) Conference Series, Vol. 7014, Ground-based and Airborne Instrumentation for Astronomy II, ed. I.~S. {McLean} \& M.~M. {Casali}, 701473, \dodoi{10.1117/12.787204}

\bibitem[{{Fabricius} {et~al.}(2012){Fabricius}, {Grupp}, {Bender}, {Drory}, {Arns}, {Barnes}, {G{\"o}ssl}, {Snigula}, {Hill}, {Hopp}, {Lang-Bardl}, {MacQueen}, {Saglia}, \& {Wullstein}}]{Fabricius_2012}
{Fabricius}, M.~H., {Grupp}, F., {Bender}, R., {et~al.} 2012, in Society of Photo-Optical Instrumentation Engineers (SPIE) Conference Series, Vol. 8446, Ground-based and Airborne Instrumentation for Astronomy IV, ed. I.~S. {McLean}, S.~K. {Ramsay}, \& H.~{Takami}, 84465K, \dodoi{10.1117/12.925177}

\bibitem[{{Falc{\'o}n-Barroso} \& {Martig}(2021)}]{Falcon_Barroso_2021}
{Falc{\'o}n-Barroso}, J., \& {Martig}, M. 2021, \aap, 646, A31, \dodoi{10.1051/0004-6361/202039624}

\bibitem[{{Falc{\'o}n-Barroso} {et~al.}(2019){Falc{\'o}n-Barroso}, {van de Ven}, {Lyubenova}, {Mendez-Abreu}, {Aguerri}, {Garc{\'\i}a-Lorenzo}, {Bekerait{\'e}}, {S{\'a}nchez}, {Husemann}, {Garc{\'\i}a-Benito}, {Gonz{\'a}lez Delgado}, {Mast}, {Walcher}, {Zibetti}, {Zhu}, {Barrera-Ballesteros}, {Galbany}, {S{\'a}nchez-Bl{\'a}zquez}, {Singh}, {van den Bosch}, {Wild}, {Bland-Hawthorn}, {Cid Fernandes}, {de Lorenzo-C{\'a}ceres}, {Gallazzi}, {Marino}, {M{\'a}rquez}, {Peletier}, {P{\'e}rez}, {P{\'e}rez}, {Roth}, {Rosales-Ortega}, {Ruiz-Lara}, {Wisotzki}, \& {Ziegler}}]{Falcon_Barroso_2019}
{Falc{\'o}n-Barroso}, J., {van de Ven}, G., {Lyubenova}, M., {et~al.} 2019, \aap, 632, A59, \dodoi{10.1051/0004-6361/201936413}

\bibitem[{{Fall}(1983)}]{Fall_1983}
{Fall}, S.~M. 1983, in Internal Kinematics and Dynamics of Galaxies, ed. E.~{Athanassoula}, Vol. 100, 391--398

\bibitem[{{Ferrarese} {et~al.}(2006){Ferrarese}, {C{\^o}t{\'e}}, {Jord{\'a}n}, {Peng}, {Blakeslee}, {Piatek}, {Mei}, {Merritt}, {Milosavljevi{\'c}}, {Tonry}, \& {West}}]{Ferrarese_2006}
{Ferrarese}, L., {C{\^o}t{\'e}}, P., {Jord{\'a}n}, A., {et~al.} 2006, \apjs, 164, 334, \dodoi{10.1086/501350}

\bibitem[{{Ferrarese} {et~al.}(2012){Ferrarese}, {C{\^o}t{\'e}}, {Cuillandre}, {Gwyn}, {Peng}, {MacArthur}, {Duc}, {Boselli}, {Mei}, {Erben}, {McConnachie}, {Durrell}, {Mihos}, {Jord{\'a}n}, {Lan{\c{c}}on}, {Puzia}, {Emsellem}, {Balogh}, {Blakeslee}, {van Waerbeke}, {Gavazzi}, {Vollmer}, {Kavelaars}, {Woods}, {Ball}, {Boissier}, {Courteau}, {Ferriere}, {Gavazzi}, {Hildebrandt}, {Hudelot}, {Huertas-Company}, {Liu}, {McLaughlin}, {Mellier}, {Milkeraitis}, {Schade}, {Balkowski}, {Bournaud}, {Carlberg}, {Chapman}, {Hoekstra}, {Peng}, {Sawicki}, {Simard}, {Taylor}, {Tully}, {van Driel}, {Wilson}, {Burdullis}, {Mahoney}, \& {Manset}}]{Ferrarese_2012}
{Ferrarese}, L., {C{\^o}t{\'e}}, P., {Cuillandre}, J.-C., {et~al.} 2012, \apjs, 200, 4, \dodoi{10.1088/0067-0049/200/1/4}

\bibitem[{Ferreras {et~al.}(2012)Ferreras, Barbera, Rosa, Vazdekis, Carvalho, Falcón-Barroso, \& Ricciardelli}]{Ferreras_2013}
Ferreras, I., Barbera, F.~L., Rosa, I. G. d.~l., {et~al.} 2012, Monthly Notices of the Royal Astronomical Society: Letters, 429, L15, \dodoi{10.1093/mnrasl/sls014}

\bibitem[{{Frigo} {et~al.}(2021){Frigo}, {Naab}, {Rantala}, {Johansson}, {Neureiter}, {Thomas}, \& {Rizzuto}}]{Frigo_2021}
{Frigo}, M., {Naab}, T., {Rantala}, A., {et~al.} 2021, \mnras, 508, 4610, \dodoi{10.1093/mnras/stab2754}

\bibitem[{{Fujita}(2004)}]{Fujita_2004}
{Fujita}, Y. 2004, \pasj, 56, 29, \dodoi{10.1093/pasj/56.1.29}

\bibitem[{{Geha} {et~al.}(2012){Geha}, {Blanton}, {Yan}, \& {Tinker}}]{Geha_2012}
{Geha}, M., {Blanton}, M.~R., {Yan}, R., \& {Tinker}, J.~L. 2012, \apj, 757, 85, \dodoi{10.1088/0004-637X/757/1/85}

\bibitem[{{Geha} {et~al.}(2006){Geha}, {Guhathakurta}, {Rich}, \& {Cooper}}]{Geha_2006}
{Geha}, M., {Guhathakurta}, P., {Rich}, R.~M., \& {Cooper}, M.~C. 2006, \aj, 131, 332, \dodoi{10.1086/498686}

\bibitem[{{Geha} {et~al.}(2002){Geha}, {Guhathakurta}, \& {van der Marel}}]{Geha_2002}
{Geha}, M., {Guhathakurta}, P., \& {van der Marel}, R.~P. 2002, \aj, 124, 3073, \dodoi{10.1086/344764}

\bibitem[{{Geha} {et~al.}(2003){Geha}, {Guhathakurta}, \& {van der Marel}}]{Geha_2003}
---. 2003, \aj, 126, 1794, \dodoi{10.1086/377624}

\bibitem[{{Geha} {et~al.}(2010){Geha}, {van der Marel}, {Guhathakurta}, {Gilbert}, {Kalirai}, \& {Kirby}}]{Geha_2010}
{Geha}, M., {van der Marel}, R.~P., {Guhathakurta}, P., {et~al.} 2010, \apj, 711, 361, \dodoi{10.1088/0004-637X/711/1/361}

\bibitem[{{Gerhard} {et~al.}(2001){Gerhard}, {Kronawitter}, {Saglia}, \& {Bender}}]{Gerhard_2001}
{Gerhard}, O., {Kronawitter}, A., {Saglia}, R.~P., \& {Bender}, R. 2001, \aj, 121, 1936, \dodoi{10.1086/319940}

\bibitem[{{Gerhard} \& {Binney}(1996)}]{Gerhard_1996}
{Gerhard}, O.~E., \& {Binney}, J.~J. 1996, MNRAS, 279, 993, \dodoi{10.1093/mnras/279.3.993}

\bibitem[{{Goessl} {et~al.}(2006){Goessl}, {Drory}, {Relke}, {Gebhardt}, {Grupp}, {Hill}, {Hopp}, {K{\"o}hler}, \& {MacQueen}}]{Goessl_2006}
{Goessl}, C.~A., {Drory}, N., {Relke}, H., {et~al.} 2006, in Society of Photo-Optical Instrumentation Engineers (SPIE) Conference Series, Vol. 6270, Society of Photo-Optical Instrumentation Engineers (SPIE) Conference Series, ed. D.~R. {Silva} \& R.~E. {Doxsey}, 627021, \dodoi{10.1117/12.671264}

\bibitem[{{G{\"o}ssl} \& {Riffeser}(2002)}]{Goessl_2002}
{G{\"o}ssl}, C.~A., \& {Riffeser}, A. 2002, \aap, 381, 1095, \dodoi{10.1051/0004-6361:20011522}

\bibitem[{{Graham} {et~al.}(2018){Graham}, {Cappellari}, {Li}, {Mao}, {Bershady}, {Bizyaev}, {Brinkmann}, {Brownstein}, {Bundy}, {Drory}, {Law}, {Pan}, {Thomas}, {Wake}, {Weijmans}, {Westfall}, \& {Yan}}]{Graham_2018}
{Graham}, M.~T., {Cappellari}, M., {Li}, H., {et~al.} 2018, \mnras, 477, 4711, \dodoi{10.1093/mnras/sty504}

\bibitem[{{Gunn} \& {Gott}(1972)}]{Gunn_1972}
{Gunn}, J.~E., \& {Gott}, J.~Richard, I. 1972, \apj, 176, 1, \dodoi{10.1086/151605}

\bibitem[{{Guo} {et~al.}(2020){Guo}, {Cortese}, {Obreschkow}, {Catinella}, {van de Sande}, {Croom}, {Brough}, {Sweet}, {Bryant}, {Medling}, {Bland-Hawthorn}, {Owers}, \& {Richards}}]{Guo_2020}
{Guo}, K., {Cortese}, L., {Obreschkow}, D., {et~al.} 2020, \mnras, 491, 773, \dodoi{10.1093/mnras/stz3042}

\bibitem[{{Hamraz} {et~al.}(2019){Hamraz}, {Peletier}, {Khosroshahi}, {Valentijn}, {den Brok}, \& {Venhola}}]{Hamraz_2019}
{Hamraz}, E., {Peletier}, R.~F., {Khosroshahi}, H.~G., {et~al.} 2019, \aap, 625, A94, \dodoi{10.1051/0004-6361/201935076}

\bibitem[{{Hayashi} {et~al.}(2020){Hayashi}, {Chiba}, \& {Ishiyama}}]{Hayashi_2020}
{Hayashi}, K., {Chiba}, M., \& {Ishiyama}, T. 2020, \apj, 904, 45, \dodoi{10.3847/1538-4357/abbe0a}

\bibitem[{{Hill} {et~al.}(2004){Hill}, {Gebhardt}, {Komatsu}, \& {MacQueen}}]{Hill_2004}
{Hill}, G.~J., {Gebhardt}, K., {Komatsu}, E., \& {MacQueen}, P.~J. 2004, in American Institute of Physics Conference Series, Vol. 743, The New Cosmology: Conference on Strings and Cosmology, ed. R.~E. {Allen}, D.~V. {Nanopoulos}, \& C.~N. {Pope}, 224--233, \dodoi{10.1063/1.1848329}

\bibitem[{{Hill} {et~al.}(2021){Hill}, {Lee}, {MacQueen}, {Kelz}, {Drory}, {Vattiat}, {Good}, {Ramsey}, {Kriel}, {Peterson}, {DePoy}, {Gebhardt}, {Marshall}, {Tuttle}, {Bauer}, {Chonis}, {Fabricius}, {Froning}, {H{\"a}user}, {Indahl}, {Jahn}, {Landriau}, {Leck}, {Montesano}, {Prochaska}, {Snigula}, {Zeimann}, {Bryant}, {Damm}, {Fowler}, {Janowiecki}, {Martin}, {Mrozinski}, {Odewahn}, {Rostopchin}, {Shetrone}, {Spencer}, {Mentuch Cooper}, {Armandroff}, {Bender}, {Dalton}, {Hopp}, {Komatsu}, {Nicklas}, {Ramsey}, {Roth}, {Schneider}, {Sneden}, \& {Steinmetz}}]{Hill_2021}
{Hill}, G.~J., {Lee}, H., {MacQueen}, P.~J., {et~al.} 2021, \aj, 162, 298, \dodoi{10.3847/1538-3881/ac2c02}

\bibitem[{{Houghton} {et~al.}(2006){Houghton}, {Magorrian}, {Sarzi}, {Thatte}, {Davies}, \& {Krajnovi{\'c}}}]{Houghton_2006}
{Houghton}, R.~C.~W., {Magorrian}, J., {Sarzi}, M., {et~al.} 2006, \mnras, 367, 2, \dodoi{10.1111/j.1365-2966.2005.09713.x}

\bibitem[{{Janz} {et~al.}(2017){Janz}, {Penny}, {Graham}, {Forbes}, \& {Davies}}]{Janz_2017}
{Janz}, J., {Penny}, S.~J., {Graham}, A.~W., {Forbes}, D.~A., \& {Davies}, R.~L. 2017, \mnras, 468, 2850, \dodoi{10.1093/mnras/stx634}

\bibitem[{{Jardel} \& {Gebhardt}(2012)}]{Jardel_2012}
{Jardel}, J.~R., \& {Gebhardt}, K. 2012, \apj, 746, 89, \dodoi{10.1088/0004-637X/746/1/89}

\bibitem[{{Jardel} {et~al.}(2013){Jardel}, {Gebhardt}, {Fabricius}, {Drory}, \& {Williams}}]{Jardel_2013_a}
{Jardel}, J.~R., {Gebhardt}, K., {Fabricius}, M.~H., {Drory}, N., \& {Williams}, M.~J. 2013, \apj, 763, 91, \dodoi{10.1088/0004-637X/763/2/91}

\bibitem[{{Jerjen} {et~al.}(2000){Jerjen}, {Kalnajs}, \& {Binggeli}}]{Jerjen_2000}
{Jerjen}, H., {Kalnajs}, A., \& {Binggeli}, B. 2000, \aap, 358, 845.
\newblock \doarXiv{astro-ph/0004248}

\bibitem[{{Jin} {et~al.}(2020){Jin}, {Zhu}, {Long}, {Mao}, {Wang}, \& {van de Ven}}]{Jin_2020}
{Jin}, Y., {Zhu}, L., {Long}, R.~J., {et~al.} 2020, \mnras, 491, 1690, \dodoi{10.1093/mnras/stz3072}

\bibitem[{{Jin} {et~al.}(2019){Jin}, {Zhu}, {Long}, {Mao}, {Xu}, {Li}, \& {van de Ven}}]{Jin_2019}
---. 2019, \mnras, 486, 4753, \dodoi{10.1093/mnras/stz1170}

\bibitem[{{Klimentowski} {et~al.}(2009){Klimentowski}, {{\L}okas}, {Kazantzidis}, {Mayer}, \& {Mamon}}]{Klimentowski_2009}
{Klimentowski}, J., {{\L}okas}, E.~L., {Kazantzidis}, S., {Mayer}, L., \& {Mamon}, G.~A. 2009, \mnras, 397, 2015, \dodoi{10.1111/j.1365-2966.2009.15046.x}

\bibitem[{{Kormendy}(1985)}]{Kormendy_1985}
{Kormendy}, J. 1985, \apj, 295, 73, \dodoi{10.1086/163350}

\bibitem[{{Kormendy}(1999)}]{Kormendy_1999}
{Kormendy}, J. 1999, in Astronomical Society of the Pacific Conference Series, Vol. 182, Galaxy Dynamics - A Rutgers Symposium, ed. D.~R. {Merritt}, M.~{Valluri}, \& J.~A. {Sellwood}, 124

\bibitem[{{Kormendy} \& {Bender}(1996)}]{Kormendy_1996}
{Kormendy}, J., \& {Bender}, R. 1996, \apjl, 464, L119, \dodoi{10.1086/310095}

\bibitem[{{Kormendy} \& {Bender}(2009)}]{Kormendy_2009_B}
---. 2009, \apjl, 691, L142, \dodoi{10.1088/0004-637X/691/2/L142}

\bibitem[{{Kormendy} \& {Bender}(2012)}]{Kormendy_2012}
---. 2012, \apjs, 198, 2, \dodoi{10.1088/0067-0049/198/1/2}

\bibitem[{{Kormendy} \& {Bender}(2013)}]{Kormendy_2013}
---. 2013, \apjl, 769, L5, \dodoi{10.1088/2041-8205/769/1/L5}

\bibitem[{{Kormendy} {et~al.}(2009){Kormendy}, {Fisher}, {Cornell}, \& {Bender}}]{Kormendy_2009}
{Kormendy}, J., {Fisher}, D.~B., {Cornell}, M.~E., \& {Bender}, R. 2009, \apjs, 182, 216, \dodoi{10.1088/0067-0049/182/1/216}

\bibitem[{{Kormendy} \& {Freeman}(2016)}]{Kormendy_2016}
{Kormendy}, J., \& {Freeman}, K.~C. 2016, \apj, 817, 84, \dodoi{10.3847/0004-637X/817/2/84}

\bibitem[{{Kowalczyk} \& {{\L}okas}(2022)}]{Kowalczyk_2022}
{Kowalczyk}, K., \& {{\L}okas}, E.~L. 2022, \aap, 659, A119, \dodoi{10.1051/0004-6361/202142212}

\bibitem[{{Kroupa}(2001)}]{Kroupa_2001}
{Kroupa}, P. 2001, \mnras, 322, 231, \dodoi{10.1046/j.1365-8711.2001.04022.x}

\bibitem[{{Kroupa}(2002)}]{Kroupa_2002}
---. 2002, Science, 295, 82, \dodoi{10.1126/science.1067524}

\bibitem[{{Kurapati} {et~al.}(2018){Kurapati}, {Chengalur}, {Pustilnik}, \& {Kamphuis}}]{Kurapati_2018}
{Kurapati}, S., {Chengalur}, J.~N., {Pustilnik}, S., \& {Kamphuis}, P. 2018, \mnras, 479, 228, \dodoi{10.1093/mnras/sty1397}

\bibitem[{{La Barbera} {et~al.}(2013){La Barbera}, {Ferreras}, {Vazdekis}, {de la Rosa}, {de Carvalho}, {Trevisan}, {Falc{\'o}n-Barroso}, \& {Ricciardelli}}]{La_Barbera_2013}
{La Barbera}, F., {Ferreras}, I., {Vazdekis}, A., {et~al.} 2013, \mnras, 433, 3017, \dodoi{10.1093/mnras/stt943}

\bibitem[{{Larson}(2005)}]{Larson_2005}
{Larson}, R.~B. 2005, \mnras, 359, 211, \dodoi{10.1111/j.1365-2966.2005.08881.x}

\bibitem[{{Larson} {et~al.}(1980){Larson}, {Tinsley}, \& {Caldwell}}]{Larson_1980}
{Larson}, R.~B., {Tinsley}, B.~M., \& {Caldwell}, C.~N. 1980, \apj, 237, 692, \dodoi{10.1086/157917}

\bibitem[{Le~Digabel(2011)}]{Digabel_2022}
Le~Digabel, S. 2011, ACM Trans. Math. Softw., 37, \dodoi{10.1145/1916461.1916468}

\bibitem[{{Li} {et~al.}(2023){Li}, {Liu}, {Zhang}, {Tian}, {Fu}, {Li}, \& {Yan}}]{Li_2023}
{Li}, J., {Liu}, C., {Zhang}, Z.-Y., {et~al.} 2023, \nat, 613, 460, \dodoi{10.1038/s41586-022-05488-1}

\bibitem[{{Liepold} {et~al.}(2020){Liepold}, {Quenneville}, {Ma}, {Walsh}, {McConnell}, {Greene}, \& {Blakeslee}}]{Liepold_2020}
{Liepold}, C.~M., {Quenneville}, M.~E., {Ma}, C.-P., {et~al.} 2020, \apj, 891, 4, \dodoi{10.3847/1538-4357/ab6f71}

\bibitem[{{Lin} \& {Faber}(1983)}]{Lin_1983}
{Lin}, D.~N.~C., \& {Faber}, S.~M. 1983, \apjl, 266, L21, \dodoi{10.1086/183971}

\bibitem[{{Lipka} \& {Thomas}(2021)}]{Lipka_2021}
{Lipka}, M., \& {Thomas}, J. 2021, \mnras, 504, 4599, \dodoi{10.1093/mnras/stab1092}

\bibitem[{{Lisker} {et~al.}(2006{\natexlab{a}}){Lisker}, {Glatt}, {Westera}, \& {Grebel}}]{Lisker_2006_b}
{Lisker}, T., {Glatt}, K., {Westera}, P., \& {Grebel}, E.~K. 2006{\natexlab{a}}, \aj, 132, 2432, \dodoi{10.1086/508414}

\bibitem[{{Lisker} {et~al.}(2006{\natexlab{b}}){Lisker}, {Grebel}, \& {Binggeli}}]{Lisker_2006_a}
{Lisker}, T., {Grebel}, E.~K., \& {Binggeli}, B. 2006{\natexlab{b}}, \aj, 132, 497, \dodoi{10.1086/505045}

\bibitem[{{Lisker} {et~al.}(2007){Lisker}, {Grebel}, {Binggeli}, \& {Glatt}}]{Lisker_2007}
{Lisker}, T., {Grebel}, E.~K., {Binggeli}, B., \& {Glatt}, K. 2007, \apj, 660, 1186, \dodoi{10.1086/513090}

\bibitem[{{{\L}okas} {et~al.}(2010){{\L}okas}, {Kazantzidis}, {Klimentowski}, {Mayer}, \& {Callegari}}]{Lokas_2010}
{{\L}okas}, E.~L., {Kazantzidis}, S., {Klimentowski}, J., {Mayer}, L., \& {Callegari}, S. 2010, \apj, 708, 1032, \dodoi{10.1088/0004-637X/708/2/1032}

\bibitem[{{Magorrian}(1999)}]{Magorrian_1999}
{Magorrian}, J. 1999, \mnras, 302, 530, \dodoi{10.1046/j.1365-8711.1999.02135.x}

\bibitem[{{Makarov} {et~al.}(2014){Makarov}, {Prugniel}, {Terekhova}, {Courtois}, \& {Vauglin}}]{Makarov_2014}
{Makarov}, D., {Prugniel}, P., {Terekhova}, N., {Courtois}, H., \& {Vauglin}, I. 2014, \aap, 570, A13, \dodoi{10.1051/0004-6361/201423496}

\bibitem[{{Mancera Pi{\~n}a} {et~al.}(2021){Mancera Pi{\~n}a}, {Posti}, {Fraternali}, {Adams}, \& {Oosterloo}}]{Mancera_Pina_2021}
{Mancera Pi{\~n}a}, P.~E., {Posti}, L., {Fraternali}, F., {Adams}, E. A.~K., \& {Oosterloo}, T. 2021, \aap, 647, A76, \dodoi{10.1051/0004-6361/202039340}

\bibitem[{{Maraston}(1998)}]{Maraston_1998}
{Maraston}, C. 1998, \mnras, 300, 872, \dodoi{10.1046/j.1365-8711.1998.01947.x}

\bibitem[{{Maraston}(2005)}]{Maraston_2005}
---. 2005, \mnras, 362, 799, \dodoi{10.1111/j.1365-2966.2005.09270.x}

\bibitem[{{Mart{\'\i}n-Navarro} {et~al.}(2015){Mart{\'\i}n-Navarro}, {Vazdekis}, {La Barbera}, {Falc{\'o}n-Barroso}, {Lyubenova}, {van de Ven}, {Ferreras}, {S{\'a}nchez}, {Trager}, {Garc{\'\i}a-Benito}, {Mast}, {Mendoza}, {S{\'a}nchez-Bl{\'a}zquez}, {Gonz{\'a}lez Delgado}, {Walcher}, \& {CALIFA Team}}]{Martin_Navarro_2015}
{Mart{\'\i}n-Navarro}, I., {Vazdekis}, A., {La Barbera}, F., {et~al.} 2015, \apjl, 806, L31, \dodoi{10.1088/2041-8205/806/2/L31}

\bibitem[{{Mart{\'\i}n-Navarro} {et~al.}(2023){Mart{\'\i}n-Navarro}, {Spiniello}, {Tortora}, {Coccato}, {D'Ago}, {Ferr{\'e}-Mateu}, {Pulsoni}, {Hartke}, {Arnaboldi}, {Hunt}, {Napolitano}, {Scognamiglio}, \& {Spavone}}]{Navarro_2023}
{Mart{\'\i}n-Navarro}, I., {Spiniello}, C., {Tortora}, C., {et~al.} 2023, \mnras, 521, 1408, \dodoi{10.1093/mnras/stad503}

\bibitem[{{Mart{\'\i}nez-Garc{\'\i}a} {et~al.}(2021){Mart{\'\i}nez-Garc{\'\i}a}, {del Pino}, {Aparicio}, {van der Marel}, \& {Watkins}}]{Martinez_2021}
{Mart{\'\i}nez-Garc{\'\i}a}, A.~M., {del Pino}, A., {Aparicio}, A., {van der Marel}, R.~P., \& {Watkins}, L.~L. 2021, \mnras, 505, 5884, \dodoi{10.1093/mnras/stab1568}

\bibitem[{{Mateo}(1998)}]{Mateo_1998}
{Mateo}, M.~L. 1998, \araa, 36, 435, \dodoi{10.1146/annurev.astro.36.1.435}

\bibitem[{{Mayer} {et~al.}(2001){Mayer}, {Governato}, {Colpi}, {Moore}, {Quinn}, {Wadsley}, {Stadel}, \& {Lake}}]{Mayer_2001}
{Mayer}, L., {Governato}, F., {Colpi}, M., {et~al.} 2001, \apj, 559, 754, \dodoi{10.1086/322356}

\bibitem[{{McDermid} {et~al.}(2015){McDermid}, {Alatalo}, {Blitz}, {Bournaud}, {Bureau}, {Cappellari}, {Crocker}, {Davies}, {Davis}, {de Zeeuw}, {Duc}, {Emsellem}, {Khochfar}, {Krajnovi{\'c}}, {Kuntschner}, {Morganti}, {Naab}, {Oosterloo}, {Sarzi}, {Scott}, {Serra}, {Weijmans}, \& {Young}}]{McDermid_2015}
{McDermid}, R.~M., {Alatalo}, K., {Blitz}, L., {et~al.} 2015, \mnras, 448, 3484, \dodoi{10.1093/mnras/stv105}

\bibitem[{{Mehrgan} {et~al.}(2019){Mehrgan}, {Thomas}, {Saglia}, {Mazzalay}, {Erwin}, {Bender}, {Kluge}, \& {Fabricius}}]{Mehrgan_2019}
{Mehrgan}, K., {Thomas}, J., {Saglia}, R., {et~al.} 2019, \apj, 887, 195, \dodoi{10.3847/1538-4357/ab5856}

\bibitem[{{Mehrgan} {et~al.}(2023){Mehrgan}, {Thomas}, {Saglia}, {Parikh}, \& {Bender}}]{Mehrgan_2023}
{Mehrgan}, K., {Thomas}, J., {Saglia}, R., {Parikh}, T., \& {Bender}, R. 2023, \apj, 948, 79, \dodoi{10.3847/1538-4357/acbf2e}

\bibitem[{{Mehrgan} {et~al.}(2024){Mehrgan}, {Thomas}, {Saglia}, {Parikh}, {Neureiter}, {Erwin}, \& {Bender}}]{Mehrgan_2024}
{Mehrgan}, K., {Thomas}, J., {Saglia}, R., {et~al.} 2024, \apj, 961, 127, \dodoi{10.3847/1538-4357/acfe09}

\bibitem[{{Michielsen} {et~al.}(2008){Michielsen}, {Boselli}, {Conselice}, {Toloba}, {Whiley}, {Arag{\'o}n-Salamanca}, {Balcells}, {Cardiel}, {Cenarro}, {Gorgas}, {Peletier}, \& {Vazdekis}}]{Michielsen_2008}
{Michielsen}, D., {Boselli}, A., {Conselice}, C.~J., {et~al.} 2008, \mnras, 385, 1374, \dodoi{10.1111/j.1365-2966.2008.12846.x}

\bibitem[{{Moore} {et~al.}(1998){Moore}, {Governato}, {Quinn}, {Stadel}, \& {Lake}}]{Moore_1998}
{Moore}, B., {Governato}, F., {Quinn}, T., {Stadel}, J., \& {Lake}, G. 1998, \apjl, 499, L5, \dodoi{10.1086/311333}

\bibitem[{{Napolitano} {et~al.}(2010){Napolitano}, {Romanowsky}, \& {Tortora}}]{Napolitano_2010}
{Napolitano}, N.~R., {Romanowsky}, A.~J., \& {Tortora}, C. 2010, \mnras, 405, 2351, \dodoi{10.1111/j.1365-2966.2010.16710.x}

\bibitem[{{Nelson} {et~al.}(2018){Nelson}, {Pillepich}, {Springel}, {Weinberger}, {Hernquist}, {Pakmor}, {Genel}, {Torrey}, {Vogelsberger}, {Kauffmann}, {Marinacci}, \& {Naiman}}]{Nelson_2018}
{Nelson}, D., {Pillepich}, A., {Springel}, V., {et~al.} 2018, \mnras, 475, 624, \dodoi{10.1093/mnras/stx3040}

\bibitem[{{Neureiter} {et~al.}(2023{\natexlab{a}}){Neureiter}, {de Nicola}, {Thomas}, {Saglia}, {Bender}, \& {Rantala}}]{Neureiter_2023_a}
{Neureiter}, B., {de Nicola}, S., {Thomas}, J., {et~al.} 2023{\natexlab{a}}, \mnras, 519, 2004, \dodoi{10.1093/mnras/stac3652}

\bibitem[{{Neureiter} {et~al.}(2023{\natexlab{b}}){Neureiter}, {Thomas}, {Rantala}, {Naab}, {Mehrgan}, {Saglia}, {de Nicola}, \& {Bender}}]{Neureiter_2023_b}
{Neureiter}, B., {Thomas}, J., {Rantala}, A., {et~al.} 2023{\natexlab{b}}, \apj, 950, 15, \dodoi{10.3847/1538-4357/accffa}

\bibitem[{{Ocvirk} {et~al.}(2006){Ocvirk}, {Pichon}, {Lan{\c{c}}on}, \& {Thi{\'e}baut}}]{Ocvirk_2006}
{Ocvirk}, P., {Pichon}, C., {Lan{\c{c}}on}, A., \& {Thi{\'e}baut}, E. 2006, \mnras, 365, 74, \dodoi{10.1111/j.1365-2966.2005.09323.x}

\bibitem[{{Ohlson} {et~al.}(2024){Ohlson}, {Seth}, {Gallo}, {Baldassare}, \& {Greene}}]{Ohlson_2024}
{Ohlson}, D., {Seth}, A.~C., {Gallo}, E., {Baldassare}, V.~F., \& {Greene}, J.~E. 2024, \aj, 167, 31, \dodoi{10.3847/1538-3881/acf7bc}

\bibitem[{{Oldham} \& {Auger}(2018)}]{Oldham_2018}
{Oldham}, L., \& {Auger}, M. 2018, \mnras, 474, 4169, \dodoi{10.1093/mnras/stx2969}

\bibitem[{{Parikh} {et~al.}(2024){Parikh}, {Saglia}, {Thomas}, {Mehrgan}, {Bender}, \& {Maraston}}]{Parikh_2024}
{Parikh}, T., {Saglia}, R., {Thomas}, J., {et~al.} 2024, arXiv e-prints, arXiv:2402.06628, \dodoi{10.48550/arXiv.2402.06628}

\bibitem[{{Parikh} {et~al.}(2018){Parikh}, {Thomas}, {Maraston}, {Westfall}, {Goddard}, {Lian}, {Meneses-Goytia}, {Jones}, {Vaughan}, {Andrews}, {Bershady}, {Bizyaev}, {Brinkmann}, {Brownstein}, {Bundy}, {Drory}, {Emsellem}, {Law}, {Newman}, {Roman-Lopes}, {Wake}, {Yan}, \& {Zheng}}]{Parikh_2018}
{Parikh}, T., {Thomas}, D., {Maraston}, C., {et~al.} 2018, \mnras, 477, 3954, \dodoi{10.1093/mnras/sty785}

\bibitem[{{Paturel} {et~al.}(2003){Paturel}, {Petit}, {Prugniel}, {Theureau}, {Rousseau}, {Brouty}, {Dubois}, \& {Cambr{\'e}sy}}]{Paturel_2003}
{Paturel}, G., {Petit}, C., {Prugniel}, P., {et~al.} 2003, \aap, 412, 45, \dodoi{10.1051/0004-6361:20031411}

\bibitem[{{Paudel} {et~al.}(2011){Paudel}, {Lisker}, \& {Kuntschner}}]{Paudel_2011}
{Paudel}, S., {Lisker}, T., \& {Kuntschner}, H. 2011, \mnras, 413, 1764, \dodoi{10.1111/j.1365-2966.2011.18256.x}

\bibitem[{{Paudel} {et~al.}(2010){Paudel}, {Lisker}, {Kuntschner}, {Grebel}, \& {Glatt}}]{Paudel_2010}
{Paudel}, S., {Lisker}, T., {Kuntschner}, H., {Grebel}, E.~K., \& {Glatt}, K. 2010, \mnras, 405, 800, \dodoi{10.1111/j.1365-2966.2010.16507.x}

\bibitem[{{Paudel} \& {Ree}(2014)}]{Paudel_2014}
{Paudel}, S., \& {Ree}, C.~H. 2014, \apjl, 796, L14, \dodoi{10.1088/2041-8205/796/1/L14}

\bibitem[{{Paudel} {et~al.}(2023){Paudel}, {Yoon}, {Yoo}, {Smith}, {Chhatkuli}, {Kumar Bachchan}, {Aryal}, {Adhikari}, {Adhikari}, {Sedain}, {Sheikh}, {Dhital}, {Giri}, \& {Baral}}]{Paudel_2023}
{Paudel}, S., {Yoon}, S.-J., {Yoo}, J., {et~al.} 2023, \apjs, 265, 57, \dodoi{10.3847/1538-4365/acbfa7}

\bibitem[{{Peebles}(1969)}]{Peebles_1969}
{Peebles}, P.~J.~E. 1969, \apj, 155, 393, \dodoi{10.1086/149876}

\bibitem[{{Poci} {et~al.}(2019){Poci}, {McDermid}, {Zhu}, \& {van de Ven}}]{Poci_2019}
{Poci}, A., {McDermid}, R.~M., {Zhu}, L., \& {van de Ven}, G. 2019, \mnras, 487, 3776, \dodoi{10.1093/mnras/stz1154}

\bibitem[{Posacki {et~al.}(2015)Posacki, Cappellari, Treu, Pellegrini, \& Ciotti}]{Posacki_2014}
Posacki, S., Cappellari, M., Treu, T., Pellegrini, S., \& Ciotti, L. 2015, Monthly Notices of the Royal Astronomical Society, 446, 493, \dodoi{10.1093/mnras/stu2098}

\bibitem[{{Posti} {et~al.}(2018){Posti}, {Fraternali}, {Di Teodoro}, \& {Pezzulli}}]{Posti_2018}
{Posti}, L., {Fraternali}, F., {Di Teodoro}, E.~M., \& {Pezzulli}, G. 2018, \aap, 612, L6, \dodoi{10.1051/0004-6361/201833091}

\bibitem[{{Prugniel} {et~al.}(2007{\natexlab{a}}){Prugniel}, {Koleva}, {Ocvirk}, {Le Borgne}, \& {Soubiran}}]{Prugniel_2007_B}
{Prugniel}, P., {Koleva}, M., {Ocvirk}, P., {Le Borgne}, D., \& {Soubiran}, C. 2007{\natexlab{a}}, in Stellar Populations as Building Blocks of Galaxies, ed. A.~{Vazdekis} \& R.~{Peletier}, Vol. 241, 68--72, \dodoi{10.1017/S1743921307007454}

\bibitem[{{Prugniel} \& {Soubiran}(2001)}]{Prugniel_2001}
{Prugniel}, P., \& {Soubiran}, C. 2001, \aap, 369, 1048, \dodoi{10.1051/0004-6361:20010163}

\bibitem[{{Prugniel} {et~al.}(2007{\natexlab{b}}){Prugniel}, {Soubiran}, {Koleva}, \& {Le Borgne}}]{Prugniel_2007}
{Prugniel}, P., {Soubiran}, C., {Koleva}, M., \& {Le Borgne}, D. 2007{\natexlab{b}}, arXiv e-prints, astro.
\newblock \doarXiv{astro-ph/0703658}

\bibitem[{{Pulsoni} {et~al.}(2023){Pulsoni}, {Gerhard}, {Fall}, {Arnaboldi}, {Ennis}, {Hartke}, {Coccato}, \& {Napolitano}}]{Pulsoni_2023}
{Pulsoni}, C., {Gerhard}, O., {Fall}, S.~M., {et~al.} 2023, \aap, 674, A96, \dodoi{10.1051/0004-6361/202346234}

\bibitem[{{Quenneville} {et~al.}(2022){Quenneville}, {Liepold}, \& {Ma}}]{Quenneville_2022}
{Quenneville}, M.~E., {Liepold}, C.~M., \& {Ma}, C.-P. 2022, \apj, 926, 30, \dodoi{10.3847/1538-4357/ac3e68}

\bibitem[{{Rantala} {et~al.}(2018){Rantala}, {Johansson}, {Naab}, {Thomas}, \& {Frigo}}]{Rantala_2018}
{Rantala}, A., {Johansson}, P.~H., {Naab}, T., {Thomas}, J., \& {Frigo}, M. 2018, \apj, 864, 113, \dodoi{10.3847/1538-4357/aada47}

\bibitem[{{Rix} {et~al.}(1992){Rix}, {Franx}, {Fisher}, \& {Illingworth}}]{Rix_1992}
{Rix}, H.-W., {Franx}, M., {Fisher}, D., \& {Illingworth}, G. 1992, \apjl, 400, L5, \dodoi{10.1086/186635}

\bibitem[{{Romanowsky} \& {Fall}(2012)}]{Romanowsky_2012}
{Romanowsky}, A.~J., \& {Fall}, S.~M. 2012, \apjs, 203, 17, \dodoi{10.1088/0067-0049/203/2/17}

\bibitem[{{Romero-G{\'o}mez} {et~al.}(2024){Romero-G{\'o}mez}, {J.}, {Peletier}, {Aguerri}, \& {Smith}}]{Romero_Gomez_2024}
{Romero-G{\'o}mez}, {J.}, {Peletier}, R.~F., {Aguerri}, J.~A.~L., \& {Smith}, R. 2024, arXiv e-prints, arXiv:2404.15519, \dodoi{10.48550/arXiv.2404.15519}

\bibitem[{Romero-Gómez {et~al.}(2023{\natexlab{a}})Romero-Gómez, Aguerri, Peletier, Mieske, van~de Ven, \& Falcón-Barroso}]{Romero_Gomez_2023_B}
Romero-Gómez, J., Aguerri, J. A.~L., Peletier, R.~F., {et~al.} 2023{\natexlab{a}}, Monthly Notices of the Royal Astronomical Society, 527, 9715, \dodoi{10.1093/mnras/stad3801}

\bibitem[{Romero-Gómez {et~al.}(2023{\natexlab{b}})Romero-Gómez, Peletier, Aguerri, Mieske, Scott, Bland-Hawthorn, Bryant, Croom, Eftekhari, Falcón-Barroso, Hilker, van~de Ven, \& Venhola}]{Romero_Gomez_2023_A}
Romero-Gómez, J., Peletier, R.~F., Aguerri, J. A.~L., {et~al.} 2023{\natexlab{b}}, Monthly Notices of the Royal Astronomical Society, 522, 130, \dodoi{10.1093/mnras/stad953}

\bibitem[{{Rubin} {et~al.}(1992){Rubin}, {Graham}, \& {Kenney}}]{Rubin_1992}
{Rubin}, V.~C., {Graham}, J.~A., \& {Kenney}, J. D.~P. 1992, \apjl, 394, L9, \dodoi{10.1086/186460}

\bibitem[{{Rybicki}(1987)}]{Rybicki_1987}
{Rybicki}, G.~B. 1987, in IAU Symposium, Vol. 127, Structure and Dynamics of Elliptical Galaxies, ed. P.~T. {de Zeeuw}, 397, \dodoi{10.1007/978-94-009-3971-4_41}

\bibitem[{{Ry{\'s}} {et~al.}(2013){Ry{\'s}}, {Falc{\'o}n-Barroso}, \& {van de Ven}}]{Rys_2013}
{Ry{\'s}}, A., {Falc{\'o}n-Barroso}, J., \& {van de Ven}, G. 2013, \mnras, 428, 2980, \dodoi{10.1093/mnras/sts245}

\bibitem[{{Ry{\'s}} {et~al.}(2015){Ry{\'s}}, {Koleva}, {Falc{\'o}n-Barroso}, {Vazdekis}, {Lisker}, {Peletier}, \& {van de Ven}}]{Rys_2015}
{Ry{\'s}}, A., {Koleva}, M., {Falc{\'o}n-Barroso}, J., {et~al.} 2015, \mnras, 452, 1888, \dodoi{10.1093/mnras/stv1364}

\bibitem[{{Salpeter}(1955)}]{Salpeter_1955}
{Salpeter}, E.~E. 1955, \apj, 121, 161, \dodoi{10.1086/145971}

\bibitem[{{Sandage} {et~al.}(1985){Sandage}, {Binggeli}, \& {Tammann}}]{Sandage_1985}
{Sandage}, A., {Binggeli}, B., \& {Tammann}, G.~A. 1985, \aj, 90, 1759, \dodoi{10.1086/113875}

\bibitem[{{Santucci} {et~al.}(2022){Santucci}, {Brough}, {van de Sande}, {McDermid}, {van de Ven}, {Zhu}, {D'Eugenio}, {Bland-Hawthorn}, {Barsanti}, {Bryant}, {Croom}, {Davies}, {Green}, {Lawrence}, {Lorente}, {Owers}, {Poci}, {Richards}, {Thater}, \& {Yi}}]{Santucci_2022}
{Santucci}, G., {Brough}, S., {van de Sande}, J., {et~al.} 2022, \apj, 930, 153, \dodoi{10.3847/1538-4357/ac5bd5}

\bibitem[{{Santucci} {et~al.}(2023){Santucci}, {Brough}, {van de Sande}, {McDermid}, {Barsanti}, {Bland-Hawthorn}, {Bryant}, {Croom}, {Lagos}, {Lawrence}, {Owers}, {van de Ven}, {Vaughan}, \& {Yi}}]{Santucci_2023}
---. 2023, \mnras, 521, 2671, \dodoi{10.1093/mnras/stad713}

\bibitem[{{Schechter} \& {Gunn}(1978)}]{Schechter_1978}
{Schechter}, P.~L., \& {Gunn}, J.~E. 1978, \aj, 83, 1360, \dodoi{10.1086/112324}

\bibitem[{{Schwarzschild}(1979)}]{Schwarzschild_1979}
{Schwarzschild}, M. 1979, \apj, 232, 236, \dodoi{10.1086/157282}

\bibitem[{{Scorza} \& {Bender}(1995)}]{Scorza_1995}
{Scorza}, C., \& {Bender}, R. 1995, \aap, 293, 20

\bibitem[{{Scott} {et~al.}(2020){Scott}, {Eftekhari}, {Peletier}, {Bryant}, {Bland-Hawthorn}, {Capaccioli}, {Croom}, {Drinkwater}, {Falc{\'o}n-Barroso}, {Hilker}, {Iodice}, {Lorente}, {Mieske}, {Spavone}, {van de Ven}, \& {Venhola}}]{Scott_2020}
{Scott}, N., {Eftekhari}, F.~S., {Peletier}, R.~F., {et~al.} 2020, \mnras, 497, 1571, \dodoi{10.1093/mnras/staa2042}

\bibitem[{{Seo} \& {Ann}(2023)}]{Seo_2023}
{Seo}, M., \& {Ann}, H.~B. 2023, \mnras, 520, 5521, \dodoi{10.1093/mnras/stad425}

\bibitem[{{Simien} \& {Prugniel}(2002)}]{Simien_2002}
{Simien}, F., \& {Prugniel}, P. 2002, \aap, 384, 371, \dodoi{10.1051/0004-6361:20020071}

\bibitem[{{Sirianni} {et~al.}(2005){Sirianni}, {Jee}, {Ben{\'\i}tez}, {Blakeslee}, {Martel}, {Meurer}, {Clampin}, {De Marchi}, {Ford}, {Gilliland}, {Hartig}, {Illingworth}, {Mack}, \& {McCann}}]{Sirianni_2005}
{Sirianni}, M., {Jee}, M.~J., {Ben{\'\i}tez}, N., {et~al.} 2005, \pasp, 117, 1049, \dodoi{10.1086/444553}

\bibitem[{{Smith} {et~al.}(2012){Smith}, {Lucey}, \& {Carter}}]{Smith_2012}
{Smith}, R.~J., {Lucey}, J.~R., \& {Carter}, D. 2012, \mnras, 426, 2994, \dodoi{10.1111/j.1365-2966.2012.21922.x}

\bibitem[{{Spavone} {et~al.}(2022){Spavone}, {Iodice}, {D'Ago}, {van de Ven}, {Morelli}, {Corsini}, {Sarzi}, {Coccato}, {Fahrion}, {Falc{\'o}n-Barroso}, {Gadotti}, {Lyubenova}, {Mart{\'\i}n-Navarro}, {McDermid}, {Pinna}, {Pizzella}, {Poci}, {de Zeeuw}, \& {Zhu}}]{Spavone_2022}
{Spavone}, M., {Iodice}, E., {D'Ago}, G., {et~al.} 2022, \aap, 663, A135, \dodoi{10.1051/0004-6361/202243290}

\bibitem[{{Sybilska} {et~al.}(2017){Sybilska}, {Lisker}, {Kuntschner}, {Vazdekis}, {van de Ven}, {Peletier}, {Falc{\'o}n-Barroso}, {Vijayaraghavan}, \& {Janz}}]{Sybilska_2017}
{Sybilska}, A., {Lisker}, T., {Kuntschner}, H., {et~al.} 2017, \mnras, 470, 815, \dodoi{10.1093/mnras/stx1138}

\bibitem[{{Thomas} {et~al.}(2003){Thomas}, {Maraston}, \& {Bender}}]{Thomas_2003}
{Thomas}, D., {Maraston}, C., \& {Bender}, R. 2003, \mnras, 339, 897, \dodoi{10.1046/j.1365-8711.2003.06248.x}

\bibitem[{{Thomas} {et~al.}(2011{\natexlab{a}}){Thomas}, {Maraston}, \& {Johansson}}]{Thomas_Maraston_2011}
{Thomas}, D., {Maraston}, C., \& {Johansson}, J. 2011{\natexlab{a}}, \mnras, 412, 2183, \dodoi{10.1111/j.1365-2966.2010.18049.x}

\bibitem[{{Thomas} \& {Lipka}(2022)}]{Thomas_2022}
{Thomas}, J., \& {Lipka}, M. 2022, \mnras, 514, 6203, \dodoi{10.1093/mnras/stac1581}

\bibitem[{{Thomas} {et~al.}(2016){Thomas}, {Ma}, {McConnell}, {Greene}, {Blakeslee}, \& {Janish}}]{Thomas_2016}
{Thomas}, J., {Ma}, C.-P., {McConnell}, N.~J., {et~al.} 2016, \nat, 532, 340, \dodoi{10.1038/nature17197}

\bibitem[{{Thomas} {et~al.}(2014){Thomas}, {Saglia}, {Bender}, {Erwin}, \& {Fabricius}}]{Thomas_2014}
{Thomas}, J., {Saglia}, R.~P., {Bender}, R., {Erwin}, P., \& {Fabricius}, M. 2014, \apj, 782, 39, \dodoi{10.1088/0004-637X/782/1/39}

\bibitem[{{Thomas} {et~al.}(2007){Thomas}, {Saglia}, {Bender}, {Thomas}, {Gebhardt}, {Magorrian}, {Corsini}, \& {Wegner}}]{Thomas_2007}
{Thomas}, J., {Saglia}, R.~P., {Bender}, R., {et~al.} 2007, \mnras, 382, 657, \dodoi{10.1111/j.1365-2966.2007.12434.x}

\bibitem[{{Thomas} {et~al.}(2009{\natexlab{a}}){Thomas}, {Saglia}, {Bender}, {Thomas}, {Gebhardt}, {Magorrian}, {Corsini}, \& {Wegner}}]{Thomas_2009}
---. 2009{\natexlab{a}}, \apj, 691, 770, \dodoi{10.1088/0004-637X/691/1/770}

\bibitem[{{Thomas} {et~al.}(2004){Thomas}, {Saglia}, {Bender}, {Thomas}, {Gebhardt}, {Magorrian}, \& {Richstone}}]{Thomas_2004}
---. 2004, \mnras, 353, 391, \dodoi{10.1111/j.1365-2966.2004.08072.x}

\bibitem[{{Thomas} {et~al.}(2009{\natexlab{b}}){Thomas}, {Jesseit}, {Saglia}, {Bender}, {Burkert}, {Corsini}, {Gebhardt}, {Magorrian}, {Naab}, {Thomas}, \& {Wegner}}]{Thomas_2009_b}
{Thomas}, J., {Jesseit}, R., {Saglia}, R.~P., {et~al.} 2009{\natexlab{b}}, \mnras, 393, 641, \dodoi{10.1111/j.1365-2966.2008.14238.x}

\bibitem[{{Thomas} {et~al.}(2011{\natexlab{b}}){Thomas}, {Saglia}, {Bender}, {Thomas}, {Gebhardt}, {Magorrian}, {Corsini}, {Wegner}, \& {Seitz}}]{Thomas_2011}
{Thomas}, J., {Saglia}, R.~P., {Bender}, R., {et~al.} 2011{\natexlab{b}}, \mnras, 415, 545, \dodoi{10.1111/j.1365-2966.2011.18725.x}

\bibitem[{{Toloba} {et~al.}(2011){Toloba}, {Boselli}, {Cenarro}, {Peletier}, {Gorgas}, {Gil de Paz}, \& {Mu{\~n}oz-Mateos}}]{Toloba_2011}
{Toloba}, E., {Boselli}, A., {Cenarro}, A.~J., {et~al.} 2011, \aap, 526, A114, \dodoi{10.1051/0004-6361/201015344}

\bibitem[{{Toloba} {et~al.}(2014){Toloba}, {Guhathakurta}, {Peletier}, {Boselli}, {Lisker}, {Falc{\'o}n-Barroso}, {Simon}, {van de Ven}, {Paudel}, {Emsellem}, {Janz}, {den Brok}, {Gorgas}, {Hensler}, {Laurikainen}, {Niemi}, {Ry{\'s}}, \& {Salo}}]{Toloba_2014}
{Toloba}, E., {Guhathakurta}, P., {Peletier}, R.~F., {et~al.} 2014, \apjs, 215, 17, \dodoi{10.1088/0067-0049/215/2/17}

\bibitem[{{Toloba} {et~al.}(2015){Toloba}, {Guhathakurta}, {Boselli}, {Peletier}, {Emsellem}, {Lisker}, {van de Ven}, {Simon}, {Falc{\'o}n-Barroso}, {Adams}, {Benson}, {Boissier}, {den Brok}, {Gorgas}, {Hensler}, {Janz}, {Laurikainen}, {Paudel}, {Ry{\'s}}, \& {Salo}}]{Toloba_2015}
{Toloba}, E., {Guhathakurta}, P., {Boselli}, A., {et~al.} 2015, \apj, 799, 172, \dodoi{10.1088/0004-637X/799/2/172}

\bibitem[{{Tortora} {et~al.}(2016){Tortora}, {La Barbera}, \& {Napolitano}}]{Tortora_2016}
{Tortora}, C., {La Barbera}, F., \& {Napolitano}, N.~R. 2016, \mnras, 455, 308, \dodoi{10.1093/mnras/stv2250}

\bibitem[{{Tortora} {et~al.}(2014){Tortora}, {Napolitano}, {Saglia}, {Romanowsky}, {Covone}, \& {Capaccioli}}]{Tortora_2014}
{Tortora}, C., {Napolitano}, N.~R., {Saglia}, R.~P., {et~al.} 2014, \mnras, 445, 162, \dodoi{10.1093/mnras/stu1712}

\bibitem[{{Tortora} {et~al.}(2019){Tortora}, {Posti}, {Koopmans}, \& {Napolitano}}]{Tortora_2019}
{Tortora}, C., {Posti}, L., {Koopmans}, L.~V.~E., \& {Napolitano}, N.~R. 2019, \mnras, 489, 5483, \dodoi{10.1093/mnras/stz2320}

\bibitem[{{van de Sande} {et~al.}(2017){van de Sande}, {Bland-Hawthorn}, {Brough}, {Croom}, {Cortese}, {Foster}, {Scott}, {Bryant}, {d'Eugenio}, {Tonini}, {Goodwin}, {Konstantopoulos}, {Lawrence}, {Medling}, {Owers}, {Richards}, {Schaefer}, \& {Yi}}]{van_de_Sande_2017}
{van de Sande}, J., {Bland-Hawthorn}, J., {Brough}, S., {et~al.} 2017, \mnras, 472, 1272, \dodoi{10.1093/mnras/stx1751}

\bibitem[{{van den Bosch} \& {de Zeeuw}(2010)}]{Bosch_2010}
{van den Bosch}, R. C.~E., \& {de Zeeuw}, P.~T. 2010, \mnras, 401, 1770, \dodoi{10.1111/j.1365-2966.2009.15832.x}

\bibitem[{{van den Bosch} \& {van de Ven}(2009)}]{Bosch_2009}
{van den Bosch}, R. C.~E., \& {van de Ven}, G. 2009, \mnras, 398, 1117, \dodoi{10.1111/j.1365-2966.2009.15177.x}

\bibitem[{{van der Marel} \& {Franx}(1993)}]{Marel_1993}
{van der Marel}, R.~P., \& {Franx}, M. 1993, \apj, 407, 525, \dodoi{10.1086/172534}

\bibitem[{{van Dokkum} {et~al.}(2017){van Dokkum}, {Conroy}, {Villaume}, {Brodie}, \& {Romanowsky}}]{van_Dokkum_2017}
{van Dokkum}, P., {Conroy}, C., {Villaume}, A., {Brodie}, J., \& {Romanowsky}, A.~J. 2017, \apj, 841, 68, \dodoi{10.3847/1538-4357/aa7135}

\bibitem[{{van Dokkum}(2008)}]{Dokkum_2008}
{van Dokkum}, P.~G. 2008, \apj, 674, 29, \dodoi{10.1086/525014}

\bibitem[{{van Dokkum} \& {Conroy}(2010)}]{van_Dokkum_2010}
{van Dokkum}, P.~G., \& {Conroy}, C. 2010, \nat, 468, 940, \dodoi{10.1038/nature09578}

\bibitem[{{van Zee} {et~al.}(2004){van Zee}, {Skillman}, \& {Haynes}}]{van_Zee_2004}
{van Zee}, L., {Skillman}, E.~D., \& {Haynes}, M.~P. 2004, \aj, 128, 121, \dodoi{10.1086/421368}

\bibitem[{{Walsh} {et~al.}(2012){Walsh}, {van den Bosch}, {Barth}, \& {Sarzi}}]{Walsh_2012}
{Walsh}, J.~L., {van den Bosch}, R. C.~E., {Barth}, A.~J., \& {Sarzi}, M. 2012, \apj, 753, 79, \dodoi{10.1088/0004-637X/753/1/79}

\bibitem[{{Wang} {et~al.}(2020){Wang}, {Cappellari}, {Peng}, \& {Graham}}]{Wang_2020}
{Wang}, B., {Cappellari}, M., {Peng}, Y., \& {Graham}, M. 2020, \mnras, 495, 1958, \dodoi{10.1093/mnras/staa1325}

\bibitem[{{Williams} \& {Schwarzschild}(1979)}]{Williams_1979}
{Williams}, T.~B., \& {Schwarzschild}, M. 1979, \apjs, 41, 209, \dodoi{10.1086/190616}

\bibitem[{{Yang} {et~al.}(2014){Yang}, {Hammer}, {Fouquet}, {Flores}, {Puech}, {Pawlowski}, \& {Kroupa}}]{Yang_2014}
{Yang}, Y., {Hammer}, F., {Fouquet}, S., {et~al.} 2014, \mnras, 442, 2419, \dodoi{10.1093/mnras/stu931}

\bibitem[{{Zeng} {et~al.}(2024){Zeng}, {Wang}, {Gao}, \& {Yang}}]{Zeng_2024}
{Zeng}, G., {Wang}, L., {Gao}, L., \& {Yang}, H. 2024, arXiv e-prints, arXiv:2404.14184, \dodoi{10.48550/arXiv.2404.14184}

\bibitem[{{Zhao}(1996)}]{Zhao_1996}
{Zhao}, H. 1996, \mnras, 278, 488, \dodoi{10.1093/mnras/278.2.488}

\bibitem[{{Zhu} {et~al.}(2024){Zhu}, {Lu}, {Cappellari}, {Li}, {Mao}, {Gao}, \& {Ge}}]{Zhu_2024}
{Zhu}, K., {Lu}, S., {Cappellari}, M., {et~al.} 2024, \mnras, 527, 706, \dodoi{10.1093/mnras/stad3213}

\bibitem[{{Zhu} {et~al.}(2018){Zhu}, {van den Bosch}, {van de Ven}, {Lyubenova}, {Falc{\'o}n-Barroso}, {Meidt}, {Martig}, {Shen}, {Li}, {Yildirim}, {Walcher}, \& {Sanchez}}]{Zhu_2018}
{Zhu}, L., {van den Bosch}, R., {van de Ven}, G., {et~al.} 2018, \mnras, 473, 3000, \dodoi{10.1093/mnras/stx2409}

\end{thebibliography}
\bibliographystyle{aasjournal}

\end{document}